\documentclass[12pt]{article}
\usepackage[english]{babel}
\usepackage{graphicx}
\usepackage{bm}
\usepackage{setspace}
\usepackage{varioref}
\usepackage{booktabs}
\usepackage{rotating}
\usepackage{xfrac}
\usepackage{adjustbox}
\usepackage{longtable} % use for tables going over more than one page
\usepackage{ragged2e}
\usepackage[osf,sc]{mathpazo} % Use the Palatino font
\usepackage{microtype} % Slightly tweak font spacing for aesthetics
\usepackage[left=0.93in,right=0.93in,bottom=0.93in,top=0.93in]{geometry} % Document margins
\usepackage[hang, small,labelfont=bf,up,textfont=it,up]{caption} % Custom captions under/above floats in tables or figures
\usepackage{booktabs} % Horizontal rules in tables
\usepackage{float} % Required for tables and figures in the multi-column environment - they need to be placed in specific locations with the [H] (e.g. \begin{table}[H])
\usepackage[colorlinks=false]{hyperref}
\hypersetup{colorlinks, citecolor=blue, urlcolor=blue, linkcolor=blue} % For hyperlinks in the PDF
\usepackage{lettrine} % The lettrine is the first enlarged letter at the beginning of the text
\usepackage{paralist} % Used for the compactitem environment which makes bullet points with less space between them
\usepackage{enumitem} 
\usepackage{amsfonts}
\usepackage{amsmath}
\usepackage{amsthm}
\usepackage{mathtools}
\usepackage{hyperref}

\usepackage{subcaption} %allows combination of figures
\usepackage{abstract} % Allows abstract customization
\usepackage{url}
\usepackage{titlesec} % Allows customization of titles
\titleformat{\section}[block]{\large\scshape\centering}{\thesection.}{1em}{} % Change the look of the section titles
\titleformat{\subsection}[block]{\large}{\thesubsection.}{1em}{} % Change the look of the section titles
\usepackage{fancyhdr} % Headers and footers
\usepackage{booktabs}
\usepackage{pdflscape}

\usepackage[english]{isodate}

\usepackage[longnamesfirst]{natbib}
\def\sym#1{\ifmmode^{#1}\else\(^{#1}\)\fi}
\usepackage{rotating,booktabs}

\usepackage{tikz}
\usetikzlibrary{snakes}
\usepackage{makeidx}
\usepackage{pst-all}
\usepackage{latexsym}
\usepackage{ifthen}
\usepackage{setspace}
\usepackage{pstricks}
\usepackage{pstricks-add}
\usepackage{theorem}
\usepackage{epsfig}
\usepackage{rotating}
\usepackage{lscape}
\usepackage{hyperref}

\usepackage{bbm}  
\usepackage{cleveref}
\usepackage{afterpage}  

\usepackage{siunitx}
\usepackage{caption}

\usepackage{algorithm,algcompatible,amsmath}
\newenvironment{tablenotes}[1][0.5]{\begin{minipage}[t]{\linewidth}\footnotesize}{\end{minipage}}
\newenvironment{figurenotes}[1][0.5]{\begin{minipage}[t]{\linewidth}\footnotesize}{\end{minipage}}
\bibpunct{(}{)}{;}{a}{,}{,}
\usepackage{etoolbox}% http://ctan.org/pkg/etoolbox
\BeforeBeginEnvironment{tabular}{\small}% Adjust tabular font size to \Huge

\usepackage{xcolor}
\pagecolor{white}

\graphicspath{{Figures_BWB/}}
\newcommand{\nominalAlpha}{0.10}
\newcommand{\nominalCov}{90\%}

\begin{document}
	\makeatletter
	\renewcommand*{\@fnsymbol}[1]{\ensuremath{\ifcase#1\or \star \or \dagger\or \ddagger\or
			\mathsection\or \mathparagraph\or \|\or **\or \dagger\dagger
			\or \ddagger\ddagger \else\@ctrerr\fi}}
	\makeatother
	
	%	\pagestyle{fancy}
	%	\chead{\small\sc Local Projections Bootstrap Inference}
	%	\rhead{\small\thepage}
	%	%\enlargethispage{0.2in}
	%	\renewcommand{\baselinestretch}{1}
	%	
	\title{Local Projections Bootstrap Inference\date{\today}
		\thanks{
			{\scriptsize  We would like to thank participants of the IAAE Conference (2023, Oslo) and the CFE meeting (2023, Berlin). Mar\'{\i}a Dolores Gadea is grateful for financial support from the  grants PID2020-114646RB-C44, PID2023-150095NB-C44, RED2018-102563-T, RED2022-134122-T and TED2021-129784B-I00 funded by MCIN/AEI/ 10.13039/501100011033. \`Oscar Jord\`a is grateful for support of the U.C. Davis Faculty Research Grant. The views expressed herein are solely the responsibility of the authors and should not be interpreted as reflecting the views of the Federal Reserve Bank
of San Francisco, or the Board of Governors of the Federal Reserve System.}}}
	\author{ Oscar Jord\'{a}
		\thanks{\scriptsize{
				Federal Reserve Bank of San Francisco, 101 Market St., San Francisco, CA 94105 (USA); and Department of Economics, University of California, Davis,
				One Shields Ave., Davis, CA 95616 (USA). e-mail:
				oscar.jorda@sf.frb.org and ojorda@ucdavis.edu}} \\
		%EndAName
		{\small Federal Reserve Bank of San Francisco and} \\ {\small University of California, Davis}
		\and Mar\'{\i}a Dolores Gadea
		\thanks{\scriptsize{
				Department of Applied Economics, University of Zaragoza. Gran V\'{\i}a, 4,
				50005 Zaragoza (Spain). Tel: +34 9767 61842 and
				e-mail: lgadea@unizar.es}} \\
		%EndAName
		{\small University of Zaragoza}}
	\vspace*{-3cm}
	{\let\newpage\relax\maketitle}
	
	\begin{abstract}
		{\small \noindent \\}
		\\
		Bootstrap procedures for local projections typically rely on assuming that the data generating process (DGP) is a finite order vector autoregression (VAR), often taken to be that implied by the local projection at horizon 1. Although  convenient, it is well documented that a VAR can be a poor approximation to impulse dynamics at horizons beyond its lag length. In this paper we assume instead that the precise form of the parametric model generating the data is not known. If one is willing to assume that the DGP is perhaps an infinite order process, a larger class of models can be accommodated and more tailored bootstrap procedures can be constructed. Using the moving average representation of the data, we construct appropriate bootstrap procedures. \\
		
		
		
	\end{abstract}

	%%%%%%%%%%%%%%%%%%%%%%%%%%%%%%%%%%%%%%%%%%
	% 				BEGIN TEXT
	%%%%%%%%%%%%%%%%%%%%%%%%%%%%%%%%%%%%%%%%%%
	
	\newpage
	\section{Introduction\label{sec-introduction}}
	
	Semi-parametric estimation of impulse responses with local projections \citep{Jorda2005} has gained considerable traction in the literature \citep[see, e.g.][]{Ramey2016, PMWolf2021}. Moreover, there have been several interesting recent developments on how to conduct inference with local projections \citep[see, e.g.][]{Jorda2009, MontielOleaPM2019, MOPM2021}. However, bootstrap procedures based on these developments typically rely on assuming that the data generating process (DGP) is a finite order vector autoregression (VAR), often taken to be that implied by the local projection at horizon 1. Although a convenient approximation, in this paper we assume that the precise form of the parametric model generating the data is not known, in keeping with the logic behind local projections, though we will assume that the model belongs to a broad class that can be characterized by infinite order moving average processes that can be well approximated.
	
	In a local projection, the residuals are usually thought to have a moving average structure of the same order as the horizon considered. However, since no assumptions about the DGP are made, in principle it is not known what this moving average might look like. One could use model-free techniques, such as the block-bootstrap \citep{Kunsch1989, LiuSingh1992, PolitisRomano1994, GWhite2004}. However, if one is willing to assume that the DGP is perhaps an infinite order process, a larger class of models can be accommodated and more tailored bootstrap procedures can be constructed.
	
	In particular, \cite{Paparoditis1996} derives bootstrap procedures where the data are assumed to be generated by an infinite order VAR. The theory relies on showing that in finite samples, the truncation of the VAR lag-order will generate valid bootstrap replicates as long as the truncation order is allowed to grow with the sample size at a particular rate---a similar result to that derived in \cite{LewisReinsel1985} and  \cite{Kuersteiner2005} to show the consistency of estimates of infinite order processes based on truncated models. However, it turns out that the \cite{Paparoditis1996} bootstrap procedure can be adapted in a convenient and intuitive way to the structure native to local projections, as we will show.
	
	Why do we need a local projections-based bootstrap procedure in light of \cite{Paparoditis1996}? We argue that the VAR truncation lag typically seen in empirical work is relatively short, thus limiting the shape that the impulse response can take \citep[see,][]{Olea2024, Olea2025}. Hence, though the truncated VAR is a useful tool to obtain approximate (centered) estimates of the residual process from which to draw bootstrap samples, a more flexible moving-average (MA) representation directly corresponding to the estimated impulse responses by local projections may be preferable.
	
	The contribution of our paper is to show how to take advantage of local projection regressions to construct bootstrap-based inference borrowing ideas from \cite{Paparoditis1996}. The key insight is that local projections can be used to estimate the first $h$ moving average terms directly, and subsequent terms can be approximated with a simple autoregression. The hope is that by directly modeling the periods were the impulse response displays its most interesting features, terms at longer horizons will have a relatively small influence that can be well approximated with an auxiliary autoregression. In practice, the concern is that finite (and short-order) VARs will provide poor approximations to the impulse response when dynamics are complex and long-lasting \citep{Olea2024, Olea2025}.
	
	In addition to providing formal justification for our procedures, simulation evidence shows that, not only this is a more intuitive way to construct the bootstrap for local projections, it has very good small sample   properties when applied to a variety of scenarios that we detail below. 
	
	%One of the results of our experiments is that pairing the \cite{Paparoditis1996} bootstrap with the Wild bootstrap \citep[see, e.g.][]{GoncalvesKilian2004} produces the best small sample   results in our simulations.
	
	\section{Consistency and asymptotic normality of impulse response estimators}
	
	We begin by reviewing well-known results from the literature to then set the stage for the local projections estimator. Because our bootstrap procedures will borrow from \cite{Paparoditis1996}, we proceed as follows. First we present standard results on infinite order processes, which form the backbone of the \cite{Paparoditis1996} bootstrap. Then we show that the same assumptions made to estimate a truncated $VAR(\infty)$ in a finite sample imply that the local projections estimator is consistent and then show that it is asymptotically normal. Based on these results and on \cite{Paparoditis1996}, we have a natural justification for the bootstrap. Hence, we then show how to construct the bootstrap for local projections by leveraging the moving-average structure of the residuals so as to keep the design of the bootstrap within the local projection framework.
	
	\subsection{The truncated VAR($\infty$)}
	
	In this section, we simply sketch the logic behind some well-known results in the literature for later use. Moreover,  these results allow us to present the class of models that we will entertain in the construction of our bootstrap procedure. Assume the DGP for the $m$-dimensional vector process $\bm y_t$ is the following \emph{invertible}, infinite-order moving average:
	\begin{align} \label{e:ma}
		\bm y_t = \sum_{j=0}^\infty B_j \bm \epsilon_{t-j}; \quad B_0 = I; \quad \sum_{j=0}^\infty ||B_j|| < \infty;
	\end{align}
	where $||B_j||^2 = tr(B_j'B_j)$ and $B(z) = \sum_{j=0}^\infty B_j z_j$ is such that $\text{det}\{B(z)\} \ne 0$ for $|z| \le 1$. \autoref{e:ma} is, of course, the impulse response representation of the data. Under these assumptions, this invertible $MA(\infty)$ can also be expressed as a $VAR(\infty)$:
	\begin{align} \label{e:var}
		\bm y_t = \sum_{j=1}^\infty A_j \bm y_{t-j} + \bm \epsilon_t; \quad \sum_{j=1}^\infty ||A_j|| < \infty; \quad \text{det}\{A(z)\} \ne 0 \, |z| \le 1.
	\end{align}
	Note that \autoref{e:ma} and \autoref{e:var} include a wide class of models, including VARMA models and several others that are usually found in the formulation of many macroeconomic models. These assumptions are common starting points in the literature \citep[see, e.g.][]{Lutkepohl2005}. However, an obvious limitation in what follows is that we will be  dealing with the class of invertible processes.
	
	Next, we state assumptions that establish the consistency of the coefficients in a $VAR(p)$ when the true model is generated by \autoref{e:ma} and hence has the $VAR(\infty)$ representation of \autoref{e:var}. Thus, we make assumptions 1--4 following \cite{LewisReinsel1985}. \citep[][makes slightly stricter assumptions in order to derive the optimal truncation lag length in finite samples]{Kuersteiner2005}. These assumptions are:
	
	\paragraph{Assumption 1} $\{\bm y_t\}$ is generated by \autoref{e:ma}.
	
	\paragraph{Assumption 2} $E|\epsilon_{it}\epsilon_{jt}\epsilon_{kt}\epsilon_{lt}| \le \gamma_4$ for $ 1 \le i, j, k, l \le m$.
	
	\paragraph{Assumption 3} The truncation lag $p$ is chosen as a function of the sample size $T$ such that $p^2/T \to 0$ as $p,T \to \infty$.
	
	\paragraph{Assumption 4} $p$ is chosen as a function of $T$ such that:
	\begin{align*}
		p^{1/2} \sum_{j = p+1}^\infty ||A_j|| \to 0 \quad \text{as} \quad p,T \to \infty.
	\end{align*}
	Then, \cite{LewisReinsel1985} show that:
	\begin{align*}
		||\hat A_j - A_j|| \overset{p}{\to} 0 \quad \text{as} \quad p,T \to \infty.
	\end{align*}
	In other words, the coefficients of the first $p$ terms of the $VAR(\infty)$ are consistently estimated. In turn, using the well-known \cite{Durbin1959} recursion, note that:
	\begin{align*}
		B_h = A_1B_{h-1} + A_2 B_{h-2} + \hdots + A_{h-1} B_1 + A_h,
	\end{align*}
where it is easy to see that the impulse response coefficients from a $VAR(p)$ will be consistent for the first $p$ periods, but not guaranteed to be consistent beyond horizon $p$ since $A_h$ is constrained to be zero for $h > p$. 
	
	As an aside we may ask why is this observation important. The reason is that the practice of specifying low-order VARs (which are preferable for forecasting purposes), are likely to generate inconsistent impulse response estimates even at relatively short horizons \citep{Olea2024, Olea2025}. Thus, when the object of interest is the impulse response function, the approximation will likely not work very well. We shall see that  local projections do not suffer from this problem to the same degree. Intuitively, a different model is used to approximate the response coefficient at each horizon, thus providing estimates that are consistent even for relatively small $p$.
	
	\subsection{The local projections estimator}
	Using assumptions 1--4 we now examine the consistency of the local projections estimator for the DGP in \autoref{e:ma}, which is in its impulse response form already. Using the same $VAR(\infty)$ as in \autoref{e:var} and recursive substitution, it is easy to see that:
	\begin{align} \label{e:lp}
		\bm y_{t+h} = \underbrace{B_{h+1} \bm y_{t-1}}_{\text{response}} + \underbrace{C_{h+2} \bm y_{t-2} + C_{h+3} \bm y_{t-3} + \hdots}_{\text{other regressors}} + \underbrace{\bm \epsilon_{t+h} + B_1 \bm \epsilon_{t+h-1}+ \hdots + \bm \epsilon_t B_h}_{\text{error term}},
	\end{align}
	where
	\begin{align*}
		C_{h+2} &= B_hA_1 + \hdots B_1 A_h + A_{h+1}, \\
		C_{h+3} &= B_hA_2 + \hdots B_1 A_{h+1} + A_{h+2}, \\
		\vdots \\
		C_{h+p} &= B_hA_{p-1} + \hdots B_1 A_{h+p-2} + A_{h+p-1}, \\
		\vdots
	\end{align*}
	Thus \autoref{e:lp} shows that the MA terms, $B_{h+1}$, of \autoref{e:ma} can be estimated with a sequence of regressions such as those in \autoref{e:lp}. In parallel fashion, consider truncating the right-hand side of the local projection at $p$, with $p$ chosen to meet Assumptions 1--4. The truncated local projection therefore becomes:
	\begin{align*}
		\bm y_{t+h} = B_{h+1} \bm y_{t-1} + C_{h+2} \bm y_{t-2} + \hdots + C_{h+p} \bm y_{t-p} + \bm u_{t+h},
	\end{align*}
	with
	\begin{align} \label{e:resid}
		\bm u_{t+h} &\,= \underbrace{\bm \epsilon_{t+h} + \{B_1 \bm \epsilon_{t+h-1} + \hdots + B_h \bm \epsilon_t \}}_{\text{previous error term}} \notag \\
		&\,+ \underbrace{\{ C_{h+p+1} \bm y_{t-p-1} + C_{h+p+2} \bm y_{t-p-2} + \hdots \}}_{\text{omitted terms due to truncation: } \bm \pi_{t-p-1}}
	\end{align}
where for ease of notation later on we refer to the last term in the summation as $\bm \pi_{t-p-1}$. Clearly the key is to show that the omitted terms due to truncation are, asymptotically, sufficiently small so as not to affect the consistency of the estimator for $B_{h+1}$. In other words, we need to show that the least-squares estimator for this truncated local projection is consistent.
	\paragraph{Consistency}
	\hrulefill
	
\noindent	Under assumptions 1--4:
	\begin{align} \label{e:consistency}
		||\hat B_{h+1} - B_{h+1}|| \overset{p} \to 0 \quad \text{as} \quad p,T \to \infty \quad \text{for} \quad  h = 1, \hdots
	\end{align}
	This is shown in Appendix A1. The intuition for this result is the following. Like the proof of consistency for a truncated $VAR(\infty)$, the key is to show that the terms $C_{h+k}$ (for $h+k >p$ and $p$ sufficiently large) become sufficiently small in the asymptotic sense. In the local projection, the truncation is on the coefficient matrices $C_{h+k}$ which are functions of the first $h$ moving average coefficients and the truncated $A_k$, which are vanishing asymptotically for the same reasons as in the proof of consistency in \cite{LewisReinsel1985}.
	
	Some remarks are worth noting. The consistency of the local projection estimator is less sensitive to the truncation lag length, $p$, than the truncation lag in the VAR. The reason is that in the VAR, the truncation lag determines the maximum horizon $h$ for which the impulse response is guaranteed to be consistent. This is not the case for the local projection for which consistency is attained for horizons $h > p$.\footnote{For finite order processes, recent work by \cite{MOPM2021} on lag-augmented local projections suggests that adding extra lags to the local projection can resolve the issues caused by the serial correlation of the residuals in the first term in brackets of \autoref{e:resid}.} In addition, note that we are interested on estimates of $B_h$, but not estimates of $C_{h+j}$ for $j = 2, \hdots, p$.

	\section{Asymptotic normality of the local projections estimator}
	The proof of asymptotic normality follows a similar approach to the proof of consistency. Again, based on \cite{LewisReinsel1985}, we make the following additional assumptions:
	
	%\paragraph{Assumption 1} $E|\epsilon_{it}\epsilon_{jt}\epsilon_{kt}\epsilon_{lt}| \le \gamma_4 < \infty \quad 1 \le i,j,k,l \le m$
	
	\paragraph{Assumption 5} $p$ is chosen such that $p^3/T \to 0$ as $p,T \to \infty$
	
	\paragraph{Assumption 6} $p$ is chosen as a function of $T$ such that
	\begin{align*}
		T^{1/2} \sum_{j = p+1}^\infty ||C_{h+j}|| \to 0 \quad \text{as} \quad p,T \to \infty
	\end{align*}
	Note that this assumption is tailored to the local projection estimator in \autoref{e:lp}. It basically says that these terms are asymptotically negligible. As a reference, the original assumption in \cite{LewisReinsel1985} is:
	\begin{align*}
		T^{1/2} \sum_{j = p+1}^\infty ||A_j|| \to 0 \quad \text{as} \quad p,T \to \infty
	\end{align*}
	though the former can be derived from the latter with a bit more work.
	
	\paragraph{Assumption 7} $\{l(p)\}$ is a sequence of $pm^2 \times 1$ vectors such that $0 < M_1 \le ||l(p)||^2 = l(p)'l(p) \le M_2 < \infty$ for $p = 1, 2, \hdots$. This assumption will be useful to construct joint hypotheses tests.
	
	Based on these assumptions, we briefly restate theorems 2, 3, and 4 in \cite{LewisReinsel1985} from which we will then derive the asymptotic normality of the local projection estimator. We therefore begin with theorem 2 in \cite{LewisReinsel1985}, which states that:
	\begin{align*}
		(T-p)^{1/2} l(k)'(\hat \alpha(p) - \alpha(p)) - (T-p)^{1/2} l(k)' vec\left[ \left\{\frac{1}{T-p} \sum_p^{T-1} \bm \epsilon_{t+1} Y_{t,p} \right\} \Gamma_p^{-1} \right] \to 0
	\end{align*}
	where $Y_{t,p} = (\bm y_t', \> \hdots, \> \bm y_{t-p+1}')'$ and $\hat \Gamma_p = (T - p)^{-1} \sum_p^{T-1} Y_{t,p}Y_{t,p}'$ and $\alpha(p) = vec(A(p))$ with $A(p) = (A_1, \> \hdots, A_p)$.
	
	Next, theorem 3 states that, based on Assumptions 2, 5, 6, and 7 and theorem 2, then
	\begin{align*}
		s_T &= (T-p)^{1/2} l(p)' vec \left[ \left\{\frac{1}{T-p} \sum_p^{T-1} \bm \epsilon_{t+1} Y_{t,p} \right\} \Gamma_p^{-1} \right] \\
		v_T^2 &= V(s_T) = l(p)' (\Gamma_p^{-1} \otimes \Sigma) l(p)
	\end{align*}
	and 
	\begin{align*}
		\frac{s_T}{v_T} \overset{d}{\to} N(0,1); \quad \Sigma = E(\bm \epsilon_t \bm \epsilon_t').
	\end{align*}
	
	\bigskip
	
	Finally, theorem 4 in \cite{LewisReinsel1985} states that using Assumptions 2, 5, 6, and 7, then
	\begin{align*}
		(T-p)^{1/2} l(p)'(\hat \alpha(p) - \alpha(p))/v_T \overset{d}{\to} N(0,1)
	\end{align*}
	
	What do these results mean for our local projection estimator? Using the notation introduced in the proof of consistency in the appendix, we can express the local projection compactly as:
	\begin{align*}
		\bm y_{t+h} = D Y_{t-1,p} + \bm u_{t+h}; 
	\end{align*}
	with $D = (B_{h+1} \> C_{h+2} \> \hdots \> C_{h+p})$, $Y_{t-1,p} = (\bm y_{t-1}', \> \hdots, \> y_{t-p}')'$ and $\bm u_{t+h} = \bm \epsilon_{t+h} + B_1 \bm \epsilon_{t+h-1} + \hdots + B_h \bm \epsilon_t + \bm \pi_{t-p-1}$ where recall that  $\bm \pi_{t-p-1} = C_{h+p+1} \bm y_{t-p-1} + \hdots$ as defined in \autoref{e:resid}. The truncation terms in $\bm \pi_{t-p-1}$ will vanish asymptotically and thus not affect the approximate asymptotic distribution.
	
	Based on the results so far, we are now in a position to state the asymptotic normality results for the local projection estimator:
	\begin{align} \label{e:anorm}
		(T-h-p)^{1/2} l(h)'(\hat \beta_{h,p} - \beta_h)/\eta_{h,T} \overset{d}{\to} N(0,I)
	\end{align}
	where $\beta_h = vec(B_h)$ and $\hat \beta_{h,p}$ is the estimate based on a local projection with $p$ lags.
	\begin{align*}
		\eta^2_{h,T} &= l(h)'(\Gamma_1^{-1} \otimes \Omega_h) l(h) \\
		\Omega_h &= \Sigma + B_1 \Sigma B_1' + \hdots + B_h \Sigma B_h'
	\end{align*}
	where $l(h)$ simply selects the coefficients of $\beta_h$ that we are interested in. Note that this result refers to the $h^{th}$ local projection. Of course, if we wanted to do hypotheses tests across horizons, then we can specify the system:
	\begin{align*}
		Y_{t,H} = (I \otimes \bm y_{t-1})\bm \beta_{1,h} + (I \otimes \bm y_{t-2}) \bm c_{2,H} + \hdots + (I \otimes \bm y_{t-p}) \bm c_{p,H} + U_{t,H}
	\end{align*}
	where $Y_{t,H} = (\bm y_t', \> \hdots, \> \bm y_{t+H}')'$; $\bm \beta_{1,h} = vec(B_1, \> \hdots, \> B_H)$ and similarly for $\bm c_{j,H}$ for $j = 2, \hdots, p$. 
	
	Since we have previously established that $||\hat B_j - B_j|| \to 0$ for $j = 1, \hdots, h$ then we can use small sample   moments to estimate the variance in small samples.
	
	\section{Asymptotic justification for the local projections bootstrap}
	
	Theorem 2.3 in \cite{Paparoditis1996} shows that the asymptotic properties of the empirical distribution $\hat F_T$ of the centered residuals $\hat{\bm \epsilon}_{p,t}$ from a truncated $VAR(p)$ is an estimator of the distribution $F$ of the true errors. In the analysis that follows, we also use these residuals to construct our bootstrap replicates using local projections. In particular, from this result and using a Mallows distance, theorem 2.4 states that if Assumption 3 is met then:
	\begin{align*}
		d_2(\hat F_T, F) \overset{p}{\to} 0
	\end{align*}
	and hence this result directly applies to the bootstrap that we describe below. Further, \cite{Bickel1981} show that convergence in $d_2$ metric implies that:
	\begin{align*}
		\hat \Sigma_p \overset{p}{\to} \Sigma; \quad \text{where} \quad \Sigma = E(\bm \epsilon_t \> \bm \epsilon_t')
	\end{align*}
	and $\hat \Sigma_p$ is the sample counterpart estimated from the residuals of the $VAR(p)$. This result justifies the use of the approximate $N(0, \hat \Sigma_p)$ in generating the centered bootstrap errors $\hat{\bm \epsilon}_t^*$.
	
	Further, theorem 2.5, which relies on Assumption 5, states that:
	\begin{enumerate}
		\item[(a)] $||\Gamma^*_p - \Gamma_p||_1 = o_p(1)$
		\item[(b)] $||{\Gamma^*}^{-1}_p - \Gamma_p^{-1}||_1 = o_p(1)$
	\end{enumerate}
	where recall that $\Gamma_p = E(Y_{t,p}\> Y_{t,p}')$ and $\Gamma^*_p$ is the sample equivalent estimated using bootstrap replicates. However, note that now we are relying on the asymptotic normality result presented in \autoref{e:anorm}.
	
	As in \cite{LewisReinsel1985}, define the bootstrap equivalent:
	\begin{align*}
		s_T^* = (T-p)^{1/2} l(p)' vec \left[ \left\{\frac{1}{T-p} \sum_p^{T-1} \bm \epsilon_{t+1}^* {Y_{t,p}^*}' \right\} {\Gamma_p^{*}}^{-1} \right].
	\end{align*}
	Theorem 3.1 in \cite{Paparoditis1996} states that for $p^{7/2}/T^{1/2} \to 0$ then:
	\begin{align*}
		(T-p)^{1/2} l(p)' (\hat a(p)^* - \hat a (p)) = s_T^* + o_p(1).
	\end{align*}
	Of course, given the asymptotic results of \autoref{e:anorm}, one can equivalently show that:
	\begin{align*}
		(T-p-h)^{1/2} l(h)' (\hat{\bm \beta}_h^* - \hat{\bm \beta}_h) = \sigma_{h,T}^* + o_p(1).
	\end{align*}
	where $\bm \beta_h = vec(B_h)$ and $V(\sigma_{h,T}) = \eta_{h,T}^2$. This result mirrors the result presented by \cite{Paparoditis1996} in theorem 3.4, which formally states that if $p^4/T^{1/2} \to 0$ then:
	\begin{gather*}
		\mathcal{L}\left( (T-p-h)^{1/2} l(h)' (\hat{\bm \beta}_h^* - \hat{\bm \beta}_h)| \bm y_1, \hdots, \bm y_T\right)  \overset{d}{\to} N\left(0, l(h)' \Omega_h l(h)\right) \\
		\Omega_h = \left( \Sigma^{-1} + \sum_{j=1}^{h} B_j \Sigma B'_{j} \right) \quad h = 1, 2, \hdots, H
	\end{gather*}
	which justifies the asymptotic validity of the bootstrap.
	
	\section{The moving average bootstrap}
	
	This section introduces our bootstrap procedure based on the results of the previous section on the asymptotic normality of local projection estimates of the moving average representation of an infinite order MA process. In order to draw the distinctions and similarities with existing methods, we begin with a brief introduction of the bootstrap procedure proposed by \cite{Paparoditis1996}. We then introduce our bootstrap procedure and discuss its features.
	
	\subsection{The VAR-based moving-average (VAR-MA) bootstrap}
	In order to motivate our bootstrap procedure, we briefly present the main results in \cite{Paparoditis1996}. The logic of his bootstrap procedure is the following. Given the asymptotic normality of the parameters of the $VAR(p)$ established in, e.g., \cite{LewisReinsel1985}, under the additional assumptions in \cite{Paparoditis1996}, the asymptotic normality of the moving average coefficients can also be established for up to the first $p$ terms \citep[see, e.g.][]{Lutkepohl2005}. Hence, the bootstrap for the moving average coefficients $B_{h}$ for $h = 1, \hdots, H$ can be constructed as follows:

	\paragraph{VAR-based MA bootstrap}
	\hrulefill
	
	\begin{enumerate}
		\item From the truncated $VAR(p)$, use the centered residuals $\hat{\bm \epsilon}_t^*$ and the moving average estimates $\hat B_{h,p}$, to generate bootstrap replicates $\{\bm y_t^*\}_{t=1}^T$ obtained from:
		\begin{align*}
			\bm y_t^* = \sum_{h = 0}^{t+s-1} \hat B_{h,p} \bm \epsilon_{t-h}^*,
		\end{align*}
		for a given $s$, where the $\bm \epsilon_{t}^*$ are drawn with replacement from the centered $\hat{\bm \epsilon}_t$, and the matrices $\hat B_{h,p}$ are calculated with the usual recursion:
		\begin{align*}
			\hat B_{h,p} = \sum_{i = 1}^h \hat B_{h-i,p} \hat A_{i,p}
		\end{align*}
		with $\hat A_{i,p} = 0$ for $h > p$ and $\hat B_{0,p} = I$. The notation $\hat B_{h,p}$ denotes that the estimate of the moving average coefficient at lag $h$ has been obtained from a truncated VAR or order $p$.
		\item Using bootstrap replicates $\{\bm y_t^*\}_{t=1}^T$, fit $VAR(p)$ models to obtain estimates of $\hat B_{h,p}^*$ for $h = 1, \hdots, H$ and estimates of $\hat \Omega_h^*$, the sample covariance matrix of $\hat {\bm \beta}_h^* = \text{vec}(\hat B_{h,p}^*)$.
		\item Store the statistics $\hat T^*_b = (\delta' \hat {\bm \beta}_{h,b} - \delta' \hat {\bm \beta}_{h,p})/(\delta' \hat \Omega_h^*\delta)^{1/2}$ for $b = 1, \hdots, B$ bootstrap replicates and where $\delta$ is an $r \times 1$ vector where $r = \text{dim}(\hat {\bm \beta}_{h,p})$ and where $\delta$ is a user-specified vector denoting the hypotheses of interest and $\hat {\bm \beta}_{h,p}$ denotes the estimates of the moving average coefficients obtained with the truncated $VAR(p)$ and the original sample, for $h = 1, \>2,\> \hdots, \>H$.
		\item Using a large number of bootstrap repetitions, approximate the distribution of statistics of interest with the empirical distribution. In particular, compute the $\alpha/2$ and $1 - \alpha/2$ quantiles of $\{\hat T^*_b\}_{b = 1}^B$, denote these $\hat q_{\alpha/2}$ and $\hat q_{1 - \alpha/2}$ respectively.
		\item Return the percentile-$t$ confidence interval:
		\begin{align*}
			\left[\delta' \hat {\bm \beta}_{h,p} - (\delta' \hat \Omega_h\delta)^{1/2}\hat q_{\alpha/2}, \, \delta' \hat {\bm \beta}_{h,p} - (\delta' \hat \Omega_h\delta)^{1/2}\hat q_{1 -\alpha/2} \right]
		\end{align*}
	\end{enumerate}
	
	\hrulefill
	
	A couple of remarks are worth making. First, recall that the consistency of the MA coefficient matrices is only guaranteed for $h \le p$. Although the asymptotic theory works with $p \to \infty$, in small samples consistency will not be guaranteed for any $\hat B_{h,p}$ for $h \le p$ and hence this will generate some error in the generation of the bootstrap replicates. Second, we could have easily constructed the covariance matrix for $\bm \beta_{1,H} = (\beta_1, \> \hdots, \> \beta_H)'$ and $V(\hat{\bm \beta}_{1,H})$ to do joint hypotheses tests across horizons. In the next section we explore an alternative way to generate the bootstrap replicates.
	
	\subsection{The local projections moving-average (LP-MA) bootstrap}
	
	The previous section provides a useful platform to introduce our methods. Using local projections, one can obtain estimates of the first $H$ coefficient matrices $B_h$ for $h = 1, \hdots, H$. However, step 1 of the procedure proposed by \cite{Paparoditis1996} and described above, may require up to $t+s-1 > H$ terms. In this section we propose a practical approach to remedy this truncation issue.
	
	By assumption, note that the data can be represented as an infinite moving average, such as:
	\begin{align} \label{e:mainf}
		\bm y_t = \bm \epsilon_t + B_1 \bm \epsilon_{t-1} + \hdots = (I + B_1L + B_2 L^2 + \hdots) \bm \epsilon_t = B(L) \bm \epsilon_t
	\end{align}
	Since we can estimate the first $H$ terms of this representation with local projections, consider a partition of the moving average lagged polynomial as follows:
	\begin{align*}
		B(L) = B_0^H(L) + B_{H+1}^\infty(L)
	\end{align*}
	where $B_0^H(L) = (I + B_1 L + \hdots + B_H L^H)$. Next, consider approximating the polynomial $B_{H+1}^\infty(L)$ with a first order autoregressive term, specifically, suppose that we can write:
	\begin{align} \label{e:maapprox}
		\bm y_t = B_0^H(L) \bm \epsilon_t + G_{H+1} \bm y_{t-(H+1)}
	\end{align}
	Using \autoref{e:mainf} to express $\bm y_t$, it is easy to see that:
	\begin{align} \label{e:trick}
		B(L) \bm \epsilon_t &= B_0^H(L) \bm \epsilon_t + G_{H+1} \bm y_{t-(H+1)} \notag \\
		(B(L) - B_0^H(L)) \bm \epsilon_t &= G_{H+1} L^{H+1} \bm y_{t} \notag \\
		B_{H+1}^\infty(L) \bm \epsilon_t & = G_{H+1} L^{H+1} B(L) \bm \epsilon_t
	\end{align}
	Hence, by equating the terms in the powers of the lagged polynomial, we arrive at the following recursion:
	\begin{align} \label{e:recursion}
		B_{H+1} &= G_{H+1} \notag \\
		B_{H+2} &= G_{H+1} B_1 \notag \\
		\vdots &= \vdots \notag \\
		B_{H+ j+1} &= G_{H+1}B_j \quad \text{for} \quad  j \ge 1
	\end{align}
	In practice, this means that one can estimate the auxiliary regression:
	\begin{align} \label{e:aux}
		(\bm y_t - \hat B_0^H(L) \hat{\bm \epsilon}_t) = G_{H+1} \bm y_{t-(H+1)} + \bm \zeta_t
	\end{align}
	to obtain  $\hat G_{H+1}$ which can then be used in the recursion shown in \autoref{e:recursion} to construct bootstrap replicates as in step 1 of the \cite{Paparoditis1996} procedure shown above.
	
	What is the justification for this recursive procedure? One could make an analogous assumption to Assumption 4 of the proof of consistency discussed above along the lines of:
	\paragraph{Assumption 8} The maximum horizon of the impulse response $H$ is chosen so that:
	\begin{align*}
		H^{1/2} \sum_{j=H+1}^\infty ||B_h|| \to 0 \quad \text{as} \quad H,T \to \infty
	\end{align*}
	to justify that the remainder terms of the impulse response are vanishingly small, and further that, based on \autoref{e:recursion} and \autoref{e:aux}:
	\begin{align*}
		||\hat B_j - B_j|| \to 0 \quad \text{for} \quad j > h \quad \text{as} \quad h,T \to \infty
	\end{align*}
	In words, under the maintained assumptions, the stationarity of $\bm y_t$ means that the moving average terms at increasingly distant horizons become vanishingly small and that, in any case, they can be approximated using a first order autoregressive approximation. In practical terms, this is a weaker assumption than the assumption of invertibility.
	
	Thus, relative to the VAR-based bootstrap procedure in \cite{Paparoditis1996}, we propose the following bootstrap procedure for local projections:
	
	\paragraph{Local projection bootstrap} 
	\hrulefill
	\begin{enumerate}
		\item Use the centered residuals, $\hat{\bm \epsilon}_t^*$ from the first local projection (which in effect is a $VAR(p)$ just as in the VAR-MA bootstrap). Further, using the estimates of the first $H$ terms $\hat B_h$ for $h=1, \hdots, H$ of the moving average representation using local projections, and using the approach based on \autoref{e:aux} and the recursion described in \autoref{e:recursion} to construct estimates for $\hat B_h$ for $ h >H$, generate bootstrap replicates $\{\bm y_t^*\}_{t=1}^T$ obtained from:
		\begin{align*}
			\bm y_t^* = \sum_{h=0}^{t+s-1} \hat B_h \bm \epsilon_{t-h}^*
		\end{align*}
		for a given $s$, where the $\bm \epsilon_t^*$ are drawn with replacement from the centered $\hat{\bm \epsilon}_t$.
		\item Using bootstrap replicates $\{\bm y_t^*\}_{t=1}^T$, estimate by local projections  $\hat B_h^*$ for $h = 1, \hdots, H$ and estimates of $\hat V(\hat \beta_h^*)$, the sample covariance matrix of $\hat{\bm \beta}_h^* \textit{}= vec(\hat B_h^*)$.
		\item Like the VAR-based procedure, store the statistics $\hat T_b^* = (\delta \hat{\bm \beta}_{h,b}^* - \delta' \hat {\bm \beta}_h)/(\delta'\hat \hat V(\hat \beta_h^*) \delta)^{1/2}$ for $b = 1, \hdots, B$ bootstrap replicates. Recall $\delta$ is a user specified vector denoting the hypotheses of interest. Note that the $\hat {\bm \beta}_h$ denote the local projection estimates from the original sample and that $\hat V(\hat \beta_h^*)$ can be calculated using the usual sample statistic based on the boostrap replicates.
		\item This step is equivalent to step 4 in the VAR-based bootstrap. That is, one computes the quantiles of the empirical distribution of $\{T_b^*\}_{b=1}^B$, denoted $\hat q_{\alpha/2}$ and $\hat q_{1-\alpha/2}$.
		\item  As in Step 5 of the VAR-based bootstrap, return the percentile-$t$ interval:
		\begin{align*}
			\left[\delta' \hat {\bm \beta}_{h}(p) - (\delta' \hat \hat V(\hat \beta_h)\delta)^{1/2}\hat q_{\alpha/2}, \, \delta' \hat {\bm \beta}_{h}(p) - (\delta' \hat \hat V(\hat \beta_h)\delta)^{1/2}\hat q_{1 -\alpha/2} \right]
		\end{align*}
		
	\end{enumerate}	
	\hrulefill

\section{Simulation results}\label{sec:sim-results}
This section evaluates the performance of the proposed methods across several data-generating processes (DGPs). We run univariate simulations for autoregressive models of order 1 and $p$ (AR(1) and AR($p$)) and for moving-average models whose coefficients are generated by a Gaussian basis function (MA($q$)–GBF(1)).\footnote{We follow the “Functional Approximation of Impulse Responses” in \citet{BarnichonMatthes2018}. We also simulated other GBF combinations (e.g., MA($q$)–GBF(2)) and multivariate designs (VAR and MA($q$)–GBF($n$) of order 2). These results are omitted for space and available upon request.} GBFs allow us to produce rich, later-horizon dynamics efficiently.

Implementing the bootstrap requires choices about how to generate pseudo-residuals $\bm{\epsilon}_t^{*}$ from centered residuals $\hat{\bm{\epsilon}}_t$. Theory permits some heteroskedasticity in the first LP regression and acknowledges that the MA structure may leave residual dependence. The Wild Bootstrap (WB) targets heteroskedasticity; block or sieve schemes address dependence; and hybrids combine both or modify WB accordingly.\footnote{See \citet{SmeekesUrbain2014a} for a review of modified wild bootstraps in unit-root testing.} We consider the standard WB \citep{GoncalvesKilian2007}; the Block Bootstrap (BB) \citep{PolitisRomano1994}, which resamples blocks of size $H$;\footnote{For an application in volatility, see \citet{HounyoGoncalvesMeddahi2017}.} the Block Wild Bootstrap (BWB) \citep{Shao2011}; the Dependent Wild Bootstrap (DWB) \citep{Shao2010}; the Autoregressive Wild Bootstrap (AWB) \citep{SmeekesUrbain2014a,FriedrichSmeekesUrbain2020}; the Sieve Bootstrap (SB); and the Sieve Wild Bootstrap (SWB) \citep{Buhlmann1997}.\footnote{Applications to unit-root and panel settings include \citet{CavaliereTaylor2009a,CavaliereTaylor2009b,SmeekesUrbain2014b}.}

Although we compared all these procedures for AR(1) and AR($p$)/MA($q$) designs, here we report only the BWB results to keep the exposition focused and because BWB proved easier to tune and more stable in our implementation. Results for DWB—whose theoretical appeal is attractive in our framework, but whose performance is more sensitive to implementation choices—are available upon request.\footnote{There is no single canonical bandwidth choice for DWB; coverage can vary across reasonable kernel/bandwidth pairs. In our experiments, plug-in and rule-of-thumb selections sometimes produced different degrees of conservatism at long horizons.}

Before turning to the simulation evidence, it is useful to contrast BWB and DWB on theoretical grounds. 
Table~\ref{tab:summary-BWB-DWB} summarizes their construction, tuning parameters, and the type of dependence each method preserves. 
Both extend the wild bootstrap to dependent data but impose dependence differently: DWB induces correlation through a kernel and a bandwidth parameter—offering flexibility but requiring careful tuning—whereas BWB resamples residuals in blocks while retaining the wild component for heteroskedasticity, with block length as its sole tuning parameter. 
This theoretical contrast provides the background for the simulation results discussed below.

\begin{table}[!ht]
	\centering
	\caption{Comparison of BWB and DWB bootstrap schemes}
	\label{tab:summary-BWB-DWB}
	\renewcommand{\arraystretch}{1.2}
	\begin{tabular}{lcc}
		\toprule
		& \textbf{BWB} & \textbf{DWB} \\
		\midrule
		Weights & Blockwise-constant $v_m^*$ & Dependent process $W_t$ \\
		Tuning parameter & Block length $l$ & Bandwidth $\ell$ \\
		Dependence preserved & Within blocks & Kernel-based, across all $t$ \\
		Typical choice & $l \propto H$ & $\ell \to \infty, \; \ell/T \to 0$ \\
		Implementation & Simple, single knob & More flexible but kernel-dependent \\
		\bottomrule
	\end{tabular}
\end{table}

Our simulations—mixed and design-dependent—do not point to a uniform winner. BWB typically delivers stable coverage and homogeneous interval lengths with modest tuning, making it a reliable default across persistence levels and for short-to-medium horizons. DWB can match or surpass BWB in highly persistent or near–unit-root settings and at long horizons, provided the bandwidth is sensibly calibrated; in that range, extending the MA recursion beyond $H$ (Method~2) helps curb truncation bias. For finite-memory MA($q$) designs both methods behave similarly, so simplicity often favors BWB. Across designs, avoid an overly small lag order in the first LP regression—SBIC is a sensible default and particularly beneficial for DWB—while BWB’s single tuning knob (block length) tends to yield flatter performance across horizons. A concise comparison appears in Table~\ref{tab:summary-BWB-DWB2}.

\begin{table}[h!]
	\caption{Summary: Dependent Wild Bootstrap (DWB) vs. Block Wild Bootstrap (BWB) }
	\label{tab:summary-BWB-DWB2}
	\begin{center}\scalebox{0.65}{
			\begin{tabular}{l|ccp{8.6cm}}
				\hline\hline
				Design & Coverage (DWB vs BWB) & Length (DWB vs BWB) & Notes \\
				\hline
				AR(1), low persistence ($\phi\approx 0$) 
				& $\approx$ & $\approx$ to BWB $\downarrow$ & Small differences overall; both close to nominal for short/medium $H$. \\[2pt]
				AR(1), medium persistence ($\phi\approx 0.5$) 
				& $\approx$ to DWB $\uparrow$ & $\approx$ & DWB tends to be more stable across $H$; differences remain modest. \\[2pt]
				AR(1), high/near–unit persistence ($\phi\approx 0.95$ or $1$) 
				& DWB $\uparrow$ & $\approx$ & DWB better preserves dependence and reduces undercoverage at long horizons; gains larger with Method~2 (recursion beyond $H$). \\[2pt]
				AR($p$), low persistence ($\sum\phi_i \in [0.3,0.9]$)
				& $\approx$ & BWB $\downarrow$ & BWB often yields slightly shorter bands; mild risk of undercoverage if blocks are too short. SBIC in first LP helps both. \\[2pt]
				AR($p$), medium persistence ($\sum\phi_i \in [0.7,0.9]$)
				& DWB $\uparrow$ & $\approx$ & Advantage for DWB grows with $H$ and with larger $P$; fixed $p{=}1$ degrades both methods. \\[2pt]
				AR($p$), high persistence ($\sum\phi_i \in [0.9,0.99]$)
				& DWB $\uparrow\uparrow$ & $\approx$ & Clear coverage edge for DWB, especially at long horizons; Method~2 mitigates truncation bias. \\[2pt]
				MA($q$) finite memory (e.g., MA(24)–GBF)
				& $\approx$ & $\approx$ & With finite impulse duration, both perform similarly; choice can be based on simplicity (BWB). \\
				\hline\hline
		\end{tabular}}
	\end{center}
	\begin{tablenotes}
		\textit{Notes}: $\uparrow$ (“higher”), $\downarrow$ (“lower”), and $\approx$ (“similar”) refer to \textbf{DWB relative to BWB}.
		Method~2 denotes extending the MA recursion beyond $H$ when generating bootstrap paths.
	\end{tablenotes}
\end{table}

Taken together, these tables provide a complementary perspective: the first highlights the theoretical construction of BWB and DWB, while the second shows how their relative performance varies across DGPs and horizons. We next provide further details on the BWB, which serves as our main bootstrap procedure in the subsequent simulations designs.
	
The block wild bootstrap (BWB; \citealp{Shao2011}) extends the wild bootstrap to dependent data by imposing blockwise-constant weights. Let $l$ be the block length. For each block $m=1,\dots,\lceil T/l\rceil$, draw an i.i.d.\ weight $v_m^*$ with $E[v_m^*]=0$ and $\mathrm{Var}(v_m^*)=1$. Assign this weight to all observations in the block,
\[
\xi_t^* = v_m^*, \qquad (m-1)l+1 \leq t \leq ml.
\]
The bootstrap residuals are then
\[
u_t^* = \xi_t^*\,\widehat{u}_t.
\]
This construction preserves the within-block dependence of $\{\widehat{u}_t\}$ while reproducing conditional heteroskedasticity through the random weights. In our implementation, the block length $l$ is linked to the forecast horizon $H$ via simple rules of thumb.

	Several small-sample bias corrections exist for LP equations (e.g., Pope \citep{Pope1990}, lag-augmentation \citep{MOPM2021}, long-difference \citep{PigerStockwell2023}). We deliberately do not use them here: our goal is to evaluate the proposed bootstrap under minimally adjusted implementations—especially in highly persistent settings—so the assessment is conservative and comparable across designs. In practice, many applications also forgo these corrections.
		
	To assess the properties of each bootstrap variant and its accuracy, the coverage is calculated with percentile-$t$ intervals (Kilian, 1999) at the \nominalCov{} nominal level ($\alpha=\nominalAlpha$). The length accuracy is calculated as the amplitude of the interval with respect to the range of the estimated LPs at each point. Finally, in all model simulations we have distinguished between two methods: (1) by only taking into account the first $H$ terms of the moving average representation (Method~1); versus (2) also including additional terms following the algorithm proposed in the previous section in \autoref{e:recursion}.
	
	We compare the coverage results obtained with the local projection bootstrap for all models to those obtained using autoregressive estimation  (AR or VAR).\footnote{To save space, we only present the results for the AR case though results for the VAR model are available upon request.} We also applied the approach proposed by \citep{Kilian1999} although without bias correction to make the results comparable across experiments. Further, we compute the bias generated using each type of bootstrap method for each iteration \textit{r} and for each horizon \textit{h} as the mean of the following equation:
	
\begin{equation}
	\left| \mathcal{R}_{\text{true}}(h)
	- \frac{1}{B} \sum_{b=1}^{B} \mathcal{R}_{b}(h) \right|,
	\qquad h = 0, 1, \ldots, H.
\end{equation}
where $B$ is the number of bootstrap replications and $\mathcal{R}(h)$ refers to the impulse response at horizon $h$. Next, we describe the different models used in our simulations.
	
\subsection{Autoregressive models}\label{sec:ar}
\begin{figure}[h!]
	\begin{center}
		\caption{Estimated impulse responses from local projections (LP) versus true AR(1) responses}
		\label{fig-IRF-ar1-estim}
		\includegraphics[scale=0.6]{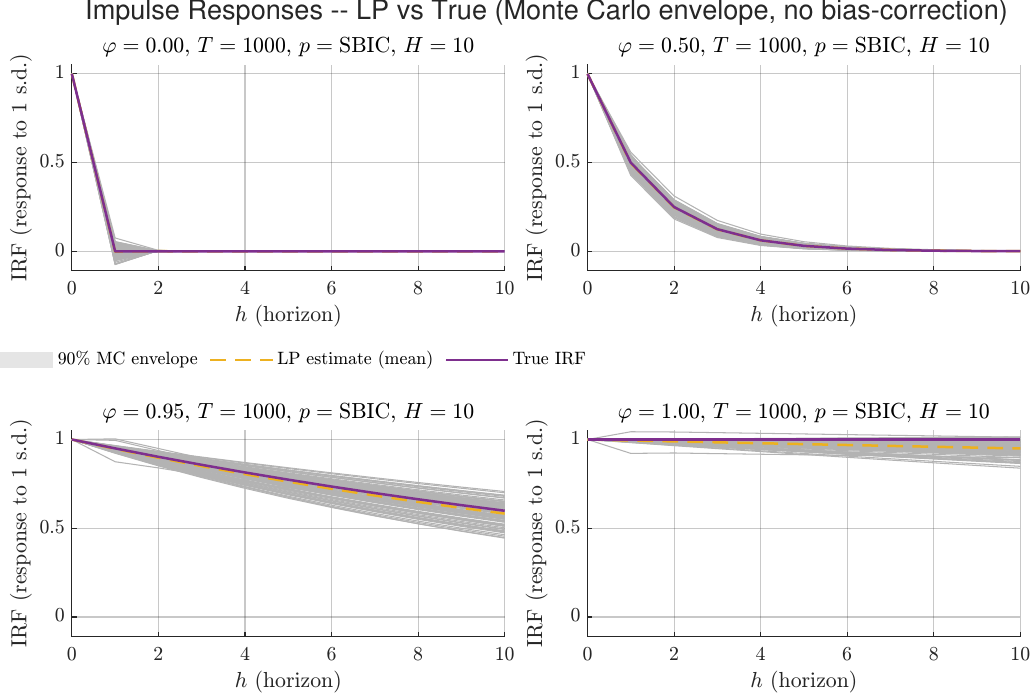}
	\end{center}
	\begin{figurenotes}
		\textit{Note}: Each panel reports results for an AR(1) process with autoregressive coefficient $\phi \in \{0,0.5,0.95,1\}$, sample size $T=1000$, lag length selected by SBIC, and maximum horizon $H=10$. 
		Thin gray lines are the parametric AR impulse responses obtained across 100 Monte Carlo replications. 
		The dashed line is the Monte Carlo mean of these estimates, the solid line is the true AR(1) IRF, and the shaded area is the \emph{pointwise} 90\% Monte Carlo envelope (5th–95th percentiles across replications), not a confidence band. 
		No Local Projections and no bootstrap are used in this figure. 
		The $x$-axis is the forecast horizon $h$ and the $y$-axis the response to a one–standard–deviation innovation.
	\end{figurenotes}
\end{figure}

We simulate data from the AR(1) model
\begin{equation}
	y_t=\phi\, y_{t-1}+\epsilon_{t},\qquad \epsilon_{t}\sim\mathcal{N}(0,1),
\end{equation}
with $t=1,\dots,T$, $T\in\{200,400,1000\}$, and $\phi\in\{0,0.5,0.95,1\}$. 
For each replication we estimate parametric AR models, with the lag length selected either based on SBIC or fixed at $p\in\{1,2,3\}$. We then compute the implied impulse responses. 
We also construct coverage statistics for Local Projections using percentile-$t$ confidence intervals based on the \emph{Block Wild Bootstrap} (BWB), referring to nominal \nominalCov{} intervals ($\alpha=\nominalAlpha$) unless noted otherwise.
For illustration, Figure~\ref{fig-IRF-ar1-estim} displays Monte Carlo envelopes (5th–95th percentiles across replications) of the parametric AR impulse responses together with the true and average responses.%
\footnote{Figure~\ref{fig-IRF-ar1-estim} is purely illustrative: the shaded area shows the 5th–95th percentile envelope across Monte Carlo replications, \emph{not} bootstrap confidence intervals. 
%Formal inference throughout the paper relies on percentile-$t$ intervals obtained via the corresponding bootstrap procedure (BWB).
}

\begin{figure}[h!]
	\begin{center}
		\caption{Impulse-response functions of AR(1) processes to a one-standard-deviation shock} 
		\label{fig-IRF-ar1}
		\includegraphics[scale=0.6]{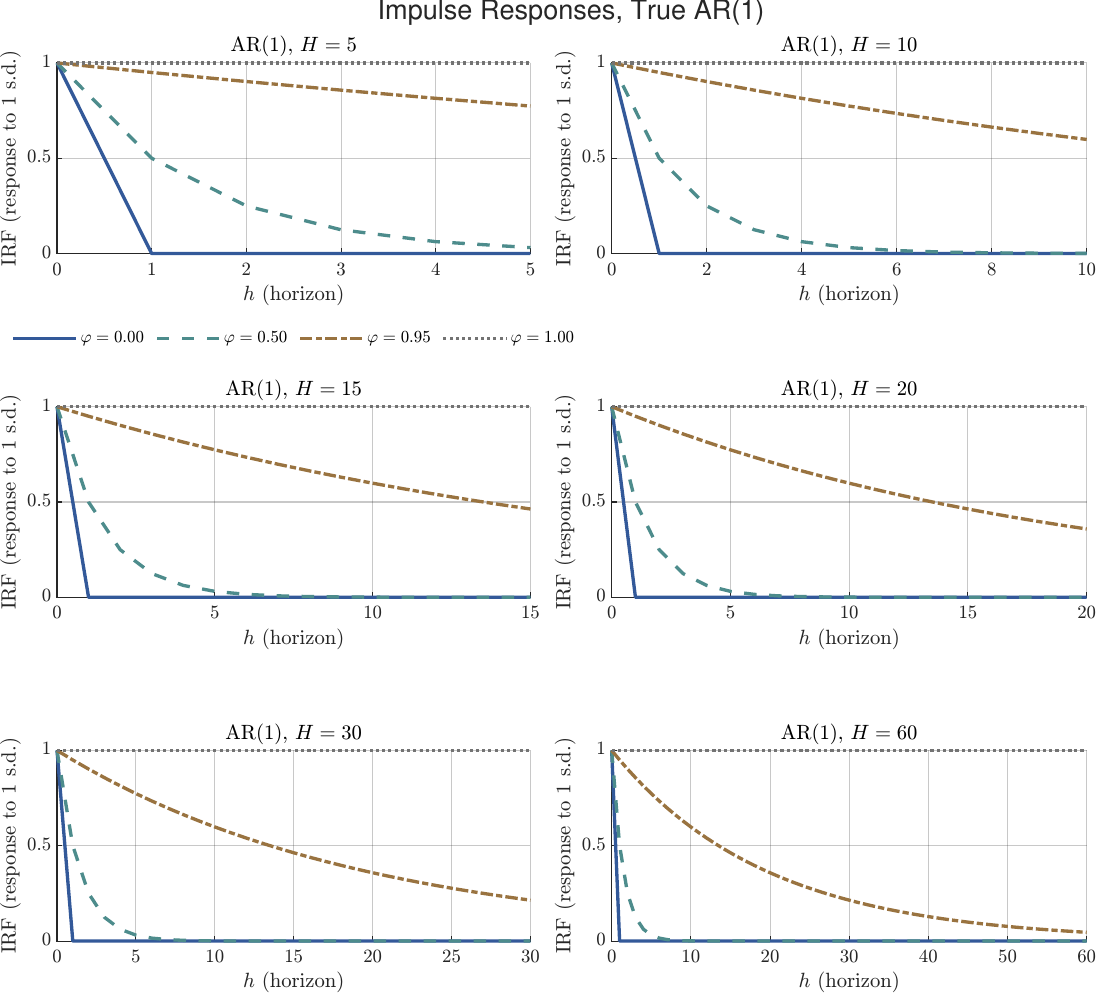}
	\end{center}
	\begin{figurenotes}
		\textit{Note}: Each panel plots the theoretical impulse response for an AR(1) process
		with autoregressive coefficient $\phi \in \{0,0.5,0.95,1\}$ up to the indicated horizon $H$.
		The vertical axis shows the response of the process to a one-standard-deviation innovation,
		the horizontal axis the forecast horizon. The scale is kept fixed across panels
		($y \in [0,1]$) to facilitate comparison across values of~$\phi$.
	\end{figurenotes}
\end{figure}

Figures~\ref{fig-IRF-ar1} and \ref{fig-IRF-ar1-estim} anchor the discussion. 
The former shows the theoretical AR(1) impulse responses for different persistence levels and horizons, keeping the vertical scale fixed across panels to facilitate comparisons. 
The latter overlays the true responses with the simulated parametric AR estimates: each gray line corresponds to one Monte Carlo replication, the dashed line is their Monte Carlo mean, and the shaded area is the 5th–95th percentile envelope.

Two features stand out. At low or moderate persistence ($\phi=0,0.5$), the mean response tracks the theoretical path closely across horizons, and dispersion remains contained even for $T=200$. 
By contrast, with high or unit-root persistence ($\phi=0.95,1$) the spread increases with the horizon; long-horizon responses are noisier and the envelopes widen, reflecting the accumulation of estimation error as $h$ grows.

While Figure~\ref{fig-IRF-ar1-estim} is only illustrative, the subsequent tables report the formal simulation results using our proposed LP--bootstrap methods, which constitute the main object of inference in this paper. Tables~\ref{tab-boot3-method1-cit-ar1} and \ref{tab-boot3-method2-cit-ar1} quantify these patterns in terms
of coverage and median interval length for Local Projections with bootstrap inference. 
For $\phi\le 0.5$, coverage lies near the nominal level across horizons and improves with $T$. 
With $\phi=0.95$ and, especially, $\phi=1$, small samples can exhibit under-coverage at medium/long horizons under Method~1; short horizons may also be sensitive when SBIC selects large orders under high persistence.\footnote{See the entries for $\phi\in\{0.95,1\}$ with SBIC at short horizons in \autoref{tab-boot3-method1-cit-ar1}. Using a small fixed $p$ often raises short-horizon coverage in small samples, and Method~2 tends to improve medium/long horizons—typically at the cost of slightly wider bands; cf.\ \autoref{tab-boot3-method2-cit-ar1}.} 
Increasing the sample to $T=400$ or $T=1000$ mitigates these issues substantially.\footnote{Improvements with $T$ are most visible at medium/long horizons; they need not be monotone at very short $h$ when SBIC picks large $p$ under high persistence.}

Lag specification in the first LP step matters primarily through a bias–variance trade-off. Fixing $p$
instead of using SBIC has little effect at short horizons in low/medium persistence, but distortions can accumulate at medium and
long horizons in high-persistence designs. The tables show that SBIC tends to curb that drift while
keeping intervals reasonably tight.\footnote{With $\phi$ close to unity and small $T$, SBIC may choose large $p$, reducing long-horizon bias yet sometimes lowering short-horizon coverage; with small fixed $p$ the pattern often reverses (better short-horizon coverage, more residual dependence at long horizons). Method~2 partly alleviates this tension.}
This is intuitive: too few lags leave serial correlation in the LP
residuals; too many lags inflate variance. BWB helps with residual dependence but cannot fully offset
either problem when $T$ is small.

Finally, Table~\ref{tab-comp-ar1} reports coverage when inference is based on the traditional autoregressive approach of \citet{Kilian1999}, without bias correction.
This benchmark illustrates the performance of AR-based intervals across persistence levels, horizons, and sample sizes.
Compared with the LP+bootstrap results in the previous table, AR intervals tend to under-cover at medium and long horizons, especially under high persistence, echoing the visual patterns in Figure~\ref{fig-IRF-ar1-estim}.

Taken together, the two tables highlight the trade–off between local–projection and autoregressive inference.
LP combined with BWB delivers coverage closer to nominal at medium and long horizons, adapting more flexibly to persistence in the data.
By contrast, the traditional AR approach of \citet{Kilian1999} tends to under–cover in those ranges, especially under high persistence.
This contrast illustrates the motivation for using LP–based inference with bootstrap refinements in subsequent sections.

\paragraph{Higher-order AR($p$) designs (expanded).}
To avoid redundancy, we do not reproduce AR($p$) figures analogous to
Figures~\ref{fig-IRF-ar1}–\ref{fig-IRF-ar1-estim}. Instead, we summarize numerical results across lag
orders, horizons, and samples in
Tables~\ref{tab-boot3-phi_min03_phi_max09-method1-cit-ARp} 
and~\ref{tab-boot3-phi_min03_phi_max09-method2-cit-ARp}, 
and Appendix Tables~\ref{tab-boot3-phi_min07_phi_max09-method1-cit-ARp}–\ref{tab-boot3-phi_min09_phi_max099-method2-cit-ARp},
and condense the main regularities in Figure~\ref{fig:ARp-coverage-T200H60}, with additional detail in
Appendix~\ref{app:simresults} (Tables \ref{tab-comp-ARp-min03-max09-nobiasc}-\ref{tab-boot3-bias-ARp_nobiasc-high}). Three robust messages emerge:

\begin{enumerate}
	\item \textit{Persistence and sample size.} In low/medium persistence
	($\sum_{i=1}^{p}\phi_i\in[0.3,0.9]$), percentile-$t$ BWB coverage is close to nominal even with $T=200$,
	and interval lengths shrink with $T$. In high persistence ($\sum_{i=1}^{p}\phi_i\in[0.9,0.99]$),
	under-coverage appears first at medium/long horizons and is most pronounced at $T=200$; moving to
	$T=400$–$1000$ restores performance.
	\item \textit{Lag choice in the first LP.} SBIC is a sensible default. Relative to small fixed $p$
	(e.g., $p=1$), SBIC improves medium/long-horizon coverage in persistent designs without materially
	inflating interval length. When the DGP order is large (e.g., $P=10$), underfitting the first-step
	can propagate residual dependence across horizons; BWB alleviates but cannot fully neutralize this in
	small samples.
	\item \textit{Bootstrap implementation (Method~1 vs.\ Method~2).} Including additional MA terms via the
	recursion (method~2) typically yields slightly more conservative long-horizon bands and modestly higher
	coverage when persistence is high or $P$ is large, at the cost of mild increases in interval length.
	In short-memory/low-$P$ designs, both methods perform similarly.
\end{enumerate}

In sum, the AR($p$) evidence reinforces the AR(1) lessons: percentile-$t$ BWB intervals are dependable
across horizons provided the first-step lag choice controls residual dependence and the sample is not too
small in highly persistent designs.

\subsection{MA($q$) models generated with a Gaussian basis function}\label{sec:ma-gbf}

We also consider MA($q$) models in which the moving–average coefficients are generated from a Gaussian
basis function (GBF) to induce richer short– and medium–run dynamics. Specifically, for
\begin{equation}
	y_t=\epsilon_t+\theta_{1}\epsilon_{t-1}+\cdots+\theta_{q}\epsilon_{t-q},\qquad
	\epsilon_t\sim\mathcal{N}(0,1),
\end{equation}
we set $q=24$ and draw the sequence $\{\theta_h\}_{h=1}^{q}$ from
\begin{equation}
	\theta_{h} = 
	\sum_{n=1}^{N} 
	a_{n} \exp\!\!\left[
	- \left( \frac{h - b_{n}}{c_{n}} \right)^{2}
	\right],
	\qquad h = 1, \ldots, q.
\end{equation}
under the ``fair1'' calibration (see Appendix~\ref{app:simresults} for details on $(a_n,b_n,c_n)$ and $N$),
with sample sizes $T\in\{200,400,1000\}$. For an MA($q$), the population impulse response to a one–standard–deviation shock
is $(1,\theta_1,\dots,\theta_q,0,0,\dots)$, i.e. it vanishes for $h>q$.

\begin{figure}[h!]
	\begin{center}
		\caption{Autoregressive (AR) estimates versus true MA(24) responses under GBF design}
		\label{fig-MA24-GBF1-ar}
		\includegraphics[scale=0.6]{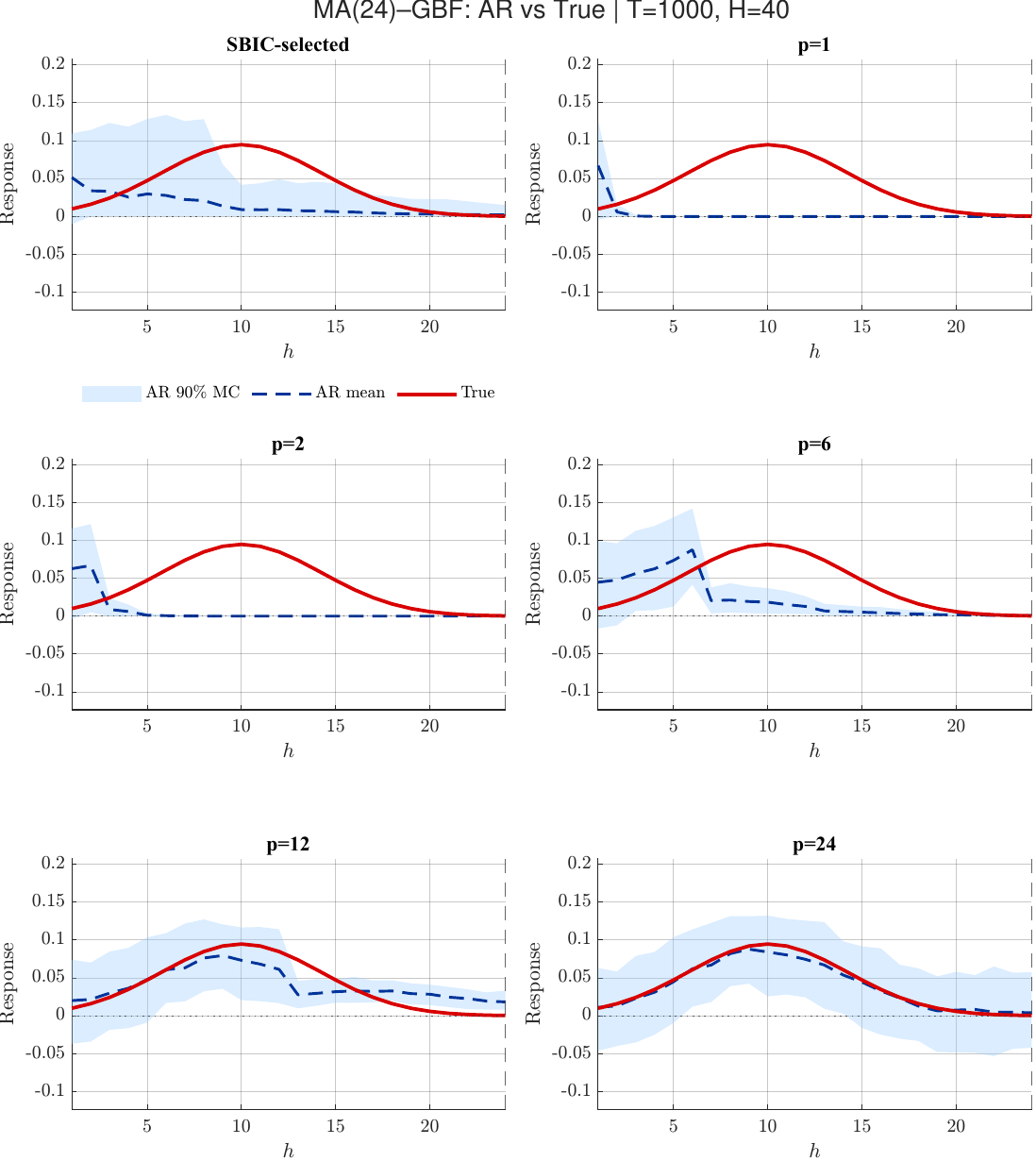}
	\end{center}
	\begin{figurenotes}
		\textit{Note}: Each panel compares the population impulse response of an MA(24) process with GBF coefficients (solid line) to AR($p$) estimates obtained from 100 Monte Carlo replications with $T=1000$. 
		The dashed line is the Monte Carlo mean of the AR estimates, and the shaded area is the 5th–95th percentile envelope across replications (not a bootstrap confidence band). 
		The vertical dashed line marks $h=q=24$; beyond this horizon the true response is zero. 
		AR approximations tend to smear the localized dynamics of the MA process into spurious persistence, producing bias around turning points and wider dispersion at medium horizons, especially when $p$ is small or fixed. 
		Panel labels indicate the AR order: SBIC-selected, 1, 2, 6, 12, and 24.
	\end{figurenotes}
\end{figure}

Figure~\ref{fig-MA-GBF-motiv} illustrates a representative GBF–generated pattern for the true IRF:
the response typically exhibits one or two local maxima and may cross zero before tapering off by $h=q$. We use
this class of designs to test whether bootstrap inference can capture sharp local features (peaks and sign
reversals) without inflating uncertainty excessively at longer horizons. Representative coverage results
appear in Tables~\ref{tab-boot3-ma24-fair1-mod1-cit}--\ref{tab-boot3-ma24-fair1-mod2-cit}; 
Figures~\ref{fig-MA24-GBF1-ar} and \ref{fig-MA24-GBF1-lp} compare AR and LP estimators against the truth.
Additional robustness checks are reported in Appendix~\ref{app:simresults}.

Two findings stand out. First, percentile-$t$ BWB intervals for LP estimates achieve coverage close to the nominal
level over most horizons once $T\ge 400$ \emph{provided the first-step lag order is not too small}.
With fixed moderate values of $p$ (e.g.\ $10,20,30,40,60$), the procedure delivers reliable inference even around peaks and sign reversals.\footnote{See
	\autoref{tab-boot3-ma24-fair1-mod1-cit}--\autoref{tab-boot3-ma24-fair1-mod2-cit}: for $T{=}1000$ and fixed $p$,
	coverage at $h\in\{10,20,40,60\}$ is typically $0.82$–$0.87$. By contrast, when SBIC selects very small $p$
	in this MA($q$) design, residual dependence remains in the first-step LP, producing under-coverage that can even worsen with $T$ (e.g., SBIC, $T{=}1000$, $h{=}10$: $0.70$; $h{=}60$: $0.41$). This SBIC-specific issue does not arise under fixed-$p$ specifications.} 
At $T=200$, coverage dips near turning points—precisely where the IRF curvature is steep and the effective sample is smallest—yet intervals remain reasonably tight and the LP mean still tracks the true shape. 
Second, because MA($q$) responses vanish for $h>q$, coverage often improves again at long horizons (the true response is essentially zero), although small samples may show mild over– or under–coverage as
the signal–to–noise ratio deteriorates.\footnote{The``rebound” at long horizons is most visible under fixed $p$; with SBIC,
	coverage may remain low if the selected order is too parsimonious for the MA($q$) environment (see the SBIC rows in
	\autoref{tab-boot3-ma24-fair1-mod1-cit}–\autoref{tab-boot3-ma24-fair1-mod2-cit}).}

Relative to AR estimation, LP is much better aligned with the finite–memory nature of the DGP:
AR approximations smear localized dynamics into artificial persistence, leading to biased responses around peaks
and systematically low coverage at medium horizons (see Figure~\ref{fig-MA24-GBF1-ar} vs.\ Figure~\ref{fig-MA24-GBF1-lp}).

Overall, the BWB with percentile-$t$ corrections provides reliable inference for IRFs with localized
features generated by GBF coefficients, particularly once $T$ reaches 400 or 1000 and the first-step LP
includes a \emph{moderate} number of lags. The most challenging regions are turning points: practitioners
should anticipate wider bands and occasional coverage shortfalls there—especially in short samples—and avoid
overly parsimonious lag choices that leave residual autocorrelation in the first-step LP.\footnote{In practice,
	it is advisable either to impose a sensible lower bound on $p$ when using SBIC in finite-memory MA($q$) settings, or to use a modest fixed $p$
	(e.g.\ $10$–$20$) to stabilize coverage across horizons; cf. the fixed-$p$ rows in
	\autoref{tab-boot3-ma24-fair1-mod1-cit}–\autoref{tab-boot3-ma24-fair1-mod2-cit}.}
	
	Summing up, the results for MA($q$)–GBF designs highlight two robust lessons. 
	First, the BWB percentile-$t$ procedure delivers reliable inference once the sample is moderately large ($T\ge 400$) and the first-step lag length is kept at sensible fixed values (e.g., $p=10$–$20$). 
	Second, across designs, the main practical pitfall arises with SBIC: while convenient in principle, automatic selection often chooses too few lags in finite-memory environments, leading to residual dependence in the first-step LP and systematic under-coverage at medium and long horizons. 
	In short samples, coverage deteriorates mainly around turning points—precisely where the IRF curvature is steep—but interval lengths remain moderate. 
	Overall, the bootstrap-based methods are well suited for designs with localized dynamics, provided practitioners guard against overly parsimonious lag specifications in the initial projection step.
	
\section{Main Simulation Insights}
\label{sec:main-insights}

The simulation exercises reported in Section~\ref{sec:sim-results}—covering univariate AR(1), higher-order AR($p$), and MA($q$) designs with Gaussian basis functions (GBF)—yield a set of consistent takeaways about the performance of the Block Wild Bootstrap (BWB) for local-projection inference. Unless otherwise noted, results are based on percentile-$t$ BWB intervals at \nominalCov{} ($\alpha=\nominalAlpha$), with no small-sample bias correction.

\begin{figure}[h!]
	\begin{center}
		\caption{Coverage of percentile-$t$ BWB confidence intervals for AR($p$) designs (by persistence regime)}
		\label{fig:ARp-coverage-T200H60}
		\includegraphics[scale=0.72]{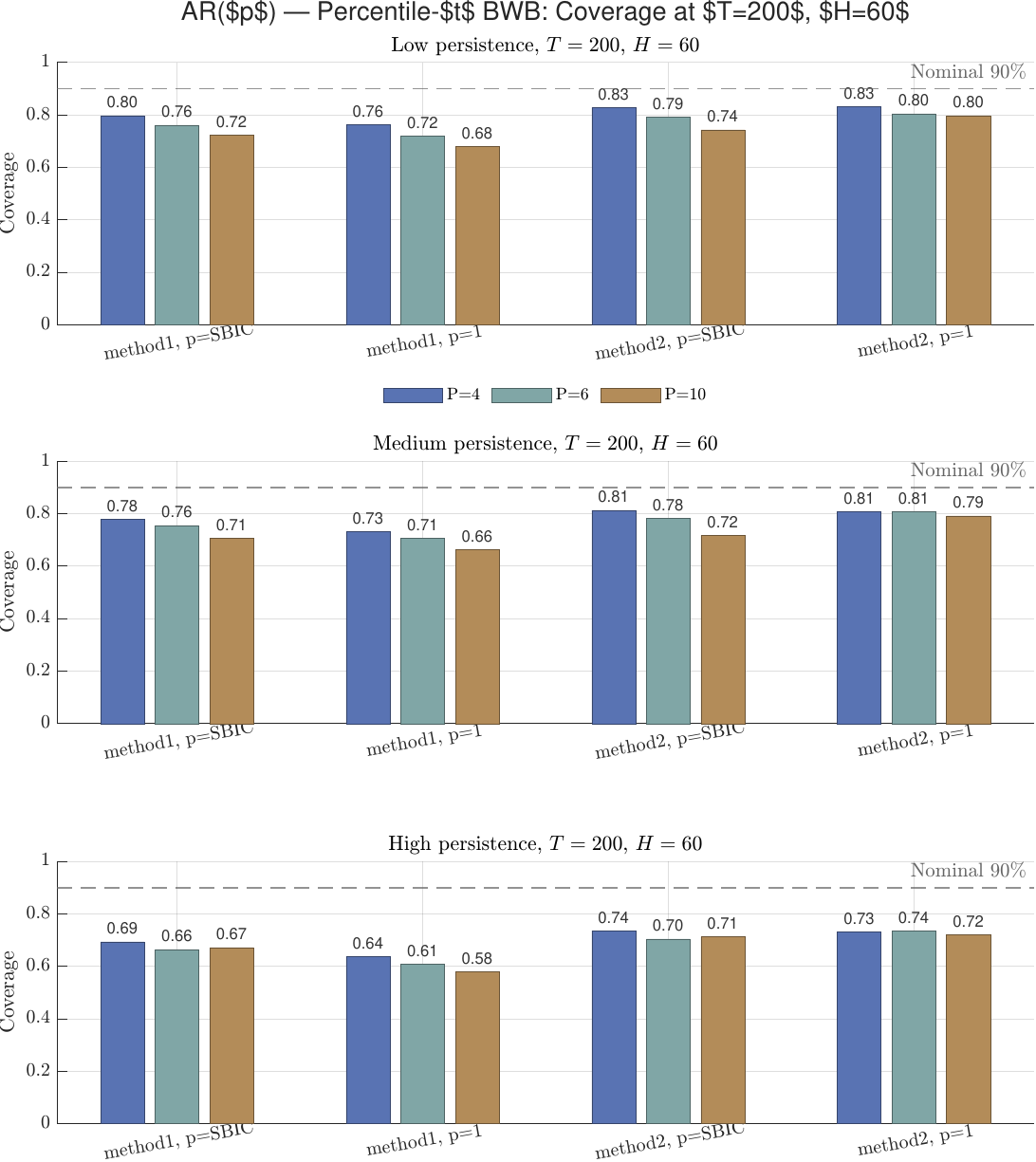}
	\end{center}
\begin{figurenotes}
	\textit{Note}: Coverage rates at horizon $H=60$ for sample size $T=200$ across three persistence regimes (rows).
	Bars compare the first-step LP lag specification (SBIC vs.\ fixed $p=1$) and the bootstrap implementation (method~1 vs.\ method~2);
	colors denote the DGP order $P\in\{4,6,10\}$. The dashed line marks nominal  \nominalCov{} coverage.
	Underlying numerical results for additional horizons and sample sizes appear in Tables~4--13 and Appendix~\ref{app:simresults}.
\end{figurenotes}
\end{figure}

\begin{figure}[h!]
	\begin{center}
		\caption{Local–projection (LP) estimates versus true MA(24) responses under GBF design}
		\label{fig-MA24-GBF1-lp}
		\includegraphics[scale=0.6]{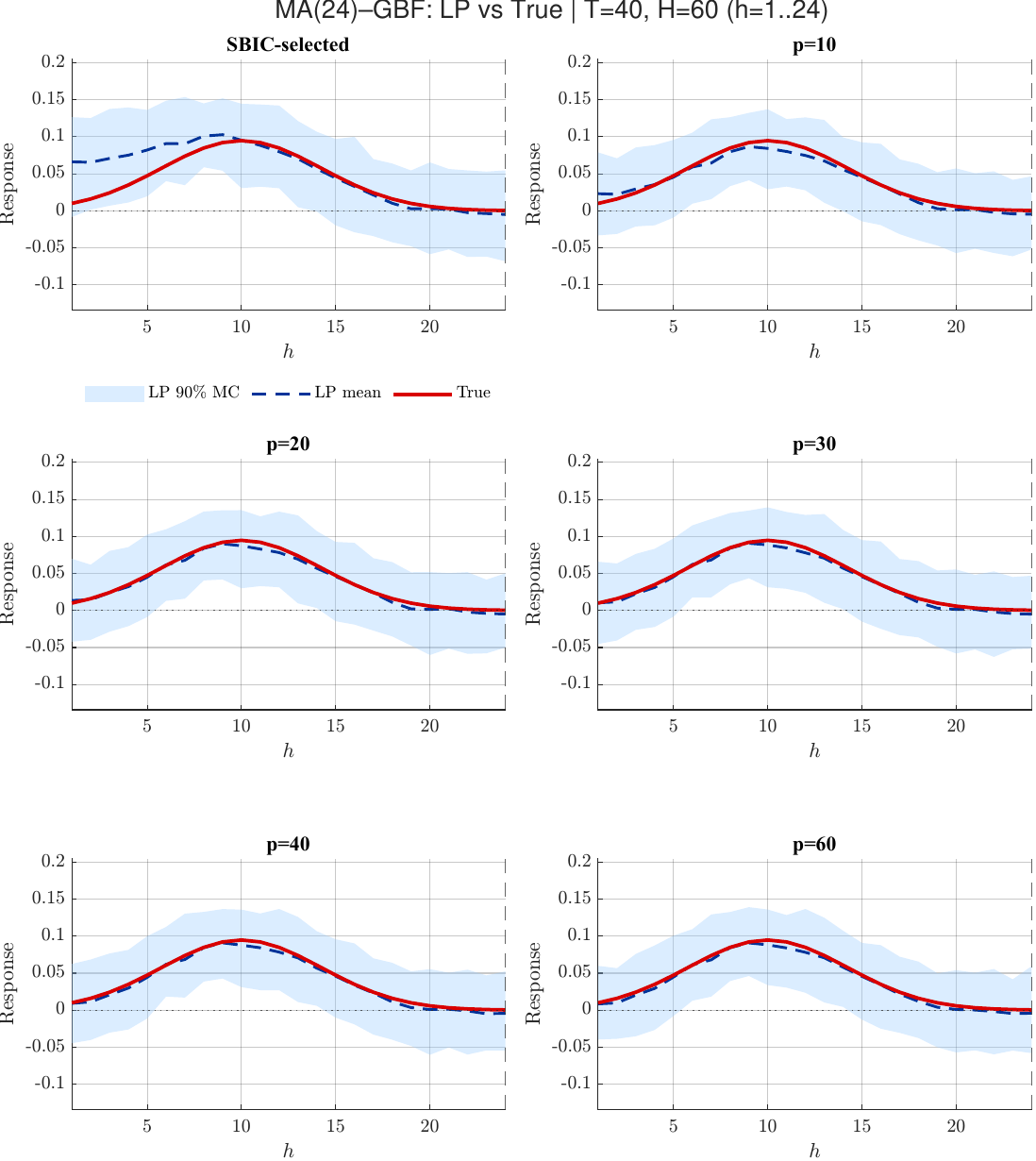}
	\end{center}
\begin{figurenotes}
	\textit{Note}: Each panel compares the population impulse response of an MA(24) process with GBF coefficients (solid line) to Local Projection (LP) estimates obtained from 100 Monte Carlo replications with $T=1000$. 
	The dashed line is the Monte Carlo mean of the LP estimates, and the shaded area is the 5th--95th percentile envelope across replications (this is \emph{not} a bootstrap confidence band). 
	The vertical dashed line marks the MA order, $h=q=24$; horizons beyond $q$ are omitted since the true response is zero. 
	The first-step LP lag length $p$ is either SBIC-selected or fixed as indicated by the panel labels: SBIC-selected, $p\!=\!10,20,30,40,60$.
\end{figurenotes}
\end{figure}

\begin{itemize}
	\item \textbf{Overall performance.} Across designs and horizons, BWB delivers coverage close to nominal with stable interval lengths. This is visible in AR(1) (\autoref{fig-IRF-ar1}; Tables~\ref{tab-boot3-method1-cit-ar1}--\ref{tab-boot3-method2-cit-ar1}), extends to AR($p$) at low to high persistence (Tables~\ref{tab-boot3-phi_min03_phi_max09-method1-cit-ARp}--\ref{tab-boot3-phi_min09_phi_max099-method2-cit-ARp}, \autoref{fig:ARp-coverage-T200H60}), and carries over to finite-memory MA($q$)–GBF designs (Tables~\ref{tab-boot3-ma24-fair1-mod1-cit}--\ref{tab-boot3-ma24-fair1-mod2-cit}, \autoref{fig-MA-GBF-motiv}, \autoref{fig-MA24-GBF1-lp}).\footnote{The contrast between SBIC and fixed-$p$ specifications arises in these MA($q$)–GBF designs, where SBIC may under-select the lag order and leave residual dependence in the first-step LP. This pattern is not a general property of SBIC and should not be extrapolated to other DGPs.} A few exceptions relate to implementation choices (first-step lagging, MA truncation vs.\ recursion) rather than to BWB per se.\footnote{For AR(1) with high/near-unit persistence, Method~1 (truncation at $H$) combined with SBIC can yield low short-horizon coverage that does not improve monotonically with $T$ (e.g., \autoref{tab-boot3-method1-cit-ar1}, $\phi{=}0.95$). This largely disappears under Method~2 (recursion beyond $H$), where short-horizon coverage is well calibrated (\autoref{tab-boot3-method2-cit-ar1}).}.
	
	\item \textbf{Horizon–persistence trade-off.} Uncertainty rises with the forecast horizon and interacts with persistence. In AR(1) with $\phi\in\{0.95,1\}$, small samples under-cover at medium/long horizons, while short horizons are generally well behaved; increasing $T$ from 200 to 400 or 1000 markedly improves coverage (Tables~\ref{tab-boot3-method1-cit-ar1}--\ref{tab-boot3-method2-cit-ar1}). A similar pattern appears in AR($p$) at high persistence and in MA($q$) near turning points of the true IRF.
	
	\item \textbf{Lag selection in the first LP step.} SBIC is a sensible default in AR environments: relative to very small fixed $p$, it mitigates residual serial correlation without over-inflating variance, improving medium/long-horizon coverage when persistence is high (\autoref{fig:ARp-coverage-T200H60} and the AR($p$) tables). For finite-memory MA($q$) designs, however, SBIC may select overly parsimonious $p$ and leave residual autocorrelation, depressing coverage—sometimes more as $T$ grows—whereas modest fixed $p$ (e.g., 10–20) stabilizes performance across horizons.\footnote{Compare SBIC vs.\ fixed-$p$ rows in \autoref{tab-boot3-ma24-fair1-mod1-cit}--\autoref{tab-boot3-ma24-fair1-mod2-cit}: with $T{=}1000$, SBIC shows lower coverage at several horizons, consistent with underfitting in the first-step LP.}
	
	\item \textbf{Finite-memory designs (MA($q$)–GBF).} LP+BWB tracks localized features (peaks and sign reversals) with coverage close to nominal once $T\ge 400$. At $T=200$, coverage may dip around turning points—where curvature is steep—yet interval lengths remain moderate and the LP mean retains the shape of the true IRF (Tables~\ref{tab-boot3-ma24-fair1-mod1-cit}--\ref{tab-boot3-ma24-fair1-mod2-cit}, \autoref{fig-MA24-GBF1-lp}). Because MA responses vanish for $h>q$, long-horizon coverage often improves again.
	
	\item \textbf{Method~1 vs.\ Method~2.} Allowing MA terms beyond $H$ via the recursion (Method~2) yields clear gains in highly persistent settings—especially at short horizons when $\phi$ is near one and SBIC is used—and small improvements elsewhere, at the cost of slightly longer intervals. Under low/medium persistence the two methods perform similarly.
	
	\item \textbf{LP vs.\ AR (truth tracking).} When benchmarked against the \emph{true} IRF, LP+BWB aligns more closely with finite-memory dynamics than simple AR-based approximations, which can smear localized features into spurious persistence (cf.\ \autoref{fig-MA24-GBF1-ar} vs.\ \autoref{fig-MA24-GBF1-lp}). The Monte Carlo envelopes shown in illustrative figures (e.g., \autoref{fig-IRF-ar1-estim}) are \emph{not} bootstrap confidence bands and are included to visualize estimator variability.
	
	\item \textbf{Practical guidance.} (i) Use BWB as the default resampling scheme for LP inference; (ii) in AR settings, select $p$ by SBIC; for finite-memory MA($q$), either impose a modest lower bound under SBIC or use a small fixed $p$ (e.g., 10–20); (iii) prefer Method~2 in highly persistent designs or when long-horizon inference matters; and (iv) expect wider bands and some under-coverage at long horizons in small samples, and prioritize $T\ge 400$ when feasible.
\end{itemize}

\noindent
\textit{Summary.} BWB paired with local projections provides a reliable and implementable route to inference on impulse responses across a variety of univariate designs and horizons. Empirical coverage is close to nominal, performance is stable across tuning choices when the first-step LP is well specified, and accuracy scales from short to medium/long horizons as sample size increases. In practice, the choice of lag length in the first-step LP has a much stronger influence on coverage performance than the distinction between Method~1 and Method~2, whose differences are generally marginal. These properties make the BWB a natural benchmark for applied work with local projections in macroeconomics and finance. 

In simulations not reported here (but available upon request), we also experimented with standard versions of the bootstrap without correcting for serial correlation. We found negligible losses in coverage as would be expected. The reason is that our bootstrap procedure includes an extra adjustment for leftover serial correlation at long lags. Of course, in practice the extra insurance provided by using the BWB procedure seems a small price to pay although in practice it may not yield very big gains.

\section{Concluding remarks}
Bootstrap inference for impulse responses estimated by local projections has often been implemented through VAR($p$)-based resampling. Since consistency of VAR-based IRFs is only guaranteed up to the lag order, this strategy can be fragile at longer horizons. We propose an alternative algorithm that exploits the moving-average representation naturally associated with local projections: bootstrap replicates are generated from a modified version of the moving-average procedure in \citet{Paparoditis1996}. Coupled with the Block Wild Bootstrap \citep{Shao2011}, the method accommodates serial dependence in the bootstrap weights while preserving the LP structure. Simulation results show that the BWB–LP approach provides reliable coverage and stable interval lengths across a wide range of designs, making it a practical and robust option for applied inference on impulse responses.

\clearpage
%----------------------------------------------------------------------------------------
%	REFERENCE LIST
%----------------------------------------------------------------------------------------
\small\singlespacing
\bibliographystyle{authordate1}
\setcounter{secnumdepth}{0}
\renewcommand\bibsection{\section{\refname}}
\bibliography{LP_Boot_refs}

\clearpage

\newpage
\section{Tables and Figures}
\subsection{AR(1)}

\clearpage
 \begin{table}[h!]\caption{Local-projection bootstrap results, AR(1), Method~1 (BWB, percentile-$t$)}\label{tab-boot3-method1-cit-ar1}\begin{center}\scalebox{0.6}{\begin{tabular}{ll|cccccc|cccccc} \hline \hline 
				&& \multicolumn{6}{c}{Coverage} & \multicolumn{6}{c}{Median interval length}\\  
				$T$& $p/H$& 5& 10& 15& 20& 30& 60& 5& 10& 15& 20& 30& 60\\ \hline 
				&& \multicolumn{12}{c}{$\phi=0$}\\  
				200&p-SBIC&0.92&0.93&0.93&0.93&0.93&0.88&1.18&1.12&1.12&1.10&1.11&1.10\\ 
				&1&0.90&0.92&0.92&0.92&0.92&0.88&1.16&1.11&1.11&1.09&1.10&1.09\\ 
				&2&0.89&0.90&0.91&0.91&0.91&0.88&1.17&1.11&1.10&1.07&1.08&1.06\\ 
				&3&0.88&0.89&0.91&0.91&0.90&0.87&1.18&1.12&1.09&1.07&1.07&1.05\\ 
				\cline{2-14} 
				400&p-SBIC&0.94&0.95&0.95&0.95&0.96&0.94&1.08&1.16&1.13&1.15&1.17&1.17\\ 
				&1&0.92&0.94&0.94&0.94&0.95&0.94&1.08&1.15&1.13&1.14&1.16&1.16\\ 
				&2&0.92&0.95&0.93&0.94&0.95&0.93&1.09&1.14&1.12&1.12&1.14&1.14\\ 
				&3&0.89&0.93&0.93&0.93&0.94&0.93&1.09&1.12&1.10&1.10&1.12&1.12\\ 
				\cline{2-14} 
				1000&p-SBIC&0.94&0.97&0.96&0.96&0.97&0.96&1.18&1.22&1.19&1.19&1.20&1.19\\ 
				&1&0.94&0.96&0.96&0.96&0.97&0.96&1.18&1.21&1.19&1.18&1.19&1.18\\ 
				&2&0.92&0.95&0.95&0.95&0.96&0.96&1.18&1.21&1.19&1.18&1.18&1.17\\ 
				&3&0.90&0.93&0.94&0.95&0.96&0.96&1.17&1.20&1.17&1.17&1.17&1.16\\ 
				\hline 
				&& \multicolumn{12}{c}{$\phi=0.5$}\\  
				200&p-SBIC&0.90&0.91&0.90&0.91&0.91&0.86&1.10&1.09&1.07&1.04&1.05&1.08\\ 
				&1&0.92&0.92&0.91&0.91&0.91&0.86&1.18&1.14&1.12&1.08&1.06&1.06\\ 
				&2&0.91&0.90&0.90&0.90&0.90&0.86&1.16&1.12&1.09&1.06&1.05&1.05\\ 
				&3&0.90&0.90&0.89&0.89&0.90&0.86&1.17&1.13&1.09&1.04&1.04&1.03\\ 
				\cline{2-14} 
				400&p-SBIC&0.93&0.94&0.94&0.94&0.94&0.93&1.15&1.16&1.12&1.12&1.14&1.15\\ 
				&1&0.92&0.94&0.94&0.94&0.94&0.93&1.14&1.17&1.12&1.12&1.13&1.13\\ 
				&2&0.90&0.93&0.93&0.94&0.94&0.93&1.15&1.15&1.11&1.11&1.11&1.11\\ 
				&3&0.90&0.92&0.92&0.92&0.93&0.92&1.16&1.15&1.10&1.10&1.10&1.10\\ 
				\cline{2-14} 
				1000&p-SBIC&0.91&0.95&0.96&0.96&0.96&0.96&1.23&1.21&1.17&1.16&1.19&1.21\\ 
				&1&0.93&0.96&0.96&0.96&0.97&0.96&1.22&1.21&1.18&1.17&1.18&1.19\\ 
				&2&0.91&0.95&0.95&0.95&0.96&0.96&1.21&1.20&1.17&1.16&1.17&1.18\\ 
				&3&0.92&0.94&0.94&0.95&0.96&0.96&1.21&1.20&1.17&1.16&1.17&1.17\\ 
				\hline 
				&& \multicolumn{12}{c}{$\phi=0.95$}\\  
				200&p-SBIC&0.39&0.62&0.71&0.73&0.75&0.68&0.92&0.91&0.88&0.87&0.83&0.78\\ 
				&1&0.75&0.83&0.85&0.84&0.83&0.72&1.21&1.19&1.11&1.03&0.97&0.85\\ 
				&2&0.65&0.80&0.83&0.82&0.81&0.72&1.12&1.14&1.09&1.02&0.96&0.84\\ 
				&3&0.63&0.76&0.81&0.81&0.79&0.70&1.06&1.09&1.05&0.98&0.93&0.82\\ 
				\cline{2-14} 
				400&p-SBIC&0.24&0.47&0.66&0.74&0.79&0.75&0.98&0.91&0.88&0.89&0.86&0.82\\ 
				&1&0.65&0.84&0.88&0.87&0.87&0.82&1.27&1.27&1.19&1.17&1.06&1.00\\ 
				&2&0.53&0.80&0.86&0.87&0.87&0.82&1.16&1.21&1.17&1.16&1.06&0.99\\ 
				&3&0.49&0.77&0.85&0.87&0.86&0.81&1.11&1.17&1.14&1.15&1.05&0.99\\ 
				\cline{2-14} 
				1000&p-SBIC&0.23&0.36&0.47&0.60&0.76&0.86&1.09&1.02&1.01&1.02&1.04&1.01\\ 
				&1&0.44&0.77&0.88&0.90&0.92&0.91&1.36&1.35&1.32&1.25&1.19&1.10\\ 
				&2&0.30&0.71&0.87&0.89&0.92&0.91&1.24&1.28&1.29&1.24&1.19&1.09\\ 
				&3&0.23&0.67&0.85&0.89&0.92&0.90&1.19&1.24&1.26&1.23&1.18&1.09\\ 
				\hline 
				&& \multicolumn{12}{c}{$\phi=1$}\\  
				200&p-SBIC&0.33&0.49&0.54&0.57&0.59&0.51&1.02&0.94&0.85&0.81&0.77&0.66\\ 
				&1&0.61&0.78&0.78&0.76&0.71&0.59&1.13&1.13&1.08&1.01&0.94&0.75\\ 
				&2&0.54&0.71&0.73&0.74&0.69&0.56&1.05&1.09&1.06&1.02&0.94&0.76\\ 
				&3&0.51&0.67&0.71&0.72&0.68&0.56&0.99&1.03&1.02&1.00&0.92&0.75\\ 
				\cline{2-14} 
				400&p-SBIC&0.14&0.29&0.39&0.46&0.53&0.58&1.09&1.03&0.98&1.00&0.89&0.75\\ 
				&1&0.43&0.67&0.78&0.80&0.78&0.70&1.29&1.30&1.27&1.24&1.14&0.90\\ 
				&2&0.33&0.58&0.74&0.78&0.78&0.69&1.17&1.22&1.20&1.21&1.12&0.89\\ 
				&3&0.28&0.52&0.70&0.76&0.76&0.69&1.11&1.16&1.15&1.16&1.09&0.88\\ 
				\cline{2-14} 
				1000&p-SBIC&0.08&0.19&0.24&0.30&0.36&0.51&1.09&0.97&0.94&0.94&0.86&0.77\\ 
				&1&0.15&0.38&0.53&0.66&0.74&0.80&1.35&1.35&1.33&1.32&1.26&1.14\\ 
				&2&0.11&0.30&0.47&0.61&0.72&0.79&1.23&1.27&1.29&1.28&1.24&1.13\\ 
				&3&0.08&0.26&0.42&0.57&0.68&0.78&1.17&1.21&1.24&1.24&1.22&1.13\\ 
											\hline \hline \end{tabular}}\end{center}
\begin{tablenotes}
	\textit{Note}: Coverage rates and median interval length for percentile-$t$ confidence intervals based on the \emph{Block Wild Bootstrap} (BWB). Results are reported for AR(1) processes with different persistence ($\phi$), sample sizes ($T$), and lag selections (SBIC or fixed $p$) in the first LP regression. Columns denote horizons $h$. Coverage is the fraction of intervals containing the true IRF. Interval length is expressed relative to the scale of the estimated impulse response.
\end{tablenotes}
\end{table}

			\clearpage
			\begin{table}[h!]\caption{AR(1): coverage for intervals targeting the true IRF vs.\ the estimated IRF}\label{tab-comp-ar1}\begin{center}\scalebox{0.65}{\begin{tabular}{ll|cccccc} \hline \hline 
							&& \multicolumn{6}{c}{Coverage}\\  
							$T$& $p/H$& 5& 10& 15& 20& 30& 60\\ \hline 
							&& \multicolumn{6}{c}{$\phi=0$}\\  
							200&p-SBIC&0.61&0.52&0.53&0.50&0.50&0.49\\ 
							&1&0.59&0.49&0.50&0.47&0.46&0.46\\ 
							&2&0.90&0.93&0.95&0.95&0.96&0.97\\ 
							&3&0.93&0.94&0.95&0.96&0.96&0.97\\ 
							\cline{2-8} 
							400&p-SBIC&0.61&0.51&0.52&0.49&0.48&0.47\\ 
							&1&0.61&0.50&0.51&0.48&0.47&0.46\\ 
							&2&0.87&0.92&0.94&0.95&0.96&0.98\\ 
							&3&0.94&0.93&0.94&0.95&0.96&0.98\\ 
							\cline{2-8} 
							1000&p-SBIC&0.61&0.50&0.52&0.48&0.47&0.47\\ 
							&1&0.60&0.50&0.51&0.48&0.47&0.46\\ 
							&2&0.84&0.89&0.92&0.93&0.94&0.95\\ 
							&3&0.92&0.90&0.91&0.92&0.94&0.96\\ 
							\hline 
							&& \multicolumn{6}{c}{$\phi=0.5$}\\  
							200&p-SBIC&0.84&0.83&0.83&0.83&0.82&0.82\\ 
							&1&0.84&0.83&0.83&0.83&0.82&0.82\\ 
							&2&0.89&0.90&0.90&0.90&0.91&0.91\\ 
							&3&0.91&0.95&0.96&0.97&0.98&0.99\\ 
							\cline{2-8} 
							400&p-SBIC&0.90&0.89&0.88&0.88&0.87&0.87\\ 
							&1&0.92&0.91&0.91&0.90&0.90&0.90\\ 
							&2&0.92&0.91&0.90&0.90&0.90&0.91\\ 
							&3&0.92&0.95&0.96&0.97&0.98&0.98\\ 
							\cline{2-8} 
							1000&p-SBIC&0.89&0.89&0.89&0.89&0.89&0.89\\ 
							&1&0.89&0.89&0.89&0.89&0.89&0.89\\ 
							&2&0.93&0.94&0.95&0.95&0.95&0.96\\ 
							&3&0.90&0.91&0.91&0.92&0.93&0.94\\ 
							\hline 
							&& \multicolumn{6}{c}{$\phi=0.95$}\\  
							200&p-SBIC&0.66&0.63&0.62&0.61&0.61&0.61\\ 
							&1&0.68&0.64&0.63&0.62&0.62&0.61\\ 
							&2&0.74&0.68&0.66&0.65&0.64&0.64\\ 
							&3&0.77&0.70&0.67&0.65&0.63&0.61\\ 
							\cline{2-8} 
							400&p-SBIC&0.80&0.78&0.77&0.77&0.76&0.76\\ 
							&1&0.80&0.78&0.77&0.77&0.77&0.77\\ 
							&2&0.87&0.83&0.81&0.79&0.78&0.78\\ 
							&3&0.86&0.81&0.79&0.78&0.77&0.77\\ 
							\cline{2-8} 
							1000&p-SBIC&0.85&0.84&0.84&0.84&0.83&0.82\\ 
							&1&0.86&0.85&0.84&0.84&0.83&0.82\\ 
							&2&0.85&0.84&0.84&0.84&0.84&0.85\\ 
							&3&0.88&0.86&0.85&0.84&0.84&0.83\\ 
							\hline 
							&& \multicolumn{6}{c}{$\phi=1$}\\  
							200&p-SBIC&0.47&0.34&0.27&0.22&0.17&0.13\\ 
							&1&0.45&0.32&0.25&0.21&0.16&0.12\\ 
							&2&0.61&0.45&0.36&0.30&0.23&0.15\\ 
							&3&0.68&0.52&0.41&0.35&0.27&0.18\\ 
							\cline{2-8} 
							400&p-SBIC&0.60&0.43&0.34&0.29&0.23&0.17\\ 
							&1&0.58&0.42&0.34&0.28&0.23&0.17\\ 
							&2&0.76&0.60&0.48&0.41&0.32&0.22\\ 
							&3&0.79&0.66&0.54&0.45&0.35&0.23\\ 
							\cline{2-8} 
							1000&p-SBIC&0.71&0.55&0.45&0.38&0.30&0.21\\ 
							&1&0.71&0.55&0.44&0.37&0.30&0.21\\ 
							&2&0.82&0.70&0.59&0.52&0.42&0.27\\ 
							&3&0.85&0.77&0.67&0.60&0.49&0.32\\ 
																\hline \hline \end{tabular}}\end{center}
				\begin{tablenotes}
\textit{Note}: Coverage probabilities for non–bias-corrected percentile-$t$ intervals (Kilian, 1999) with $\alpha=0.1$ (i.e., 90\% bands), computed under the AR benchmark.
Rows vary $T$ and the lag specification (SBIC or fixed $p$); columns report horizons $h$.
“Coverage” is the fraction of intervals containing the \emph{true} IRF.
Values around $0.50$–$0.60$ correspond to coverage when the \emph{target} is the \emph{estimated} IRF rather than the true IRF; these are not comparable to the nominal \nominalCov{} rate and are shown only for reference.
\end{tablenotes}
\end{table}

\clearpage

	\subsection{AR(p) models}

 \begin{table}[h!]\caption{AR($p$), low persistence: percentile-$t$ BWB coverage and median interval length (Method~1)}\label{tab-boot3-phi_min03_phi_max09-method1-cit-ARp}\begin{center}\scalebox{0.8}{\begin{tabular}{ll|cccccc|cccccc} \hline \hline 
				&& \multicolumn{6}{c}{Coverage} & \multicolumn{6}{c}{median interval length}\\  
				$T$& $p/H$& 5& 10& 15& 20& 30& 60& 5& 10& 15& 20& 30& 60\\ \hline 
				&& \multicolumn{12}{c}{P=4}\\  
				200&p-SBIC&0.84&0.87&0.85&0.86&0.84&0.80&0.23&0.29&0.32&0.36&0.44&0.58\\ 
				&1&0.63&0.69&0.72&0.73&0.74&0.76&0.17&0.22&0.26&0.32&0.39&0.60\\ 
				&P true&0.84&0.87&0.85&0.86&0.84&0.79&0.23&0.29&0.32&0.35&0.44&0.58\\ 
				\cline{2-14} 
				400&p-SBIC&0.80&0.86&0.88&0.90&0.90&0.88&0.17&0.21&0.24&0.28&0.34&0.52\\ 
				&1&0.59&0.61&0.65&0.68&0.71&0.77&0.11&0.16&0.19&0.24&0.33&0.53\\ 
				&P true&0.80&0.86&0.88&0.89&0.90&0.88&0.17&0.21&0.24&0.28&0.34&0.52\\ 
				\cline{2-14} 
				1000&p-SBIC&0.75&0.84&0.87&0.90&0.92&0.92&0.11&0.14&0.16&0.19&0.23&0.37\\ 
				&1&0.48&0.57&0.59&0.61&0.65&0.74&0.07&0.10&0.13&0.16&0.21&0.36\\ 
				&P true&0.75&0.84&0.87&0.90&0.91&0.92&0.10&0.14&0.16&0.19&0.23&0.37\\ 
				\hline 
				&& \multicolumn{12}{c}{P=6}\\  
				200&p-SBIC&0.81&0.83&0.82&0.81&0.81&0.76&0.23&0.28&0.30&0.34&0.40&0.51\\ 
				&1&0.53&0.57&0.62&0.65&0.68&0.72&0.15&0.20&0.24&0.28&0.35&0.56\\ 
				&P true&0.82&0.84&0.82&0.82&0.81&0.76&0.23&0.27&0.31&0.35&0.40&0.52\\ 
				\cline{2-14} 
				400&p-SBIC&0.75&0.82&0.81&0.84&0.86&0.84&0.16&0.21&0.23&0.26&0.30&0.41\\ 
				&1&0.46&0.47&0.52&0.56&0.60&0.67&0.11&0.15&0.17&0.20&0.26&0.38\\ 
				&P true&0.76&0.81&0.81&0.84&0.86&0.84&0.17&0.21&0.23&0.26&0.30&0.41\\ 
				\cline{2-14} 
				1000&p-SBIC&0.66&0.76&0.81&0.83&0.87&0.90&0.11&0.14&0.16&0.18&0.21&0.28\\ 
				&1&0.37&0.41&0.42&0.45&0.50&0.61&0.07&0.10&0.12&0.14&0.17&0.26\\ 
				&P true&0.66&0.76&0.81&0.83&0.87&0.90&0.11&0.14&0.16&0.18&0.21&0.28\\ 
				\hline 
				&& \multicolumn{12}{c}{P=10}\\  
				200&p-SBIC&0.74&0.76&0.76&0.74&0.76&0.72&0.20&0.23&0.27&0.29&0.33&0.46\\ 
				&1&0.45&0.43&0.49&0.53&0.58&0.68&0.14&0.18&0.22&0.26&0.32&0.51\\ 
				&P true&0.76&0.76&0.77&0.74&0.77&0.73&0.20&0.23&0.27&0.29&0.33&0.46\\ 
				\cline{2-14} 
				400&p-SBIC&0.72&0.71&0.74&0.78&0.80&0.81&0.15&0.17&0.20&0.22&0.26&0.35\\ 
				&1&0.38&0.39&0.40&0.46&0.50&0.58&0.11&0.13&0.16&0.18&0.22&0.31\\ 
				&P true&0.73&0.71&0.75&0.78&0.80&0.81&0.15&0.17&0.20&0.22&0.26&0.35\\ 
				\cline{2-14} 
				1000&p-SBIC&0.65&0.65&0.69&0.73&0.78&0.85&0.10&0.11&0.14&0.15&0.18&0.26\\ 
				&1&0.34&0.29&0.33&0.38&0.43&0.50&0.07&0.09&0.11&0.12&0.15&0.21\\ 
				&P true&0.66&0.64&0.69&0.73&0.78&0.85&0.10&0.11&0.14&0.15&0.18&0.26\\  
				\hline \hline \end{tabular}}\end{center}
\begin{tablenotes}
	\textit{Note}: Coverage probabilities and median interval length for percentile-$t$ confidence intervals based on the Block Wild Bootstrap (BWB).
	Rows vary the first-step LP lag choice (SBIC vs.\ fixed $p$) and the DGP order ($P\in\{4,6,10\}$); columns report horizons $h$.
	Low persistence is defined as $\sum_{i=1}^{p}\phi_i\in[0.3,0.9]$.
	Interval length is reported relative to the scale of the estimated response at each $h$.
\end{tablenotes}
\end{table}

\begin{table}[h!]\caption{AR($p$), low persistence: percentile-$t$ BWB coverage and median interval length (Method~2)}\label{tab-boot3-phi_min03_phi_max09-method2-cit-ARp}\begin{center}\scalebox{0.8}{\begin{tabular}{ll|cccccc|cccccc} \hline \hline 
				&& \multicolumn{6}{c}{Coverage} & \multicolumn{6}{c}{Median interval length}\\  
				$T$& $p/H$& 5& 10& 15& 20& 30& 60& 5& 10& 15& 20& 30& 60\\ \hline 
				&& \multicolumn{12}{c}{P=4}\\  
				200&p-SBIC&0.84&0.87&0.87&0.86&0.85&0.83&0.23&0.29&0.33&0.37&0.45&0.62\\ 
				&1&0.71&0.74&0.76&0.78&0.81&0.83&0.18&0.24&0.30&0.36&0.45&0.71\\ 
				&P true&0.83&0.87&0.86&0.86&0.85&0.82&0.23&0.29&0.33&0.37&0.45&0.62\\ 
				\cline{2-14} 
				400&p-SBIC&0.80&0.87&0.87&0.90&0.89&0.89&0.17&0.22&0.24&0.28&0.35&0.54\\ 
				&1&0.64&0.67&0.70&0.74&0.76&0.83&0.12&0.18&0.21&0.26&0.37&0.60\\ 
				&P true&0.80&0.87&0.87&0.90&0.89&0.89&0.17&0.22&0.24&0.28&0.34&0.54\\ 
				\cline{2-14} 
				1000&p-SBIC&0.76&0.83&0.87&0.90&0.91&0.92&0.11&0.14&0.16&0.19&0.23&0.37\\ 
				&1&0.55&0.62&0.63&0.66&0.70&0.80&0.07&0.11&0.14&0.17&0.23&0.38\\ 
				&P true&0.76&0.83&0.87&0.90&0.91&0.92&0.11&0.14&0.16&0.19&0.23&0.37\\ 
				\hline 
				&& \multicolumn{12}{c}{P=6}\\  
				200&p-SBIC&0.79&0.82&0.82&0.83&0.82&0.79&0.23&0.27&0.31&0.35&0.41&0.54\\ 
				&1&0.57&0.66&0.72&0.74&0.76&0.80&0.16&0.23&0.28&0.34&0.45&0.73\\ 
				&P true&0.80&0.82&0.82&0.83&0.82&0.79&0.23&0.27&0.31&0.35&0.41&0.54\\ 
				\cline{2-14} 
				400&p-SBIC&0.73&0.82&0.81&0.83&0.85&0.84&0.16&0.21&0.23&0.26&0.30&0.42\\ 
				&1&0.51&0.57&0.61&0.65&0.70&0.77&0.11&0.17&0.20&0.26&0.33&0.51\\ 
				&P true&0.74&0.81&0.81&0.83&0.85&0.84&0.17&0.21&0.23&0.26&0.30&0.42\\ 
				\cline{2-14} 
				1000&p-SBIC&0.62&0.75&0.80&0.81&0.86&0.89&0.11&0.14&0.16&0.17&0.21&0.28\\ 
				&1&0.40&0.45&0.51&0.55&0.59&0.71&0.08&0.11&0.14&0.17&0.21&0.33\\ 
				&P true&0.63&0.75&0.80&0.81&0.86&0.88&0.11&0.14&0.16&0.17&0.21&0.28\\ 
				\hline 
				&& \multicolumn{12}{c}{P=10}\\  
				200&p-SBIC&0.75&0.73&0.74&0.74&0.76&0.74&0.20&0.23&0.27&0.29&0.34&0.47\\ 
				&1&0.49&0.53&0.62&0.68&0.75&0.80&0.15&0.21&0.29&0.35&0.47&0.76\\ 
				&P true&0.76&0.73&0.75&0.74&0.76&0.74&0.20&0.23&0.27&0.29&0.34&0.47\\ 
				\cline{2-14} 
				400&p-SBIC&0.71&0.69&0.71&0.77&0.79&0.81&0.15&0.17&0.20&0.22&0.26&0.36\\ 
				&1&0.41&0.47&0.52&0.58&0.66&0.76&0.11&0.15&0.19&0.24&0.31&0.50\\ 
				&P true&0.71&0.69&0.71&0.77&0.79&0.81&0.15&0.17&0.20&0.22&0.27&0.36\\ 
				\cline{2-14} 
				1000&p-SBIC&0.65&0.62&0.65&0.70&0.77&0.85&0.10&0.11&0.13&0.15&0.18&0.26\\ 
				&1&0.34&0.36&0.42&0.48&0.56&0.69&0.07&0.11&0.14&0.17&0.21&0.35\\ 
				&P true&0.65&0.62&0.64&0.70&0.77&0.85&0.10&0.11&0.13&0.15&0.18&0.26\\ 
				\hline \hline \end{tabular}}\end{center}
\begin{tablenotes}
	\textit{Note}: Same design and reporting as the method~1 table but using method~2 (recursion adds MA terms beyond $H$).
	Low persistence is $\sum_{i=1}^{p}\phi_i\in[0.3,0.9]$.
\end{tablenotes}
\end{table}

			\clearpage
			\subsection{MA(q)-GBF(1) univariate}		
			A small experiment to motivate this model:
			
\begin{figure}[h!]
	\begin{center}
		\caption{MA(24) impulse response generated by a Gaussian basis function (GBF)}
		\label{fig-MA-GBF-motiv}
		\includegraphics[scale=0.6]{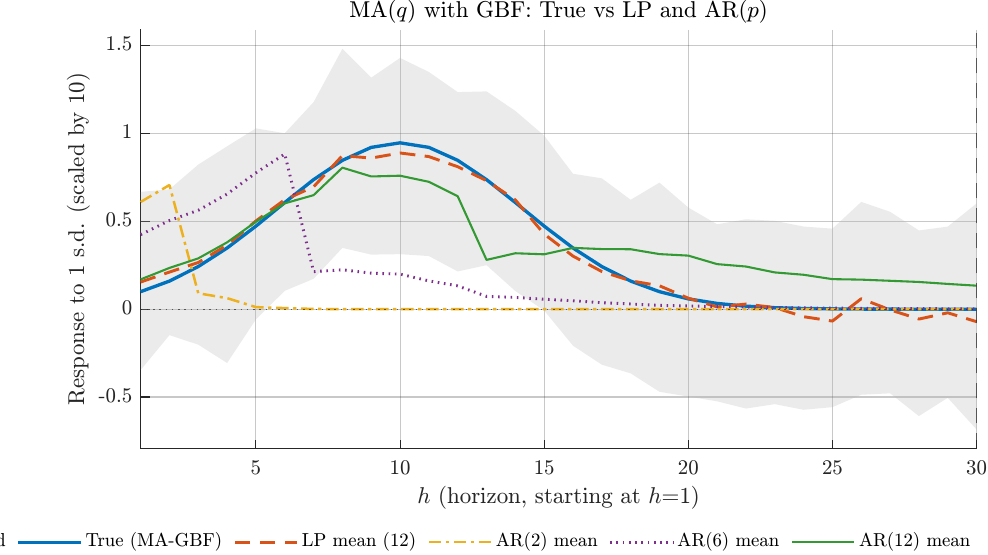}
	\end{center}
\begin{figurenotes}
	\textit{Note}: Population IRF for an MA(24) with GBF coefficients (solid), with LP and AR($p$) estimates superimposed. The IRF may show local peaks and sign changes, and it is exactly zero for horizons beyond the MA order ($h>24$; vertical marker). We use this pattern to test whether the methods capture peaks and zero crossings reliably.
	\end{figurenotes}\end{figure}

\clearpage

			\begin{table}[h!]\caption{Coverage and median interval length: MA(24)–GBF, percentile-$t$ BWB (Method~1)}\label{tab-boot3-ma24-fair1-mod1-cit}\begin{center}\scalebox{0.8}{\begin{tabular}{ll|cccc|cccc} \hline \hline 
			&& \multicolumn{4}{c}{Coverage} & \multicolumn{4}{c}{Median interval length}\\  
			$T$& $p/H$& 10& 20& 40& 60& 10& 20& 40& 60\\ \hline 
		200&p-SBIC&0.82&0.81&0.75&0.66&0.98&0.93&1.16&1.04\\ 
		&10&0.80&0.82&0.77&0.72&0.95&0.86&0.85&0.78\\ 
		&20&0.80&0.81&0.77&0.73&0.96&1.00&0.83&0.78\\ 
		&30&0.82&0.83&0.78&0.76&1.00&0.90&0.99&0.94\\ 
		&40&0.82&0.84&0.78&0.75&1.04&0.94&1.00&0.95\\ 
		&60&0.84&0.87&0.83&0.82&1.13&1.07&1.15&1.12\\ 
		\cline{2-10} 
		400&p-SBIC&0.80&0.75&0.67&0.62&1.13&1.24&1.27&1.29\\ 
		&10&0.81&0.83&0.82&0.80&0.95&0.93&1.00&1.01\\ 
		&20&0.83&0.83&0.81&0.79&0.99&0.92&1.00&1.01\\ 
		&30&0.82&0.84&0.82&0.80&1.12&0.95&0.97&0.95\\ 
		&40&0.82&0.85&0.82&0.81&1.10&1.03&0.99&0.96\\ 
		&60&0.83&0.86&0.84&0.81&1.26&1.03&0.88&0.85\\ 
		\cline{2-10} 
		1000&p-SBIC&0.70&0.51&0.47&0.41&1.21&1.35&1.49&1.52\\ 
		&10&0.82&0.86&0.84&0.83&1.01&1.01&1.21&1.25\\ 
		&20&0.83&0.87&0.86&0.86&1.04&1.03&1.17&1.19\\ 
		&30&0.81&0.85&0.86&0.86&1.03&1.03&1.21&1.20\\ 
		&40&0.82&0.87&0.87&0.87&1.07&1.01&1.13&1.11\\ 
		&60&0.81&0.87&0.86&0.86&1.11&1.09&1.05&1.04\\ 
		\hline \hline \end{tabular}}\end{center}
\begin{tablenotes}
\textit{Note}: Coverage probabilities and median interval length for percentile-$t$ confidence
intervals based on the Block Wild Bootstrap (BWB) under an MA(24) data–generating process with
coefficients generated by a Gaussian basis function (``fair1'' calibration; see Appendix~\ref{app:simresults}).
The first LP regression either selects lags by SBIC or fixes $p$; columns report forecast horizons $h$.
Coverage is the fraction of intervals containing the true impulse response; interval length is reported
relative to the scale of the estimated response at each $h$. Results are aggregated over 100 Monte Carlo replications.
\end{tablenotes}
\end{table}

			\begin{table}[h!]\caption{Coverage and median interval length: MA(24)–GBF, percentile-$t$ BWB (Method~2)}\label{tab-boot3-ma24-fair1-mod2-cit}\begin{center}\scalebox{0.8}{\begin{tabular}{ll|cccc|cccc} \hline \hline 
			&& \multicolumn{4}{c}{Coverage} & \multicolumn{4}{c}{Median interval length}\\  
		$T$& $p/H$& 10& 20& 40& 60& 10& 20& 40& 60\\ \hline 
		200&p-SBIC&0.82&0.81&0.76&0.68&1.01&0.95&1.18&1.09\\ 
		&10&0.80&0.82&0.78&0.73&0.95&0.87&0.86&0.80\\ 
		&20&0.80&0.82&0.78&0.73&0.96&1.00&0.84&0.78\\ 
		&30&0.82&0.83&0.78&0.76&1.00&0.90&0.99&0.95\\ 
		&40&0.82&0.84&0.78&0.76&1.04&0.94&1.00&0.95\\ 
		&60&0.84&0.87&0.83&0.82&1.12&1.07&1.15&1.12\\ 
		\cline{2-10} 
		400&p-SBIC&0.78&0.75&0.68&0.62&1.19&1.26&1.28&1.31\\ 
		&10&0.81&0.83&0.82&0.80&0.94&0.93&1.01&1.01\\ 
		&20&0.82&0.84&0.82&0.79&0.99&0.91&1.00&1.01\\ 
		&30&0.82&0.84&0.81&0.80&1.12&0.95&0.96&0.96\\ 
		&40&0.83&0.85&0.82&0.82&1.10&1.03&0.99&0.96\\ 
		&60&0.83&0.86&0.84&0.81&1.26&1.03&0.88&0.85\\ 
		\cline{2-10} 
		1000&p-SBIC&0.65&0.51&0.46&0.42&1.27&1.36&1.49&1.53\\ 
		&10&0.82&0.86&0.85&0.84&1.01&1.01&1.22&1.25\\ 
		&20&0.83&0.87&0.86&0.87&1.04&1.03&1.18&1.19\\ 
		&30&0.81&0.85&0.86&0.86&1.03&1.03&1.21&1.20\\ 
		&40&0.82&0.87&0.87&0.87&1.07&1.01&1.12&1.11\\ 
		&60&0.81&0.87&0.86&0.86&1.11&1.09&1.05&1.04\\ 
			\hline \hline \end{tabular}}\end{center}
	\begin{tablenotes}
\textit{Note}: Coverage probabilities and median interval length for percentile-$t$ confidence
intervals based on the Block Wild Bootstrap (BWB) under an MA(24) data–generating process with
coefficients generated by a Gaussian basis function (``fair1'' calibration; see Appendix~\ref{app:simresults}).
The first LP regression either selects lags by SBIC or fixes $p$; columns report forecast horizons $h$.
Coverage is the fraction of intervals containing the true impulse response; interval length is reported
relative to the scale of the estimated response at each $h$. Results are aggregated over 100 Monte Carlo replications.
\end{tablenotes}
\end{table}

	\clearpage
	
	%----------------------------------------------------------------------------------------
	%	APPENDIX
	%----------------------------------------------------------------------------------------
	\newpage
	\appendix
	\setcounter{equation}{0}
	\renewcommand{\theequation}{A\arabic{equation}}
	\section*{Appendices}
	\section{Technical proofs: Consistency of Local Projections for infinite order DGPs} \label{app:proofs}
	For convenience, recall that we had obtained the following expression for a local projection of an infinite order process (see \autoref{e:resid}):
	\begin{align} \label{e:alp}
		\bm y_{t+h} = B_{h+1} \bm y_{t-1} + C_{h+2} \bm y_{t-2} + \hdots + C_{h+p} \bm y_{t-p} + \bm u_{t+h},
	\end{align}
	with
	\begin{align} 
		\bm u_{t+h} = \underbrace{\bm \epsilon_{t+h} + B_1 \bm \epsilon_{t+h-1} + \hdots + B_h \bm \epsilon_t }_{\text{previous error term}} + \underbrace{ C_{h+p+1} \bm y_{t-p-1} + C_{h+p+2} \bm y_{t-p-2} + \hdots }_{\text{omitted terms due to truncation}}
	\end{align}
	
	Define,
	\begin{align*}
		\underset{m \times mp}{D} \equiv (\underset{m \times m}{B_{h+1}}, \> C_{h+2},\> \hdots, \>\underset{m \times m}{C_{h+p}}) \quad \text{and} \quad \underset{mp \times 1}{Y_{t-1,p}} = (\underset{1 \times m}{\bm y_{t-1}'}, \> \hdots, \> \bm y_{t-p}')'
	\end{align*}
	and hence rewrite \autoref{e:alp} as:
	\begin{align} \label{e: lpcompact}
		\bm y_{t+h} = D Y_{t-1,p} + \bm u_{t+h}.
	\end{align}
	Note, as was indicated in the main text, that:
	\begin{align*}
		\begin{cases}
			C_{h+2} &= B_hA_1 + \hdots + B_1 A_h + A_{h+1} \\
			\vdots & \\
			C_{h+k} &= B_hA_{k-1} + \hdots + B_1 A_{h+k-2} + A_{h+k-1} \quad k \ge 2
		\end{cases}.
	\end{align*}
	The OLS estimator can therefore be written as:
	\begin{align*}
		\hat D &= \left( \frac{1}{T-(h+p)} \sum_p^{T-h} \bm y_{t+h} Y_{t-1,p}'\right)\underbrace{\left(\frac{1}{T-(h+p)} \sum_p^{T-h} Y_{t-1,p}Y_{t-1,p}'\right)^{-1}}_{\hat Q^{-1}} \\
		\hat D - D &= \left( \frac{1}{T-(h+p)} \sum_p^{T-h} \bm u_{t+h} Y_{t-1,p}'\right)\hat Q^{-1}
	\end{align*}
	It is relatively straightforward, as explained by \cite{LewisReinsel1985}, to determine that $\hat Q \overset{p}{\to} Q$, hence we will focus on understanding the properties of the first term. In particular,
	\begin{align*}
		\frac{1}{T-(h+p)} \sum_p^{T-h} \bm u_{t+h} Y_{t-1,p}' & = 
		\frac{1}{T-(h+p)} \sum_p^{T-h} \left[ C_{h+p+1} \bm y_{t-p-1} + \hdots \right]Y_{t-1,p}' \\
		&\>+ \frac{1}{T-(h+p)} \sum_p^{T-h} \left[\bm \epsilon_{t+h} + B_1 \bm \epsilon_{t+h-1} + \hdots + B_h \bm \epsilon_{t} \right]Y_{t-1,p}'.
	\end{align*}
	It is easy to see, given the maintained assumptions, that:
	\begin{align*}
		\frac{B_j}{T-(h+p)} \sum_p^{T-h} \bm \epsilon_{t+h-j} Y_{t-1,p}' \overset{p}{\to} 0 \quad \text{for} \quad j = 0, 1, \hdots, h \quad \text{with} \quad B_0 = I.
	\end{align*}
	Define 
	\begin{align*}
		\frac{1}{T-(h+p)} \sum_p^{T-h} \bm \epsilon_{t+h-j} Y_{t-1,p}' \equiv \hat \Psi_{h-j,p}.
	\end{align*}
	We want to show that $||B_j \hat \Psi_{h-j,p}|| \overset{p}{\to} 0$. Two well known inequalities will be useful to prove this result: $||AB||^2 \le ||A||_1^2 ||B||^2;$ and $ ||AB||^2 \le ||A||^2 ||B||^2$ where:
	\begin{align*} 
		||C||^2_1 = sup_{l \ne 0} \frac{l'C'C'}{l'l},
	\end{align*}
	that is, the largest eigenvalue of $C'C$. When $C$ is square, $||C||_1^2$ is the square of the largest, in absolute value, eigenvalue of $C$. Recall that $||B_j||^2 = tr(B_j'B_j)$. Hence note that:
	\begin{align} \label{e:ineq1}
		||B_j \hat \Psi_{h-j-p}||^2 \le ||B_j||^2 || \hat \Psi_{h-j-p}||^2_1.
	\end{align}
	Under the maintained assumption that $\sum_{j=0}^{\infty} ||B_j|| < \infty$, we know that $||B_j|| < \infty$. Next note that,
	\begin{align} \label{e:ineq2}
		||\hat \Psi_{h-j-p}||_1 \le || \Psi_{h-j-p}||_1 + ||\hat \Psi_{h-j-p} - \Psi_{h-j-p} ||_1.
	\end{align}
	Next we need to establish that $||\hat \Psi_{h-j-p} - \Psi_{h-j-p} || \overset{p}{\to} 0$. If $p$ is chosen such that $p^2/T \to 0$ as $p,T \to \infty$, which is true by assumption, then \cite{Hannan2009} establishes that $||\hat \Psi_{h-j-k} - \Psi_{h-j-k}|| \overset{p}{\to} 0$ since:
	\begin{align*}
		E\left( ||\hat \Psi_{h-j-k} - \Psi_{h-j-k}||_1^2\right) \le E \left( ||\hat \Psi_{h-j-k} - \Psi_{h-j-k}||^2\right) \le \frac{\lambda m k}{T - h - p} \to 0; \quad |\lambda| < \infty
	\end{align*}
	by Assumption 3, as stated above. Note that to simplify the derivations, it is convenient to assume that $H$, that is, the longest horizon used to plot the impulse response, is a fixed number rather than growing with the sample. To understand this result, note that $||\Psi_{h-j-p}||$ is the matrix of population moments with typical element given by $E(\bm \epsilon_{t+h-j} \> \bm y_{t-i}) = 0$ for $ h \ge 0$, $ h-j \ge i$. Hence, from \autoref{e:ineq2}:
	\begin{align*}
		||\hat \Psi_{h-j-p}||_1 \le \underbrace{||\Psi_{h-j-p}||_1}_{\to 0} + \underbrace{||\hat \Psi_{h-j-p} - \Psi_{h-j-p}||}_{\to 0} \to 0.
	\end{align*}
	Hence, going back to \autoref{e:ineq1}, it is easy to see that:
	\begin{align*}
		||B_j \hat \Psi_{h-j-p}||^2 \le ||B_j||^2 ||\hat \Psi_{h-j-p}||_1^2 \to 0,
	\end{align*}
	as we wanted to show. Next, we need to deal with the term
	\begin{align*}
		\frac{1}{T-(h+p)} \sum_p^{T-h} \left[ C_{h+p+1} \bm y_{t-p-1} + \hdots \right] Y_{t-1,p}'.
	\end{align*}
	More specifically, we want to characterize:
	\begin{align*}
		\frac{C_{h+p+j}}{T-{(h+p)}} \sum_p^{T-h} \bm y_{t-p-(j+1)} Y_{t-1,p}' \quad j = 0,1,\hdots
	\end{align*}
	Define for later use
	\begin{align*}
		\hat \Gamma_{p+j+2}^{2p+j+1} = \left( \hat \Gamma_{p+j+2}, \hat \Gamma_{p+j+3}, \>\hdots, \> \hat \Gamma_{2p+j+1} \right).
	\end{align*}
	Moreover, recall that we previously defined:
	\begin{align*}
		C_{h+p+j} \equiv B_h A_{p+j-1} + \hdots + B_1 A_{h+p+j-2} + A_{h+p+j-1}; \quad j = 0,1,\> \hdots
	\end{align*}
	Hence,
	\begin{align} \label{e:longc}
		\sum_{j=0}^\infty ||C_{h+p+j}|| &= \sum_{j=0}^\infty ||B_h A_{p+j-1} + \hdots + B_1 A_{h+p+j-2} + A_{h+p+j-1}|| \notag \\
		&\le \sum_{j=0}^\infty ||B_h A_{p+j-1}|| + \hdots + \sum_{j=0}^\infty ||B_1 A_{h+p+j-2}|| + \sum_{j=0}^\infty ||A_{p+j-1}|| \notag \\
		&= ||B_h||_1 \sum_{j=0}^\infty ||A_{p+j-1}|| + \hdots + ||B_1||_1 \sum_{j=0}^\infty || A_{h+p+j-2}| + \sum_{j=0}^\infty ||A_{p+j-1}||.
	\end{align}
	We know that $||B_j||_1$ are uniformly bounded and also that, by assumption,
	\begin{align*}
		p^{1/2} \sum_{j=1}^\infty ||A_{p+j}|| \to 0 \quad p,T \to \infty
	\end{align*}
	hence \autoref{e:longc} scaled by $p^{1/2}$ is converging to zero as $T \to \infty$. The only issue that remains to be shown is that $\hat \Gamma_{p+j+2}^{2p+j+1}$ is bounded, but previously we showed that 
	\begin{align*}
		E\left( ||\hat \Gamma_p - \Gamma_p||^2_1\right) \le E\left( ||\hat \Gamma_p - \Gamma_p||^2\right) \le \lambda \frac{mp}{T-p} \to 0 \quad \text{as} \quad T \to \infty \quad \text{since} \quad \frac{p^2}{T} \to 0 \quad \text{as} \quad p,T \to \infty.
	\end{align*}
	Moreover, as $p \to \infty$, $\Gamma_p \to 0$ so that $\hat \Gamma_p$ is uniformly bounded and therefore
	\begin{align*}
		\Big\|  \frac{1}{T-(h+p)}  \sum_p^{T-h} \bm u_{t+h} Y_{t-1,p}'\Big\| \to 0.
	\end{align*}
	Finally, we need to show that $||\hat Q||$ is uniformly bounded. However, note that
	\begin{align*}
		\hat Q = \frac{1}{T-(h+p)}  \sum_p^{T-h} Y_{t-1,p}Y_{t-1,p}' = 
		\begin{pmatrix}
			\hat \Gamma_0 & \hdots & \hat \Gamma_{p-1} \\
			\hat \Gamma_1 & \hdots & \hat \Gamma_{p} \\
			\vdots & \hdots & \vdots \\
			\hat \Gamma_{-p+1} & \hdots & \hat \Gamma_{0} 
		\end{pmatrix} \equiv \hat \Gamma(0, k-1)
	\end{align*}
	and hence $|| \hat Q|| \to || \Gamma(0,k-1)|| < \infty$ using similar steps  based on the assumptions for consistency stated in text, thus proving consistency of the LP in \autoref{e:alp}.
	
\section{Bootstrap algorithm}\label{app:bootalg}
\begin{algorithm}[H]
	\caption{Confidence intervals for local projections using the bootstrap (BWB; MA-based resampling)}
	\begin{algorithmic}
		\STATE \textbf{Step 1 -- Simulate data.} Generate a process $\bm y_t$ under the desired DGP(s).
		
		\STATE \textbf{Step 2 -- Estimate local projections.} For horizons $h=0,\dots,H$, estimate LPs of $\bm y_{t+h}$ on $p$ lags of $\bm y_t$ (first-step lag length chosen by SBIC or fixed):
		\begin{align}
			\bm y_t &= B_1^{(1)} \bm y_{t-1} + \dots + B_p^{(1)} \bm y_{t-p} + \bm{\varepsilon}_t, \\
			\vdots \\
			\bm y_{t+h} &= B_1^{(h+1)} \bm y_{t-1} + \dots + B_p^{(h+1)} \bm y_{t-p} + \bm{\varepsilon}_{t+h}.
		\end{align}
		
		\STATE \textbf{Step 3 -- Fitted component and centered residuals.} With $\{ \hat B_{h,p}^{LP} \}_{h=0}^{H}$ (set $\hat B_{0,p}^{LP}=I$), form
		$\hat{\bm y}_t = \sum_{h=0}^{H} \hat B_{h,p}^{LP}\, \hat{\bm{\varepsilon}}_{t-h}$ and $\hat{\bm v}_t = \bm y_t - \hat{\bm y}_t$.
		
		\STATE \textbf{Step 4 -- Auxiliary regression.} Estimate $\hat C_{h+1}$ from
		$\hat{\bm v}_t = C_{h+1}\, \bm y_{t-H+1} + \bm\xi_t$.
		
		\STATE \textbf{Step 5 -- Recursion for extra MA terms.} Using the recursion in the appendix, generate additional coefficients
		$\{ \hat B_{H+1,p}^{LP}, \ldots, \hat B_{s,p}^{LP} \}$ with $s = T - H$ (only needed for Method~2).
		
		\STATE \textbf{Step 6 -- Block Wild Bootstrap (BWB) weights.} Partition the time index into consecutive, non-overlapping blocks of length $l$.
		Draw i.i.d.\ block weights $\{v_m^*\}$ with $E[v_m^*]=0$ and $\mathrm{Var}(v_m^*)=1$ (e.g., Rademacher or $\mathcal{N}(0,1)$) and set $w_t \equiv v_m^*$ for all $t$ in block $m$.
		Define block-wild residuals $\hat{\bm{\varepsilon}}_t^{*} = w_t\, \hat{\bm{\varepsilon}}_t$.
		Unless otherwise noted, choose $l$ as a simple function of $H$ (e.g., $l \in \{H, \lfloor 1.5H \rfloor\}$).
		
		\STATE \textbf{Step 7 -- Generate bootstrap series.} Construct $\{\bm y_t^*\}_{t=1}^{T}$ via:
		\begin{enumerate}
			\item \textit{Method~1}: $\;\bm y_t^* = \sum_{h=0}^{H} \hat B_{h,p}^{LP}\, \hat{\bm{\varepsilon}}_{t-h}^*$,
			\item \textit{Method~2}: $\;\bm y_t^* = \sum_{h=0}^{T} \hat B_{h,p}^{LP}\, \hat{\bm{\varepsilon}}_{t-h}^*$,
		\end{enumerate}
		where $\hat B_{0,p}^{LP}=I$ (or $1$ univariate) and $\{\hat B_{h,p}^{LP}\}_{h\ge 1}$ are the LP coefficients with $p$ lags.

		\algstore{myalg}
		\end{algorithmic}
	\end{algorithm}

\begin{algorithm}[H]                     
\begin{algorithmic} 
\algrestore{myalg}
		
		\STATE \textbf{Step 8 -- Re-estimate LP on the bootstrap sample.} Using $\{\bm y_t^*\}$, re-estimate LPs with $p$ lags to obtain $\{\hat B_{h,p}^{LP*}\}_{h=1}^{H}$ and the covariance $\hat\Sigma_H^{LP*}$ of $\hat{\bm\beta}_H^{LP*} = \mathrm{vec}(\hat B_{1,p}^{LP*},\dots,\hat B_{H,p}^{LP*})$.
		
		\STATE \textbf{Step 9 -- Studentized statistics.} For each horizon $h$ and replication $b$, store
		\[
		T_{b,h}^{LP*} \;=\; \frac{\delta_h' \big(\hat{\bm\beta}_{H,b}^{LP*} - \hat{\bm\beta}_{H}^{LP}\big)}{\sqrt{\delta_h' \hat\Sigma_H^{LP*} \delta_h}},
		\qquad b=1,\dots,B,
		\]
		where $\delta_h$ selects the element (or linear combination) of interest.
		
		\STATE \textbf{Step 10 -- Quantiles.} Approximate the distribution of $T_{b,h}^{LP*}$ and compute the $\alpha/2$ and $1-\alpha/2$ quantiles, $\hat q_{\alpha/2}^{LP}$ and $\hat q_{1-\alpha/2}^{LP}$.
		
		\STATE \textbf{Step 11 -- Percentile-$t$ confidence intervals.} For each $h$,
		\[
		\Big[\, \delta_h' \hat{\bm\beta}_{h}^{LP} - \sqrt{\delta_h' \hat\Sigma_H^{LP} \delta_h}\; \hat q_{\alpha/2}^{LP} \;,\;
		\delta_h' \hat{\bm\beta}_{h}^{LP} - \sqrt{\delta_h' \hat\Sigma_H^{LP} \delta_h}\; \hat q_{1-\alpha/2}^{LP} \,\Big].
		\]
		
		\STATE \textbf{Step 12 -- Reporting.} Optionally report Efron-type percentile intervals. For tables, compute (i) coverage at each $h$ and (ii) median interval length relative to the scale of the estimated response; then aggregate as in the main text.
	\end{algorithmic}
\end{algorithm}

\clearpage

\section{Additional simulation results} \label{app:simresults}
\setcounter{figure}{0}
\setcounter{table}{0}
\renewcommand{\thefigure}{A-\arabic{figure}}
\renewcommand{\thetable}{A-\arabic{table}}

\subsubsection{AR(1)}

 \begin{table}[h!]\caption{Local-projection bootstrap results, AR(1), method~2 (BWB, percentile-$t$)}\label{tab-boot3-method2-cit-ar1}\begin{center}\scalebox{0.6}{\begin{tabular}{ll|cccccc|cccccc} \hline \hline 
				&& \multicolumn{6}{c}{Coverage} & \multicolumn{6}{c}{Median interval length}\\  
				$T$& $p/H$& 5& 10& 15& 20& 30& 60& 5& 10& 15& 20& 30& 60\\ \hline 
				&& \multicolumn{12}{c}{$\phi=0$}\\  
				200&p-SBIC&0.92&0.93&0.93&0.93&0.93&0.89&1.18&1.13&1.13&1.12&1.13&1.13\\ 
				&1&0.91&0.92&0.92&0.92&0.93&0.89&1.17&1.12&1.12&1.12&1.12&1.12\\ 
				&2&0.89&0.90&0.91&0.92&0.92&0.89&1.17&1.12&1.11&1.09&1.09&1.09\\ 
				&3&0.88&0.89&0.91&0.91&0.91&0.88&1.18&1.13&1.10&1.08&1.08&1.07\\ 
				\cline{2-14} 
				400&p-SBIC&0.95&0.96&0.95&0.95&0.96&0.94&1.09&1.16&1.14&1.15&1.18&1.19\\ 
				&1&0.93&0.94&0.94&0.94&0.96&0.94&1.09&1.16&1.13&1.15&1.17&1.17\\ 
				&2&0.92&0.95&0.94&0.94&0.95&0.93&1.09&1.14&1.12&1.13&1.15&1.15\\ 
				&3&0.90&0.93&0.93&0.93&0.94&0.93&1.09&1.13&1.11&1.11&1.13&1.14\\ 
				\cline{2-14} 
				1000&p-SBIC&0.94&0.97&0.96&0.96&0.97&0.96&1.18&1.22&1.20&1.19&1.21&1.19\\ 
				&1&0.94&0.96&0.96&0.96&0.97&0.96&1.18&1.22&1.19&1.19&1.20&1.18\\ 
				&2&0.91&0.95&0.95&0.96&0.96&0.96&1.19&1.21&1.19&1.18&1.19&1.17\\ 
				&3&0.90&0.94&0.94&0.95&0.96&0.96&1.18&1.20&1.18&1.17&1.17&1.17\\ 
				\hline 
				&& \multicolumn{12}{c}{$\phi=0.5$}\\  
				200&p-SBIC&0.93&0.92&0.93&0.92&0.92&0.88&1.11&1.12&1.09&1.06&1.08&1.12\\ 
				&1&0.93&0.91&0.92&0.92&0.92&0.88&1.19&1.16&1.14&1.09&1.09&1.10\\ 
				&2&0.91&0.90&0.91&0.91&0.91&0.88&1.16&1.14&1.11&1.07&1.07&1.08\\ 
				&3&0.89&0.90&0.90&0.89&0.90&0.87&1.16&1.14&1.10&1.06&1.06&1.06\\ 
				\cline{2-14} 
				400&p-SBIC&0.92&0.95&0.95&0.94&0.94&0.94&1.17&1.17&1.13&1.14&1.15&1.18\\ 
				&1&0.92&0.94&0.94&0.94&0.95&0.94&1.15&1.17&1.13&1.13&1.14&1.14\\ 
				&2&0.90&0.93&0.93&0.94&0.94&0.93&1.15&1.16&1.11&1.11&1.12&1.12\\ 
				&3&0.90&0.91&0.92&0.92&0.93&0.92&1.16&1.15&1.10&1.10&1.11&1.11\\ 
				\cline{2-14} 
				1000&p-SBIC&0.95&0.96&0.96&0.97&0.97&0.96&1.26&1.22&1.18&1.17&1.20&1.22\\ 
				&1&0.93&0.96&0.96&0.96&0.97&0.96&1.22&1.22&1.18&1.17&1.18&1.20\\ 
				&2&0.92&0.95&0.95&0.95&0.96&0.96&1.22&1.21&1.18&1.17&1.17&1.19\\ 
				&3&0.92&0.94&0.94&0.95&0.96&0.96&1.22&1.21&1.17&1.16&1.17&1.18\\ 
				\hline 
				&& \multicolumn{12}{c}{$\phi=0.95$}\\  
				200&p-SBIC&0.89&0.87&0.84&0.82&0.77&0.74&1.00&1.02&1.01&0.95&0.89&0.87\\ 
				&1&0.90&0.88&0.87&0.86&0.83&0.77&1.19&1.17&1.14&1.08&1.06&0.97\\ 
				&2&0.88&0.87&0.84&0.84&0.81&0.76&1.19&1.16&1.12&1.07&1.04&0.97\\ 
				&3&0.88&0.84&0.84&0.83&0.79&0.75&1.17&1.12&1.07&1.02&1.00&0.94\\ 
				\cline{2-14} 
				400&p-SBIC&0.93&0.91&0.90&0.89&0.86&0.79&1.05&1.03&1.00&1.01&0.93&0.89\\ 
				&1&0.92&0.92&0.92&0.89&0.87&0.84&1.22&1.19&1.15&1.16&1.10&1.07\\ 
				&2&0.93&0.91&0.92&0.89&0.87&0.83&1.22&1.19&1.15&1.16&1.10&1.06\\ 
				&3&0.92&0.91&0.91&0.89&0.86&0.83&1.23&1.19&1.14&1.14&1.09&1.06\\ 
				\cline{2-14} 
				1000&p-SBIC&0.97&0.96&0.94&0.91&0.92&0.89&1.09&1.15&1.14&1.13&1.13&1.06\\ 
				&1&0.97&0.96&0.95&0.92&0.93&0.91&1.29&1.23&1.23&1.19&1.17&1.13\\ 
				&2&0.95&0.95&0.94&0.92&0.92&0.91&1.28&1.23&1.22&1.18&1.17&1.13\\ 
				&3&0.94&0.94&0.93&0.92&0.92&0.91&1.30&1.23&1.22&1.18&1.16&1.12\\ 
				\hline 
				&& \multicolumn{12}{c}{$\phi=1$}\\  
				200&p-SBIC&0.87&0.89&0.86&0.81&0.75&0.59&1.02&0.98&0.97&0.93&0.90&0.75\\ 
				&1&0.82&0.87&0.84&0.80&0.76&0.63&1.23&1.19&1.13&1.06&1.03&0.88\\ 
				&2&0.80&0.84&0.82&0.80&0.74&0.61&1.25&1.20&1.13&1.07&1.02&0.88\\ 
				&3&0.79&0.82&0.81&0.78&0.72&0.61&1.21&1.17&1.11&1.05&0.99&0.87\\ 
				\cline{2-14} 
				400&p-SBIC&0.86&0.90&0.90&0.91&0.85&0.76&0.82&1.00&1.05&1.13&1.07&0.90\\ 
				&1&0.82&0.86&0.87&0.86&0.82&0.72&1.54&1.47&1.39&1.31&1.21&1.01\\ 
				&2&0.83&0.85&0.87&0.87&0.83&0.72&1.55&1.44&1.37&1.30&1.18&0.99\\ 
				&3&0.81&0.85&0.86&0.86&0.82&0.72&1.53&1.42&1.37&1.29&1.17&0.98\\ 
				\cline{2-14} 
				1000&p-SBIC&0.92&0.91&0.89&0.93&0.89&0.85&0.54&0.61&0.71&0.84&0.83&0.84\\ 
				&1&0.77&0.81&0.83&0.87&0.86&0.83&1.94&1.63&1.57&1.48&1.35&1.18\\ 
				&2&0.77&0.82&0.82&0.87&0.86&0.83&2.04&1.65&1.56&1.49&1.36&1.18\\ 
				&3&0.78&0.81&0.81&0.86&0.85&0.82&1.98&1.65&1.56&1.50&1.37&1.18\\ 
				\hline \hline \end{tabular}}\end{center}
\begin{tablenotes}
	\textit{Note}: Same design as Table~\ref{tab-boot3-method1-cit-ar1}, but Method~2 augments the MA construction beyond $H$ using the recursion proposed in Appendix~\ref{app:bootalg}. Median interval lengths are reported relative to the scale of the estimated response.
\end{tablenotes}
\end{table}

\begin{table}[h!]\caption{Bias of bootstrap impulse-response estimates: AR(1) with BWB)}\label{tab-boot3-bias-AR1_nobiasc}\begin{center}\scalebox{0.8}{\begin{tabular}{ll|cccccc|cccccc|cccccc} \hline \hline 
				&& \multicolumn{12}{c}{LP} & \multicolumn{6}{c}{AR}\\  
				&& \multicolumn{6}{c}{Method~1} & \multicolumn{6}{c}{Method~2}&&&&&&\\  
				$T$& $p/H$& 5& 10& 15& 20& 30& 60& 5& 10& 15& 20& 30& 60& 5& 10& 15& 20& 30& 60\\ \hline 
				&& \multicolumn{18}{c}{$\phi=0$}\\  
				200&p-SBIC&0.05&0.06&0.07&0.08&0.08&0.10&0.05&0.06&0.07&0.08&0.08&0.10&0.02&0.01&0.01&0.01&0.00&0.00\\ 
				&1&0.13&0.12&0.11&0.10&0.09&0.10&0.13&0.12&0.11&0.11&0.09&0.10&0.00&0.00&0.00&0.00&0.00&0.00\\ 
				&2&0.13&0.12&0.11&0.10&0.09&0.09&0.13&0.12&0.11&0.10&0.09&0.09&0.00&0.00&0.00&0.00&0.00&0.00\\ 
				&3&0.12&0.12&0.11&0.10&0.09&0.09&0.13&0.12&0.11&0.10&0.09&0.09&0.00&0.00&0.00&0.00&0.00&0.00\\ 
				\cline{2-20} 
				400&p-SBIC&0.68&0.65&0.59&0.57&0.48&0.07&0.69&0.66&0.60&0.57&0.48&0.07&0.70&0.67&0.62&0.57&0.44&0.00\\ 
				&1&0.07&0.06&0.06&0.06&0.06&0.06&0.07&0.06&0.06&0.06&0.06&0.06&0.00&0.00&0.00&0.00&0.00&0.00\\ 
				&2&0.07&0.06&0.06&0.06&0.06&0.06&0.07&0.06&0.06&0.06&0.06&0.06&0.00&0.00&0.00&0.00&0.00&0.00\\ 
				&3&0.07&0.06&0.06&0.06&0.06&0.06&0.06&0.06&0.06&0.06&0.06&0.06&0.00&0.00&0.00&0.00&0.00&0.00\\ 
				\cline{2-20} 
				1000&p-SBIC&0.55&0.54&0.53&0.49&0.33&0.04&0.59&0.58&0.57&0.52&0.36&0.04&0.55&0.53&0.50&0.46&0.37&0.00\\ 
				&1&0.04&0.04&0.04&0.04&0.04&0.04&0.04&0.04&0.04&0.04&0.04&0.04&0.00&0.00&0.00&0.00&0.00&0.00\\ 
				&2&0.04&0.04&0.04&0.04&0.04&0.04&0.04&0.04&0.04&0.04&0.04&0.04&0.00&0.00&0.00&0.00&0.00&0.00\\ 
				&3&0.04&0.04&0.04&0.04&0.04&0.04&0.04&0.04&0.04&0.04&0.04&0.04&0.00&0.00&0.00&0.00&0.00&0.00\\ 
				\hline 
				&& \multicolumn{18}{c}{$\phi=0.5$}\\  
				200&p-SBIC&0.12&0.11&0.10&0.10&0.10&0.12&0.12&0.11&0.11&0.10&0.10&0.12&0.00&0.00&0.00&0.00&0.00&0.00\\ 
				&1&0.11&0.11&0.10&0.10&0.09&0.10&0.11&0.11&0.10&0.09&0.09&0.11&0.00&0.00&0.00&0.00&0.00&0.00\\ 
				&2&0.11&0.11&0.10&0.09&0.09&0.10&0.11&0.11&0.10&0.09&0.09&0.10&0.00&0.00&0.00&0.01&0.01&0.01\\ 
				&3&0.11&0.10&0.09&0.09&0.08&0.10&0.11&0.10&0.09&0.09&0.08&0.10&0.00&0.01&0.01&0.01&0.01&0.01\\ 
				\cline{2-20} 
				400&p-SBIC&0.06&0.06&0.06&0.06&0.07&0.08&0.06&0.06&0.06&0.06&0.07&0.08&0.00&0.00&0.00&0.00&0.00&0.00\\ 
				&1&0.08&0.08&0.08&0.07&0.07&0.07&0.08&0.07&0.07&0.07&0.06&0.07&0.00&0.00&0.00&0.00&0.00&0.00\\ 
				&2&0.08&0.08&0.08&0.07&0.06&0.07&0.08&0.07&0.07&0.07&0.06&0.07&0.00&0.00&0.00&0.00&0.00&0.00\\ 
				&3&0.08&0.07&0.07&0.07&0.06&0.07&0.07&0.07&0.07&0.06&0.06&0.07&0.00&0.00&0.00&0.00&0.00&0.00\\ 
				\cline{2-20} 
				1000&p-SBIC&0.04&0.04&0.04&0.04&0.04&0.05&0.04&0.04&0.04&0.04&0.04&0.05&0.00&0.00&0.00&0.00&0.00&0.00\\ 
				&1&0.04&0.04&0.05&0.04&0.04&0.04&0.04&0.04&0.05&0.04&0.04&0.04&0.00&0.00&0.00&0.00&0.00&0.00\\ 
				&2&0.04&0.04&0.05&0.04&0.04&0.04&0.04&0.04&0.05&0.04&0.04&0.04&0.00&0.00&0.00&0.00&0.00&0.00\\ 
				&3&0.04&0.04&0.04&0.04&0.04&0.04&0.04&0.04&0.04&0.04&0.04&0.04&0.00&0.00&0.00&0.00&0.00&0.00\\ 
				\hline 
				&& \multicolumn{18}{c}{$\phi=0.95$}\\  
				200&p-SBIC&0.12&0.13&0.14&0.15&0.18&0.26&0.11&0.12&0.13&0.14&0.17&0.27&0.01&0.02&0.04&0.05&0.08&0.11\\ 
				&1&0.17&0.16&0.14&0.15&0.18&0.25&0.17&0.16&0.14&0.16&0.18&0.25&0.10&0.10&0.10&0.10&0.11&0.11\\ 
				&2&0.17&0.15&0.14&0.15&0.17&0.24&0.17&0.16&0.14&0.15&0.18&0.24&0.10&0.10&0.11&0.11&0.11&0.11\\ 
				&3&0.19&0.17&0.14&0.16&0.18&0.23&0.18&0.17&0.15&0.16&0.18&0.23&0.10&0.10&0.10&0.10&0.11&0.11\\ 
				\cline{2-20} 
				400&p-SBIC&0.09&0.10&0.11&0.11&0.13&0.19&0.07&0.08&0.09&0.09&0.12&0.19&0.01&0.01&0.02&0.03&0.05&0.07\\ 
				&1&0.32&0.29&0.28&0.27&0.20&0.17&0.31&0.28&0.28&0.27&0.19&0.17&0.05&0.05&0.05&0.06&0.06&0.07\\ 
				&2&0.32&0.29&0.28&0.27&0.20&0.17&0.31&0.28&0.27&0.27&0.19&0.16&0.05&0.05&0.06&0.06&0.06&0.07\\ 
				&3&0.31&0.28&0.27&0.26&0.19&0.16&0.29&0.27&0.26&0.25&0.18&0.16&0.05&0.05&0.05&0.06&0.06&0.07\\ 
				\cline{2-20} 
				1000&p-SBIC&0.06&0.07&0.08&0.07&0.07&0.11&0.04&0.05&0.05&0.05&0.06&0.12&0.00&0.01&0.01&0.02&0.03&0.04\\ 
				&1&0.11&0.11&0.11&0.10&0.09&0.10&0.09&0.10&0.10&0.09&0.08&0.10&0.07&0.07&0.06&0.06&0.05&0.04\\ 
				&2&0.11&0.11&0.11&0.09&0.09&0.10&0.09&0.09&0.10&0.09&0.08&0.10&0.07&0.07&0.06&0.06&0.05&0.04\\ 
				&3&0.11&0.11&0.10&0.09&0.08&0.10&0.09&0.09&0.09&0.08&0.08&0.10&0.08&0.07&0.07&0.06&0.05&0.04\\ 
				\hline 
				&& \multicolumn{18}{c}{$\phi=1$}\\  
				200&p-SBIC&0.19&0.20&0.20&0.25&0.35&0.78&0.17&0.17&0.18&0.22&0.34&0.76&0.10&0.11&0.13&0.16&0.24&0.59\\ 
				&1&0.67&0.66&0.64&0.65&0.67&0.73&0.68&0.67&0.65&0.67&0.70&0.74&0.70&0.70&0.69&0.68&0.65&0.59\\ 
				&2&0.68&0.66&0.64&0.65&0.67&0.72&0.68&0.67&0.65&0.67&0.70&0.73&0.70&0.70&0.69&0.68&0.65&0.59\\ 
				&3&0.68&0.67&0.64&0.65&0.67&0.71&0.69&0.67&0.65&0.67&0.70&0.72&0.71&0.70&0.69&0.68&0.66&0.59\\ 
				\cline{2-20} 
				400&p-SBIC&0.30&0.31&0.33&0.35&0.34&0.59&0.27&0.26&0.27&0.29&0.27&0.54&0.06&0.07&0.08&0.10&0.15&0.41\\ 
				&1&0.60&0.60&0.61&0.60&0.50&0.49&0.63&0.62&0.64&0.63&0.53&0.54&0.55&0.54&0.53&0.52&0.49&0.41\\ 
				&2&0.60&0.60&0.61&0.60&0.50&0.49&0.63&0.62&0.64&0.62&0.53&0.54&0.55&0.54&0.53&0.52&0.49&0.41\\ 
				&3&0.59&0.59&0.61&0.59&0.50&0.49&0.61&0.61&0.63&0.62&0.53&0.54&0.55&0.54&0.53&0.52&0.50&0.41\\ 
				\cline{2-20} 
				1000&p-SBIC&0.12&0.15&0.18&0.20&0.25&0.44&0.09&0.10&0.10&0.11&0.13&0.29&0.09&0.08&0.08&0.08&0.10&0.23\\ 
				&1&0.38&0.39&0.40&0.40&0.38&0.31&0.37&0.38&0.38&0.37&0.35&0.31&0.43&0.42&0.41&0.40&0.36&0.23\\ 
				&2&0.38&0.40&0.40&0.40&0.38&0.32&0.37&0.38&0.38&0.37&0.34&0.30&0.43&0.42&0.41&0.40&0.36&0.23\\ 
				&3&0.39&0.41&0.41&0.41&0.39&0.31&0.38&0.38&0.38&0.37&0.35&0.30&0.44&0.43&0.42&0.40&0.36&0.23\\ 
					\hline \hline \end{tabular}}\end{center}
\begin{tablenotes}
	\textit{Note}: Average absolute bias at each horizon $h$ for AR(1) designs, comparing local projections (LP) with Method~1 (truncation at $H$) and Method~2 (recursion beyond $H$), and autoregressive (AR) estimation.
	Rows vary sample size $T$ and first-step lag specification (SBIC or fixed $p$).
	Bias is computed as $\left| \text{IRF}_{\text{true},h} - \frac{1}{B}\sum_{b=1}^B \text{IRF}_{\text{boot},h}^{(b)} \right|$ and then averaged across replications.
\end{tablenotes}
\end{table}

\clearpage
\subsection{AR(p) models}

\begin{table}[h!]\caption{AR($p$), medium persistence: percentile-$t$ BWB coverage and median interval length (Method~1)}\label{tab-boot3-phi_min07_phi_max09-method1-cit-ARp}\begin{center}\scalebox{0.8}{\begin{tabular}{ll|cccccc|cccccc} \hline \hline 
				&& \multicolumn{6}{c}{Coverage} & \multicolumn{6}{c}{Median interval length}\\  
				$T$& $p/H$& 5& 10& 15& 20& 30& 60& 5& 10& 15& 20& 30& 60\\ \hline 
				&& \multicolumn{12}{c}{P=4}\\  
				200&p-SBIC&0.84&0.87&0.85&0.86&0.84&0.80&0.23&0.29&0.32&0.36&0.44&0.58\\ 
				&1&0.63&0.69&0.72&0.73&0.74&0.76&0.17&0.22&0.26&0.32&0.39&0.60\\ 
				&P true&0.84&0.87&0.85&0.86&0.84&0.79&0.23&0.29&0.32&0.35&0.44&0.58\\ 
				\cline{2-14} 
				400&p-SBIC&0.80&0.86&0.88&0.90&0.90&0.88&0.17&0.21&0.24&0.28&0.34&0.52\\ 
				&1&0.59&0.61&0.65&0.68&0.71&0.77&0.11&0.16&0.19&0.24&0.33&0.53\\ 
				&P true&0.80&0.86&0.88&0.89&0.90&0.88&0.17&0.21&0.24&0.28&0.34&0.52\\ 
				\cline{2-14} 
				1000&p-SBIC&0.75&0.84&0.87&0.90&0.92&0.92&0.11&0.14&0.16&0.19&0.23&0.37\\ 
				&1&0.48&0.57&0.59&0.61&0.65&0.74&0.07&0.10&0.13&0.16&0.21&0.36\\ 
				&P true&0.75&0.84&0.87&0.90&0.91&0.92&0.10&0.14&0.16&0.19&0.23&0.37\\ 
				\hline 
				&& \multicolumn{12}{c}{P=6}\\  
				200&p-SBIC&0.81&0.83&0.82&0.81&0.81&0.76&0.23&0.28&0.30&0.34&0.40&0.51\\ 
				&1&0.53&0.57&0.62&0.65&0.68&0.72&0.15&0.20&0.24&0.28&0.35&0.56\\ 
				&P true&0.82&0.84&0.82&0.82&0.81&0.76&0.23&0.27&0.31&0.35&0.40&0.52\\ 
				\cline{2-14} 
				400&p-SBIC&0.75&0.82&0.81&0.84&0.86&0.84&0.16&0.21&0.23&0.26&0.30&0.41\\ 
				&1&0.46&0.47&0.52&0.56&0.60&0.67&0.11&0.15&0.17&0.20&0.26&0.38\\ 
				&P true&0.76&0.81&0.81&0.84&0.86&0.84&0.17&0.21&0.23&0.26&0.30&0.41\\ 
				\cline{2-14} 
				1000&p-SBIC&0.66&0.76&0.81&0.83&0.87&0.90&0.11&0.14&0.16&0.18&0.21&0.28\\ 
				&1&0.37&0.41&0.42&0.45&0.50&0.61&0.07&0.10&0.12&0.14&0.17&0.26\\ 
				&P true&0.66&0.76&0.81&0.83&0.87&0.90&0.11&0.14&0.16&0.18&0.21&0.28\\ 
				\hline 
				&& \multicolumn{12}{c}{P=10}\\  
				200&p-SBIC&0.74&0.76&0.76&0.74&0.76&0.72&0.20&0.23&0.27&0.29&0.33&0.46\\ 
				&1&0.45&0.43&0.49&0.53&0.58&0.68&0.14&0.18&0.22&0.26&0.32&0.51\\ 
				&P true&0.76&0.76&0.77&0.74&0.77&0.73&0.20&0.23&0.27&0.29&0.33&0.46\\ 
				\cline{2-14} 
				400&p-SBIC&0.72&0.71&0.74&0.78&0.80&0.81&0.15&0.17&0.20&0.22&0.26&0.35\\ 
				&1&0.38&0.39&0.40&0.46&0.50&0.58&0.11&0.13&0.16&0.18&0.22&0.31\\ 
				&P true&0.73&0.71&0.75&0.78&0.80&0.81&0.15&0.17&0.20&0.22&0.26&0.35\\ 
				\cline{2-14} 
				1000&p-SBIC&0.65&0.65&0.69&0.73&0.78&0.85&0.10&0.11&0.14&0.15&0.18&0.26\\ 
				&1&0.34&0.29&0.33&0.38&0.43&0.50&0.07&0.09&0.11&0.12&0.15&0.21\\ 
				&P true&0.66&0.64&0.69&0.73&0.78&0.85&0.10&0.11&0.14&0.15&0.18&0.26\\ 
				\hline \hline \end{tabular}}\end{center}
\begin{tablenotes}
	\textit{Note}: Coverage and median interval length for percentile-$t$ BWB intervals under medium persistence,
	$\sum_{i=1}^{p}\phi_i\in[0.7,0.9]$. The first-step LP lag length is SBIC-selected or fixed ($p=1$); the DGP order is $P\in\{4,6,10\}$.
	Columns denote horizons $h$; interval length is scaled to the response at each $h$.
\end{tablenotes}
\end{table}

\begin{table}[h!]\caption{AR($p$), medium persistence: percentile-$t$ BWB coverage and median interval length (Method~2)}\label{tab-boot3-phi_min07_phi_max09-method2-cit-ARp}\begin{center}\scalebox{0.8}{\begin{tabular}{ll|cccccc|cccccc} \hline \hline 
				&& \multicolumn{6}{c}{Coverage} & \multicolumn{6}{c}{Median interval length}\\  
				$T$& $p/H$& 5& 10& 15& 20& 30& 60& 5& 10& 15& 20& 30& 60\\ \hline 
				&& \multicolumn{12}{c}{P=4}\\  
				200&p-SBIC&0.84&0.87&0.87&0.86&0.85&0.83&0.23&0.29&0.33&0.37&0.45&0.62\\ 
				&1&0.71&0.74&0.76&0.78&0.81&0.83&0.18&0.24&0.30&0.36&0.45&0.71\\ 
				&P true&0.83&0.87&0.86&0.86&0.85&0.82&0.23&0.29&0.33&0.37&0.45&0.62\\ 
				\cline{2-14} 
				400&p-SBIC&0.80&0.87&0.87&0.90&0.89&0.89&0.17&0.22&0.24&0.28&0.35&0.54\\ 
				&1&0.64&0.67&0.70&0.74&0.76&0.83&0.12&0.18&0.21&0.26&0.37&0.60\\ 
				&P true&0.80&0.87&0.87&0.90&0.89&0.89&0.17&0.22&0.24&0.28&0.34&0.54\\ 
				\cline{2-14} 
				1000&p-SBIC&0.76&0.83&0.87&0.90&0.91&0.92&0.11&0.14&0.16&0.19&0.23&0.37\\ 
				&1&0.55&0.62&0.63&0.66&0.70&0.80&0.07&0.11&0.14&0.17&0.23&0.38\\ 
				&P true&0.76&0.83&0.87&0.90&0.91&0.92&0.11&0.14&0.16&0.19&0.23&0.37\\ 
				\hline 
				&& \multicolumn{12}{c}{P=6}\\  
				200&p-SBIC&0.79&0.82&0.82&0.83&0.82&0.79&0.23&0.27&0.31&0.35&0.41&0.54\\ 
				&1&0.57&0.66&0.72&0.74&0.76&0.80&0.16&0.23&0.28&0.34&0.45&0.73\\ 
				&P true&0.80&0.82&0.82&0.83&0.82&0.79&0.23&0.27&0.31&0.35&0.41&0.54\\ 
				\cline{2-14} 
				400&p-SBIC&0.73&0.82&0.81&0.83&0.85&0.84&0.16&0.21&0.23&0.26&0.30&0.42\\ 
				&1&0.51&0.57&0.61&0.65&0.70&0.77&0.11&0.17&0.20&0.26&0.33&0.51\\ 
				&P true&0.74&0.81&0.81&0.83&0.85&0.84&0.17&0.21&0.23&0.26&0.30&0.42\\ 
				\cline{2-14} 
				1000&p-SBIC&0.62&0.75&0.80&0.81&0.86&0.89&0.11&0.14&0.16&0.17&0.21&0.28\\ 
				&1&0.40&0.45&0.51&0.55&0.59&0.71&0.08&0.11&0.14&0.17&0.21&0.33\\ 
				&P true&0.63&0.75&0.80&0.81&0.86&0.88&0.11&0.14&0.16&0.17&0.21&0.28\\ 
				\hline 
				&& \multicolumn{12}{c}{P=10}\\  
				200&p-SBIC&0.75&0.73&0.74&0.74&0.76&0.74&0.20&0.23&0.27&0.29&0.34&0.47\\ 
				&1&0.49&0.53&0.62&0.68&0.75&0.80&0.15&0.21&0.29&0.35&0.47&0.76\\ 
				&P true&0.76&0.73&0.75&0.74&0.76&0.74&0.20&0.23&0.27&0.29&0.34&0.47\\ 
				\cline{2-14} 
				400&p-SBIC&0.71&0.69&0.71&0.77&0.79&0.81&0.15&0.17&0.20&0.22&0.26&0.36\\ 
				&1&0.41&0.47&0.52&0.58&0.66&0.76&0.11&0.15&0.19&0.24&0.31&0.50\\ 
				&P true&0.71&0.69&0.71&0.77&0.79&0.81&0.15&0.17&0.20&0.22&0.27&0.36\\ 
				\cline{2-14} 
				1000&p-SBIC&0.65&0.62&0.65&0.70&0.77&0.85&0.10&0.11&0.13&0.15&0.18&0.26\\ 
				&1&0.34&0.36&0.42&0.48&0.56&0.69&0.07&0.11&0.14&0.17&0.21&0.35\\ 
				&P true&0.65&0.62&0.64&0.70&0.77&0.85&0.10&0.11&0.13&0.15&0.18&0.26\\ 
				\hline \hline \end{tabular}}\end{center}
\begin{tablenotes}
	\textit{Note}: Same set-up as the method~1 table but using method~2 (recursion beyond $H$).
	Medium persistence is $\sum_{i=1}^{p}\phi_i\in[0.7,0.9]$.
\end{tablenotes}
\end{table}

\begin{table}[h!]\caption{AR($p$), high persistence: percentile-$t$ BWB coverage and median interval length (Method~1}\label{tab-boot3-phi_min09_phi_max099-method1-cit-ARp}\begin{center}\scalebox{0.8}{\begin{tabular}{ll|cccccc|cccccc} \hline \hline 
				&& \multicolumn{6}{c}{Coverage} & \multicolumn{6}{c}{Median interval length}\\  
				$T$& $p/H$& 5& 10& 15& 20& 30& 60& 5& 10& 15& 20& 30& 60\\ \hline 
				&& \multicolumn{12}{c}{P=4}\\  
				200&p-SBIC&0.68&0.77&0.81&0.78&0.76&0.69&0.19&0.23&0.29&0.32&0.35&0.50\\ 
				&1&0.50&0.55&0.58&0.60&0.63&0.64&0.13&0.14&0.19&0.25&0.30&0.49\\ 
				&P true&0.68&0.76&0.82&0.78&0.77&0.69&0.19&0.23&0.29&0.32&0.35&0.50\\ 
				\cline{2-14} 
				400&p-SBIC&0.61&0.70&0.79&0.83&0.82&0.81&0.15&0.18&0.23&0.28&0.30&0.42\\ 
				&1&0.39&0.45&0.50&0.57&0.60&0.67&0.12&0.15&0.17&0.24&0.28&0.42\\ 
				&P true&0.60&0.70&0.79&0.83&0.82&0.81&0.15&0.18&0.23&0.28&0.30&0.42\\ 
				\cline{2-14} 
				1000&p-SBIC&0.50&0.62&0.78&0.83&0.87&0.90&0.10&0.13&0.18&0.21&0.23&0.37\\ 
				&1&0.31&0.33&0.40&0.45&0.53&0.66&0.07&0.08&0.12&0.16&0.20&0.34\\ 
				&P true&0.50&0.63&0.77&0.83&0.87&0.90&0.10&0.13&0.18&0.21&0.24&0.37\\ 
				\hline 
				&& \multicolumn{12}{c}{P=6}\\  
				200&p-SBIC&0.67&0.66&0.71&0.72&0.72&0.66&0.19&0.21&0.23&0.25&0.28&0.38\\ 
				&1&0.49&0.50&0.57&0.55&0.58&0.61&0.12&0.13&0.15&0.17&0.24&0.35\\ 
				&P true&0.68&0.65&0.70&0.71&0.72&0.66&0.19&0.20&0.22&0.25&0.28&0.38\\ 
				\cline{2-14} 
				400&p-SBIC&0.57&0.61&0.67&0.74&0.76&0.78&0.14&0.16&0.17&0.20&0.24&0.36\\ 
				&1&0.40&0.43&0.48&0.52&0.55&0.60&0.09&0.10&0.12&0.16&0.22&0.34\\ 
				&P true&0.58&0.61&0.67&0.73&0.75&0.77&0.14&0.16&0.17&0.20&0.24&0.36\\ 
				\cline{2-14} 
				1000&p-SBIC&0.50&0.51&0.61&0.72&0.78&0.86&0.09&0.11&0.13&0.15&0.18&0.24\\ 
				&1&0.34&0.32&0.36&0.41&0.47&0.57&0.07&0.07&0.09&0.11&0.15&0.24\\ 
				&P true&0.50&0.50&0.60&0.72&0.78&0.86&0.09&0.11&0.13&0.15&0.18&0.24\\ 
				\hline 
				&& \multicolumn{12}{c}{P=10}\\  
				200&p-SBIC&0.71&0.65&0.67&0.69&0.71&0.67&0.20&0.24&0.26&0.28&0.32&0.42\\ 
				&1&0.44&0.43&0.50&0.52&0.53&0.58&0.17&0.20&0.23&0.26&0.31&0.47\\ 
				&P true&0.73&0.66&0.68&0.70&0.73&0.68&0.20&0.24&0.26&0.28&0.32&0.42\\ 
				\cline{2-14} 
				400&p-SBIC&0.66&0.59&0.64&0.68&0.72&0.77&0.14&0.18&0.19&0.22&0.24&0.31\\ 
				&1&0.38&0.34&0.39&0.43&0.49&0.56&0.13&0.15&0.16&0.19&0.23&0.34\\ 
				&P true&0.66&0.59&0.64&0.68&0.72&0.78&0.14&0.18&0.19&0.22&0.24&0.30\\ 
				\cline{2-14} 
				1000&p-SBIC&0.59&0.50&0.57&0.62&0.71&0.80&0.09&0.11&0.13&0.14&0.17&0.23\\ 
				&1&0.30&0.24&0.26&0.30&0.35&0.46&0.08&0.10&0.11&0.13&0.15&0.21\\ 
				&P true&0.60&0.50&0.57&0.62&0.71&0.80&0.09&0.11&0.13&0.14&0.17&0.23\\ 
				\hline \hline \end{tabular}}\end{center}
\begin{tablenotes}
	\textit{Note}: Coverage and median interval length for percentile-$t$ BWB intervals under high persistence,
	$\sum_{i=1}^{p}\phi_i\in[0.9,0.99]$. Rows vary first-step LP lag choice (SBIC vs.\ fixed $p$) and the DGP order $P$; columns are horizons $h$.
	Interval length is reported relative to the scale of the estimated response at each $h$.
\end{tablenotes}
\end{table}

\begin{table}[h!]\caption{AR($p$), high persistence: percentile-$t$ BWB coverage and median interval length (Method~2)}\label{tab-boot3-phi_min09_phi_max099-method2-cit-ARp}\begin{center}\scalebox{0.8}{\begin{tabular}{ll|cccccc|cccccc} \hline \hline 
				&& \multicolumn{6}{c}{Coverage} & \multicolumn{6}{c}{Median interval length}\\  
				$T$& $p/H$& 5& 10& 15& 20& 30& 60& 5& 10& 15& 20& 30& 60\\ \hline 
				&& \multicolumn{12}{c}{P=4}\\  
				200&p-SBIC&0.75&0.78&0.81&0.78&0.77&0.74&0.20&0.24&0.30&0.33&0.38&0.56\\ 
				&1&0.58&0.62&0.69&0.69&0.72&0.73&0.15&0.16&0.23&0.30&0.36&0.60\\ 
				&P true&0.74&0.78&0.82&0.78&0.78&0.74&0.20&0.23&0.30&0.33&0.38&0.56\\ 
				\cline{2-14} 
				400&p-SBIC&0.65&0.73&0.79&0.84&0.83&0.81&0.16&0.18&0.23&0.28&0.31&0.44\\ 
				&1&0.49&0.56&0.62&0.69&0.70&0.75&0.15&0.17&0.21&0.28&0.34&0.50\\ 
				&P true&0.65&0.72&0.79&0.84&0.83&0.81&0.16&0.18&0.23&0.28&0.31&0.44\\ 
				\cline{2-14} 
				1000&p-SBIC&0.54&0.65&0.79&0.84&0.87&0.90&0.10&0.13&0.18&0.21&0.23&0.38\\ 
				&1&0.38&0.43&0.52&0.57&0.63&0.75&0.08&0.10&0.14&0.19&0.23&0.41\\ 
				&P true&0.54&0.65&0.78&0.84&0.87&0.90&0.10&0.13&0.18&0.21&0.23&0.38\\ 
				\hline 
				&& \multicolumn{12}{c}{P=6}\\  
				200&p-SBIC&0.66&0.67&0.72&0.71&0.74&0.70&0.19&0.21&0.23&0.26&0.29&0.42\\ 
				&1&0.55&0.56&0.62&0.65&0.69&0.74&0.14&0.15&0.18&0.21&0.29&0.48\\ 
				&P true&0.67&0.67&0.72&0.70&0.73&0.70&0.19&0.21&0.23&0.25&0.29&0.41\\ 
				\cline{2-14} 
				400&p-SBIC&0.59&0.62&0.71&0.74&0.77&0.79&0.14&0.16&0.18&0.21&0.24&0.38\\ 
				&1&0.47&0.51&0.59&0.62&0.66&0.72&0.11&0.12&0.15&0.21&0.28&0.42\\ 
				&P true&0.58&0.61&0.70&0.73&0.76&0.79&0.14&0.16&0.18&0.20&0.24&0.38\\ 
				\cline{2-14} 
				1000&p-SBIC&0.48&0.52&0.66&0.72&0.77&0.86&0.09&0.11&0.13&0.16&0.18&0.24\\ 
				&1&0.37&0.38&0.46&0.53&0.60&0.69&0.08&0.09&0.11&0.15&0.20&0.32\\ 
				&P true&0.50&0.52&0.66&0.72&0.77&0.86&0.09&0.11&0.13&0.16&0.18&0.24\\ 
				\hline 
				&& \multicolumn{12}{c}{P=10}\\  
				200&p-SBIC&0.71&0.65&0.68&0.70&0.72&0.71&0.20&0.24&0.26&0.29&0.32&0.45\\ 
				&1&0.50&0.52&0.59&0.63&0.68&0.72&0.19&0.24&0.28&0.34&0.43&0.72\\ 
				&P true&0.73&0.65&0.68&0.71&0.73&0.71&0.20&0.24&0.26&0.29&0.32&0.45\\ 
				\cline{2-14} 
				400&p-SBIC&0.69&0.58&0.65&0.67&0.72&0.77&0.14&0.17&0.20&0.22&0.24&0.31\\ 
				&1&0.44&0.41&0.51&0.55&0.63&0.72&0.15&0.18&0.21&0.25&0.33&0.52\\ 
				&P true&0.69&0.58&0.65&0.67&0.72&0.77&0.14&0.17&0.20&0.22&0.24&0.31\\ 
				\cline{2-14} 
				1000&p-SBIC&0.59&0.50&0.54&0.61&0.69&0.80&0.09&0.12&0.13&0.14&0.17&0.23\\ 
				&1&0.34&0.32&0.37&0.43&0.51&0.66&0.10&0.12&0.15&0.18&0.23&0.39\\ 
				&P true&0.60&0.50&0.55&0.61&0.69&0.80&0.09&0.12&0.13&0.14&0.17&0.23\\ 
				\hline \hline \end{tabular}}\end{center}
\begin{tablenotes}
	\textit{Note}: Same design as the Method~1 table but using Method~2 (recursion beyond $H$).
	High persistence is $\sum_{i=1}^{p}\phi_i\in[0.9,0.99]$.
\end{tablenotes}
\end{table}

\clearpage

\begin{table}[h!]\caption{AR($p$): coverage for confidence intervals targeting the true IRF vs.\ the estimated IRF (low persistence)}\label{tab-comp-ARp-min03-max09-nobiasc}\begin{center}\scalebox{1}{\begin{tabular}{ll|cccccc} \hline \hline 
				&& \multicolumn{6}{c}{Coverage}\\  
				$T$& $p/H$& 5& 10& 15& 20& 30& 60\\ \hline 
				&& \multicolumn{6}{c}{P=4}\\  
				200&p-SBIC&0.86&0.86&0.86&0.86&0.86&0.85\\ 
				&1&0.24&0.17&0.14&0.12&0.10&0.07\\ 
				&P true&0.89&0.89&0.90&0.90&0.90&0.91\\ 
				\cline{2-8} 
				400&p-SBIC&0.89&0.87&0.87&0.86&0.86&0.86\\ 
				&1&0.15&0.11&0.09&0.08&0.07&0.06\\ 
				&P true&0.90&0.90&0.89&0.89&0.89&0.90\\ 
				\cline{2-8} 
				1000&p-SBIC&0.88&0.88&0.88&0.88&0.88&0.88\\ 
				&1&0.12&0.09&0.07&0.06&0.05&0.03\\ 
				&P true&0.89&0.90&0.90&0.90&0.90&0.90\\ 
				\hline 
				&& \multicolumn{6}{c}{P=6}\\  
				200&p-SBIC&0.83&0.82&0.82&0.82&0.82&0.82\\ 
				&1&0.21&0.16&0.12&0.10&0.07&0.04\\ 
				&P true&0.87&0.87&0.87&0.87&0.87&0.87\\ 
				\cline{2-8} 
				400&p-SBIC&0.87&0.85&0.85&0.85&0.85&0.84\\ 
				&1&0.15&0.11&0.08&0.07&0.06&0.04\\ 
				&P true&0.89&0.88&0.88&0.88&0.87&0.87\\ 
				\cline{2-8} 
				1000&p-SBIC&0.86&0.85&0.85&0.86&0.86&0.86\\ 
				&1&0.10&0.08&0.06&0.05&0.04&0.02\\ 
				&P true&0.87&0.87&0.87&0.87&0.87&0.88\\ 
				\hline 
				&& \multicolumn{6}{c}{P=10}\\  
				200&p-SBIC&0.81&0.82&0.81&0.81&0.81&0.82\\ 
				&1&0.17&0.15&0.12&0.10&0.07&0.04\\ 
				&P true&0.85&0.87&0.86&0.86&0.86&0.86\\ 
				\cline{2-8} 
				400&p-SBIC&0.88&0.89&0.88&0.88&0.88&0.88\\ 
				&1&0.13&0.10&0.09&0.07&0.05&0.03\\ 
				&P true&0.91&0.91&0.90&0.90&0.90&0.89\\ 
				\cline{2-8} 
				1000&p-SBIC&0.85&0.86&0.86&0.87&0.87&0.87\\ 
				&1&0.08&0.07&0.06&0.05&0.03&0.02\\ 
				&P true&0.89&0.89&0.88&0.89&0.89&0.89\\ 
				\hline \hline \end{tabular}}\end{center}
\begin{tablenotes}
	\textit{Note}: Coverage probabilities for confidence intervals when the target is the \emph{true} impulse response versus the \emph{estimated} IRF function under AR($p$) designs with low persistence ($\sum_{i=1}^{p}\phi_i \in [0.3,0.9]$).
	Rows vary the first-step lag choice in AR estimation (SBIC, fixed $p=1$, or $P$ equal to the DGP order); blocks report DGP orders $P\in\{4,6,10\}$; columns are horizons $h$.
	Coverage is computed using non–bias-corrected percentile-$t$ intervals (Kilian, 1999) with $\alpha=0.1$.
\end{tablenotes}
\end{table}

\begin{table}[h!]\caption{AR($p$): coverage for confidence intervals targeting the true IRF vs.\ the estimated IRF (medium persistence)}\label{tab-comp-ARp-min07-max09-nobiasc}\begin{center}\scalebox{1}{\begin{tabular}{ll|cccccc} \hline \hline 
				&& \multicolumn{6}{c}{Coverage}\\  
				$T$& $p/H$& 5& 10& 15& 20& 30& 60\\ \hline 
				&& \multicolumn{6}{c}{P=4}\\  
				200&p-SBIC&0.81&0.81&0.82&0.81&0.81&0.82\\ 
				&1&0.27&0.24&0.18&0.15&0.14&0.10\\ 
				&P true&0.84&0.84&0.84&0.84&0.84&0.86\\ 
				\cline{2-8} 
				400&p-SBIC&0.90&0.88&0.86&0.86&0.86&0.85\\ 
				&1&0.19&0.16&0.12&0.10&0.09&0.07\\ 
				&P true&0.90&0.88&0.87&0.87&0.87&0.87\\ 
				\cline{2-8} 
				1000&p-SBIC&0.87&0.87&0.87&0.86&0.85&0.84\\ 
				&1&0.12&0.11&0.08&0.07&0.06&0.04\\ 
				&P true&0.89&0.88&0.88&0.87&0.87&0.86\\ 
				\hline 
				&& \multicolumn{6}{c}{P=6}\\  
				200&p-SBIC&0.82&0.81&0.81&0.80&0.80&0.80\\ 
				&1&0.22&0.18&0.15&0.12&0.10&0.07\\ 
				&P true&0.84&0.83&0.84&0.84&0.84&0.84\\ 
				\cline{2-8} 
				400&p-SBIC&0.86&0.86&0.86&0.85&0.85&0.84\\ 
				&1&0.17&0.13&0.11&0.09&0.08&0.06\\ 
				&P true&0.89&0.89&0.89&0.88&0.89&0.89\\ 
				\cline{2-8} 
				1000&p-SBIC&0.87&0.87&0.88&0.88&0.87&0.88\\ 
				&1&0.09&0.08&0.07&0.06&0.05&0.04\\ 
				&P true&0.88&0.88&0.89&0.89&0.89&0.89\\ 
				\hline 
				&& \multicolumn{6}{c}{P=10}\\  
				200&p-SBIC&0.79&0.79&0.79&0.79&0.78&0.79\\ 
				&1&0.19&0.16&0.13&0.11&0.08&0.05\\ 
				&P true&0.83&0.84&0.84&0.84&0.83&0.84\\ 
				\cline{2-8} 
				400&p-SBIC&0.88&0.87&0.86&0.85&0.85&0.85\\ 
				&1&0.15&0.12&0.10&0.07&0.05&0.03\\ 
				&P true&0.90&0.89&0.88&0.88&0.88&0.87\\ 
				\cline{2-8} 
				1000&p-SBIC&0.84&0.86&0.86&0.87&0.87&0.88\\ 
				&1&0.09&0.08&0.06&0.05&0.04&0.02\\ 
				&P true&0.88&0.88&0.88&0.89&0.89&0.89\\ 			\hline \hline \end{tabular}}\end{center}
\begin{tablenotes}
	\textit{Note}: Coverage probabilities comparing intervals for the \emph{true} IRF to those for the \emph{estimated} IRF under AR($p$) designs with medium persistence ($\sum_{i=1}^{p}\phi_i \in [0.7,0.9]$).
	Rows show AR lag specifications (SBIC, fixed $p=1$, or $P$ equal to the DGP order); blocks correspond to $P\in\{4,6,10\}$; columns are horizons $h$.
	Intervals use non–bias-corrected percentile-$t$ (Kilian, 1999) with $\alpha=0.1$.
\end{tablenotes}
\end{table}

\begin{table}[h!]\caption{AR($p$): coverage for confidence intervals targeting the true IRF vs.\ the estimated IRF (high persistence)}\label{tab-comp-ARp-min09-max099-nobiasc}\begin{center}\scalebox{1}{\begin{tabular}{ll|cccccc} \hline \hline 
				&& \multicolumn{6}{c}{Coverage}\\  
				$T$& $p/H$& 5& 10& 15& 20& 30& 60\\ \hline 
				&& \multicolumn{6}{c}{P=4}\\  
				200&p-SBIC&0.77&0.72&0.69&0.66&0.63&0.60\\ 
				&1&0.29&0.29&0.26&0.23&0.19&0.18\\ 
				&P true&0.79&0.74&0.69&0.66&0.62&0.59\\ 
				\cline{2-8} 
				400&p-SBIC&0.85&0.83&0.81&0.80&0.78&0.74\\ 
				&1&0.23&0.25&0.27&0.25&0.22&0.20\\ 
				&P true&0.88&0.86&0.83&0.81&0.79&0.75\\ 
				\cline{2-8} 
				1000&p-SBIC&0.84&0.84&0.83&0.83&0.82&0.81\\ 
				&1&0.12&0.18&0.21&0.19&0.17&0.14\\ 
				&P true&0.86&0.85&0.84&0.83&0.82&0.81\\ 
				\hline 
				&& \multicolumn{6}{c}{P=6}\\  
				200&p-SBIC&0.80&0.75&0.72&0.70&0.68&0.65\\ 
				&1&0.23&0.25&0.24&0.22&0.20&0.18\\ 
				&P true&0.82&0.77&0.73&0.71&0.69&0.67\\ 
				\cline{2-8} 
				400&p-SBIC&0.86&0.81&0.79&0.77&0.75&0.72\\ 
				&1&0.16&0.21&0.21&0.20&0.18&0.14\\ 
				&P true&0.89&0.83&0.81&0.78&0.76&0.73\\ 
				\cline{2-8} 
				1000&p-SBIC&0.86&0.84&0.83&0.82&0.81&0.80\\ 
				&1&0.10&0.13&0.13&0.14&0.13&0.11\\ 
				&P true&0.87&0.85&0.83&0.83&0.82&0.81\\ 
				\hline 
				&& \multicolumn{6}{c}{P=10}\\  
				200&p-SBIC&0.80&0.76&0.74&0.71&0.67&0.64\\ 
				&1&0.20&0.21&0.18&0.16&0.13&0.09\\ 
				&P true&0.81&0.78&0.77&0.73&0.69&0.66\\ 
				\cline{2-8} 
				400&p-SBIC&0.88&0.86&0.83&0.81&0.78&0.74\\ 
				&1&0.14&0.16&0.15&0.14&0.12&0.09\\ 
				&P true&0.90&0.88&0.85&0.82&0.78&0.74\\ 
				\cline{2-8} 
				1000&p-SBIC&0.85&0.85&0.85&0.84&0.83&0.83\\ 
				&1&0.09&0.12&0.11&0.10&0.10&0.07\\ 
				&P true&0.87&0.87&0.86&0.85&0.84&0.83\\ 
				\hline \hline \end{tabular}}\end{center}
\begin{tablenotes}
	\textit{Note}: Coverage probabilities for intervals aimed at the \emph{true} IRF versus the \emph{estimated} IRF under AR($p$) designs with high persistence ($\sum_{i=1}^{p}\phi_i \in [0.9,0.99]$).
	Rows vary AR lag selection (SBIC, fixed $p=1$, or $P$ equal to the DGP order); blocks indicate DGP order $P$; columns are horizons $h$.
	Intervals are non–bias-corrected percentile-$t$ (Kilian, 1999), $\alpha=0.1$.
\end{tablenotes}
\end{table}

\begin{table}[h!]\caption{Bias with bootstrapping local projections, AR(p) model, low persistence (block wild bootstrap, BWB)}\label{tab-boot3-bias-ARp_nobiasc_low}\begin{center}\scalebox{0.8}{\begin{tabular}{ll|cccccc|cccccc|cccccc} \hline \hline 
				&& \multicolumn{12}{c}{LP} & \multicolumn{6}{c}{AR}\\  
				&& \multicolumn{6}{c}{Method~1} & \multicolumn{6}{c}{Method~2}&&&&&&\\  
				$T$& $p/H$& 5& 10& 15& 20& 30& 60& 5& 10& 15& 20& 30& 60& 5& 10& 15& 20& 30& 60\\ \hline 
				&& \multicolumn{18}{c}{P=4}\\  
				200&p-SBIC&0.09&0.09&0.10&0.11&0.12&0.14&0.09&0.09&0.10&0.11&0.12&0.14&0.06&0.06&0.06&0.05&0.05&0.04\\ 
				&1&0.16&0.16&0.15&0.16&0.17&0.24&0.15&0.15&0.15&0.16&0.17&0.25&0.03&0.05&0.07&0.08&0.10&0.12\\ 
				&P true&0.15&0.15&0.13&0.13&0.12&0.14&0.15&0.14&0.13&0.12&0.12&0.14&0.01&0.01&0.01&0.02&0.02&0.04\\ 
				\cline{2-20} 
				400&p-SBIC&0.13&0.13&0.14&0.14&0.13&0.10&0.13&0.13&0.14&0.14&0.13&0.10&0.06&0.06&0.06&0.06&0.05&0.03\\ 
				&1&0.08&0.09&0.10&0.10&0.12&0.19&0.08&0.09&0.10&0.10&0.13&0.20&0.04&0.06&0.07&0.08&0.10&0.13\\ 
				&P true&0.08&0.08&0.08&0.08&0.08&0.10&0.08&0.08&0.08&0.08&0.09&0.10&0.01&0.01&0.01&0.01&0.02&0.03\\ 
				\cline{2-20} 
				1000&p-SBIC&0.09&0.09&0.09&0.08&0.08&0.06&0.09&0.09&0.09&0.08&0.08&0.06&0.05&0.05&0.04&0.04&0.04&0.02\\ 
				&1&0.06&0.07&0.08&0.09&0.11&0.17&0.06&0.07&0.08&0.09&0.11&0.18&0.04&0.06&0.08&0.09&0.11&0.13\\ 
				&P true&0.08&0.08&0.08&0.07&0.07&0.06&0.08&0.08&0.07&0.07&0.07&0.06&0.01&0.01&0.01&0.01&0.01&0.02\\ 
				\hline 
				&& \multicolumn{18}{c}{P=6}\\  
				200&p-SBIC&0.14&0.14&0.13&0.13&0.13&0.15&0.14&0.14&0.13&0.12&0.13&0.15&0.01&0.01&0.02&0.02&0.03&0.06\\ 
				&1&0.14&0.15&0.16&0.17&0.18&0.28&0.15&0.15&0.16&0.17&0.19&0.29&0.06&0.08&0.10&0.11&0.13&0.17\\ 
				&P true&0.18&0.17&0.17&0.17&0.16&0.15&0.18&0.18&0.18&0.17&0.17&0.15&0.08&0.06&0.06&0.05&0.05&0.06\\ 
				\cline{2-20} 
				400&p-SBIC&0.08&0.08&0.08&0.08&0.08&0.11&0.08&0.08&0.08&0.07&0.08&0.11&0.01&0.01&0.01&0.02&0.02&0.04\\ 
				&1&0.09&0.10&0.11&0.12&0.15&0.24&0.09&0.10&0.12&0.13&0.16&0.26&0.08&0.10&0.11&0.12&0.14&0.18\\ 
				&P true&0.13&0.12&0.11&0.11&0.10&0.11&0.13&0.12&0.11&0.11&0.10&0.11&0.08&0.06&0.05&0.05&0.04&0.04\\ 
				\cline{2-20} 
				1000&p-SBIC&0.06&0.05&0.06&0.05&0.06&0.07&0.06&0.06&0.06&0.05&0.06&0.07&0.00&0.01&0.01&0.01&0.01&0.03\\ 
				&1&0.07&0.08&0.09&0.10&0.13&0.24&0.07&0.08&0.10&0.12&0.15&0.25&0.06&0.09&0.10&0.12&0.14&0.18\\ 
				&P true&0.22&0.19&0.16&0.14&0.10&0.07&0.23&0.20&0.17&0.14&0.11&0.07&0.08&0.06&0.05&0.04&0.03&0.03\\ 
				\hline 
				&& \multicolumn{18}{c}{P=10}\\  
				200&p-SBIC&0.14&0.13&0.13&0.14&0.13&0.15&0.14&0.14&0.14&0.14&0.13&0.15&0.03&0.03&0.04&0.04&0.05&0.08\\ 
				&1&0.15&0.16&0.18&0.20&0.23&0.32&0.15&0.17&0.19&0.21&0.24&0.33&0.11&0.13&0.15&0.17&0.19&0.23\\ 
				&P true&0.33&0.29&0.26&0.24&0.18&0.15&0.36&0.32&0.29&0.27&0.20&0.15&0.17&0.16&0.15&0.14&0.12&0.08\\ 
				\cline{2-20} 
				400&p-SBIC&0.09&0.08&0.09&0.09&0.09&0.12&0.08&0.08&0.09&0.09&0.09&0.12&0.04&0.04&0.04&0.04&0.04&0.05\\ 
				&1&0.11&0.12&0.14&0.15&0.19&0.28&0.10&0.12&0.14&0.16&0.19&0.31&0.10&0.12&0.14&0.16&0.20&0.24\\ 
				&P true&0.25&0.22&0.20&0.18&0.14&0.12&0.27&0.25&0.23&0.21&0.18&0.12&0.17&0.16&0.15&0.13&0.11&0.05\\ 
				\cline{2-20} 
				1000&p-SBIC&0.06&0.06&0.07&0.06&0.07&0.08&0.06&0.06&0.07&0.07&0.07&0.08&0.02&0.02&0.02&0.02&0.03&0.04\\ 
				&1&0.10&0.11&0.14&0.15&0.18&0.27&0.10&0.12&0.15&0.16&0.20&0.29&0.09&0.11&0.14&0.16&0.19&0.25\\ 
				&P true&0.30&0.26&0.24&0.20&0.14&0.08&0.34&0.31&0.28&0.25&0.19&0.08&0.17&0.15&0.14&0.12&0.10&0.03\\ 
				\hline \hline 
					\end{tabular}}\end{center}
		\begin{tablenotes}
			\textit{Note}: Average absolute bias of bootstrap impulse–response estimates for AR($p$) designs under
			low persistence, $\sum_{i=1}^{p}\phi_i \in [0.3,0.9]$. Entries report, for each horizon $h$, the Monte Carlo
			mean of
			\[
			\left| \mathrm{IRF}_{\text{true},h} - \frac{1}{B}\sum_{b=1}^{B} \mathrm{IRF}^{(b)}_{\text{boot},h} \right|,
			\]
			with bootstrap series generated using the Block Wild Bootstrap (BWB), without small sample  bias correction.
			Rows vary sample size $T$, the first-step LP lag choice (SBIC vs.\ fixed $p$), and the DGP order $P$ as indicated.
			Where multiple estimators are shown, LP results are reported for Method~1 (truncation at $H$) and, where applicable,
			Method~2 (recursion beyond $H$); “AR” denotes the autoregressive benchmark. Bias is measured in the same units as the
			response (shock normalized to one standard deviation).
		\end{tablenotes}	
		
\end{table}

\begin{table}[h!]\caption{Bias with bootstrapping local projections, AR(p) model, medium persistence (block wild bootstrap, BWB)}\label{tab-boot3-bias-ARp_nobiasc_medium}\begin{center}\scalebox{0.8}{\begin{tabular}{ll|cccccc|cccccc|cccccc} \hline \hline 
				&& \multicolumn{12}{c}{LP} & \multicolumn{6}{c}{AR}\\  
				&& \multicolumn{6}{c}{Method~1} & \multicolumn{6}{c}{Method~2}&&&&&&\\  
				$T$& $p/H$& 5& 10& 15& 20& 30& 60& 5& 10& 15& 20& 30& 60& 5& 10& 15& 20& 30& 60\\ \hline 
				&& \multicolumn{18}{c}{P=4}\\  
				200&p-SBIC&0.11&0.10&0.12&0.12&0.13&0.15&0.11&0.11&0.11&0.12&0.13&0.15&0.07&0.07&0.07&0.07&0.06&0.04\\ 
				&1&0.11&0.12&0.13&0.14&0.16&0.25&0.11&0.12&0.13&0.14&0.17&0.27&0.06&0.07&0.09&0.10&0.11&0.13\\ 
				&P true&0.13&0.13&0.13&0.13&0.14&0.15&0.13&0.13&0.13&0.13&0.14&0.15&0.05&0.04&0.03&0.03&0.03&0.04\\ 
				\cline{2-20} 
				400&p-SBIC&0.13&0.12&0.13&0.13&0.13&0.11&0.13&0.13&0.12&0.13&0.13&0.11&0.07&0.07&0.06&0.06&0.05&0.03\\ 
				&1&0.09&0.10&0.10&0.11&0.12&0.21&0.09&0.10&0.11&0.11&0.13&0.21&0.04&0.06&0.08&0.09&0.10&0.13\\ 
				&P true&0.16&0.14&0.13&0.12&0.11&0.11&0.16&0.15&0.13&0.13&0.11&0.11&0.05&0.04&0.03&0.03&0.02&0.03\\ 
				\cline{2-20} 
				1000&p-SBIC&0.11&0.10&0.11&0.09&0.08&0.07&0.10&0.10&0.10&0.09&0.08&0.07&0.04&0.04&0.04&0.03&0.03&0.02\\ 
				&1&0.07&0.08&0.08&0.09&0.11&0.18&0.07&0.08&0.09&0.10&0.12&0.19&0.04&0.06&0.08&0.09&0.11&0.14\\ 
				&P true&0.12&0.11&0.10&0.09&0.09&0.07&0.13&0.11&0.10&0.10&0.10&0.07&0.05&0.03&0.02&0.02&0.02&0.02\\ 
				\hline 
				&& \multicolumn{18}{c}{P=6}\\  
				200&p-SBIC&0.12&0.12&0.12&0.12&0.13&0.16&0.12&0.12&0.12&0.12&0.13&0.16&0.02&0.03&0.03&0.03&0.04&0.06\\ 
				&1&0.50&0.52&0.51&0.50&0.46&0.28&0.50&0.51&0.50&0.49&0.45&0.30&0.31&0.32&0.33&0.32&0.30&0.17\\ 
				&P true&0.69&0.56&0.52&0.40&0.30&0.16&1.05&0.92&0.88&0.75&0.62&0.16&0.77&0.72&0.57&0.47&0.38&0.06\\ 
				\cline{2-20} 
				400&p-SBIC&0.09&0.08&0.09&0.09&0.08&0.11&0.09&0.08&0.08&0.09&0.08&0.11&0.02&0.02&0.02&0.02&0.03&0.04\\ 
				&1&0.37&0.36&0.36&0.35&0.31&0.24&0.35&0.35&0.35&0.34&0.30&0.26&0.21&0.22&0.23&0.23&0.23&0.17\\ 
				&P true&0.64&0.51&0.45&0.32&0.20&0.11&1.00&0.86&0.79&0.63&0.47&0.11&0.80&0.73&0.58&0.46&0.38&0.04\\ 
				\cline{2-20} 
				1000&p-SBIC&0.06&0.06&0.06&0.06&0.06&0.07&0.06&0.06&0.06&0.06&0.06&0.07&0.01&0.02&0.01&0.02&0.02&0.03\\ 
				&1&0.32&0.30&0.30&0.28&0.28&0.23&0.27&0.26&0.26&0.25&0.25&0.24&0.10&0.12&0.14&0.15&0.17&0.18\\ 
				&P true&0.67&0.53&0.48&0.34&0.22&0.07&1.06&0.93&0.87&0.73&0.57&0.07&0.90&0.80&0.61&0.46&0.40&0.03\\ 
				\hline 
				&& \multicolumn{18}{c}{P=10}\\  
				200&p-SBIC&0.50&0.50&0.49&0.47&0.41&0.16&0.49&0.49&0.48&0.46&0.40&0.16&0.28&0.27&0.27&0.25&0.22&0.09\\ 
				&1&0.14&0.16&0.18&0.19&0.22&0.31&0.14&0.16&0.18&0.20&0.24&0.33&0.13&0.14&0.16&0.17&0.20&0.24\\ 
				&P true&0.15&0.14&0.14&0.15&0.14&0.16&0.15&0.14&0.14&0.15&0.14&0.16&0.12&0.11&0.10&0.10&0.09&0.08\\ 
				\cline{2-20} 
				400&p-SBIC&0.36&0.34&0.33&0.31&0.25&0.12&0.34&0.33&0.32&0.30&0.24&0.13&0.17&0.17&0.16&0.15&0.13&0.06\\ 
				&1&0.12&0.14&0.16&0.16&0.19&0.29&0.12&0.14&0.16&0.17&0.19&0.31&0.09&0.11&0.13&0.15&0.18&0.25\\ 
				&P true&0.28&0.26&0.22&0.19&0.15&0.12&0.28&0.26&0.22&0.20&0.15&0.12&0.12&0.10&0.09&0.09&0.08&0.06\\ 
				\cline{2-20} 
				1000&p-SBIC&0.31&0.28&0.27&0.24&0.21&0.08&0.26&0.24&0.23&0.20&0.18&0.08&0.06&0.06&0.06&0.06&0.06&0.04\\ 
				&1&0.09&0.11&0.13&0.14&0.17&0.27&0.09&0.11&0.13&0.15&0.18&0.29&0.09&0.11&0.13&0.15&0.19&0.25\\ 
				&P true&0.29&0.26&0.23&0.20&0.15&0.08&0.29&0.27&0.24&0.21&0.16&0.08&0.12&0.10&0.09&0.08&0.06&0.04\\ 
				\hline \hline \end{tabular}}\end{center}
			\begin{tablenotes}
				\textit{Note}: Average absolute bias of bootstrap impulse–response estimates for AR($p$) designs under
				medium persistence, $\sum_{i=1}^{p}\phi_i \in [0.7,0.9]$. Entries report, for each horizon $h$, the Monte Carlo
				mean of
				\[
				\left| \mathrm{IRF}_{\text{true},h} - \frac{1}{B}\sum_{b=1}^{B} \mathrm{IRF}^{(b)}_{\text{boot},h} \right|,
				\]
				with bootstrap series generated using the Block Wild Bootstrap (BWB), without small sample  bias correction.
				Rows vary sample size $T$, the first-step LP lag choice (SBIC vs.\ fixed $p$), and the DGP order $P$ as indicated.
				Where multiple estimators are shown, LP results are reported for Method~1 (truncation at $H$) and, where applicable,
				Method~2 (recursion beyond $H$); “AR” denotes the autoregressive benchmark. Bias is measured in the same units as the
				response (shock normalized to one standard deviation).
				\end{tablenotes}
			\end{table}

\begin{table}[h!]\caption{Bias with bootstrapping local projections, AR(p) model, high persistence (block wild bootstrap, BWB)}\label{tab-boot3-bias-ARp_nobiasc-high}\begin{center}\scalebox{0.8}{\begin{tabular}{ll|cccccc|cccccc|cccccc} \hline \hline 
				&& \multicolumn{12}{c}{LP} & \multicolumn{6}{c}{AR}\\  
				&& \multicolumn{6}{c}{Method~1} & \multicolumn{6}{c}{Method~2}&&&&&&\\  
				$T$& $p/H$& 5& 10& 15& 20& 30& 60& 5& 10& 15& 20& 30& 60& 5& 10& 15& 20& 30& 60\\ \hline 
				&& \multicolumn{18}{c}{P=4}\\  
				200&p-SBIC&0.17&0.21&0.20&0.22&0.22&0.26&0.16&0.21&0.21&0.21&0.23&0.26&0.08&0.11&0.12&0.13&0.14&0.13\\ 
				&1&0.64&0.64&0.60&0.60&0.48&0.35&0.65&0.66&0.61&0.61&0.50&0.39&0.25&0.25&0.26&0.28&0.28&0.26\\ 
				&P true&1.11&0.89&0.76&0.72&0.42&0.26&1.47&1.24&1.09&1.03&0.64&0.25&0.78&0.76&0.74&0.63&0.37&0.13\\ 
				\cline{2-20} 
				400&p-SBIC&0.22&0.23&0.21&0.20&0.20&0.18&0.21&0.22&0.20&0.18&0.19&0.18&0.17&0.15&0.15&0.15&0.13&0.09\\ 
				&1&0.30&0.29&0.27&0.29&0.29&0.28&0.31&0.29&0.28&0.30&0.30&0.30&0.12&0.15&0.16&0.18&0.22&0.25\\ 
				&P true&1.19&0.94&0.77&0.72&0.38&0.18&1.19&0.94&0.76&0.71&0.37&0.18&0.85&0.83&0.80&0.67&0.35&0.09\\ 
				\cline{2-20} 
				1000&p-SBIC&0.18&0.18&0.17&0.17&0.15&0.12&0.16&0.16&0.16&0.15&0.14&0.12&0.16&0.16&0.15&0.13&0.10&0.05\\ 
				&1&0.35&0.32&0.28&0.28&0.26&0.23&0.34&0.31&0.28&0.29&0.26&0.23&0.12&0.13&0.15&0.16&0.20&0.24\\ 
				&P true&1.12&0.91&0.75&0.71&0.39&0.12&1.13&0.92&0.75&0.71&0.41&0.12&0.90&0.89&0.84&0.69&0.34&0.05\\ 
				\hline 
				&& \multicolumn{18}{c}{P=6}\\  
				200&p-SBIC&0.63&0.64&0.59&0.58&0.45&0.26&0.65&0.65&0.60&0.59&0.45&0.26&0.22&0.21&0.21&0.22&0.19&0.14\\ 
				&1&0.41&0.43&0.44&0.45&0.48&0.47&0.40&0.42&0.42&0.44&0.49&0.47&0.18&0.19&0.20&0.22&0.24&0.30\\ 
				&P true&0.79&0.67&0.60&0.57&0.45&0.26&0.83&0.71&0.64&0.61&0.48&0.26&0.22&0.21&0.18&0.16&0.15&0.14\\ 
				\cline{2-20} 
				400&p-SBIC&0.30&0.28&0.26&0.26&0.25&0.19&0.30&0.29&0.27&0.27&0.26&0.19&0.09&0.10&0.10&0.10&0.11&0.10\\ 
				&1&0.29&0.31&0.34&0.34&0.31&0.39&0.29&0.31&0.33&0.35&0.33&0.44&0.17&0.19&0.21&0.21&0.24&0.30\\ 
				&P true&0.50&0.40&0.34&0.32&0.23&0.19&0.51&0.41&0.35&0.34&0.24&0.19&0.22&0.19&0.16&0.13&0.12&0.10\\ 
				\cline{2-20} 
				1000&p-SBIC&0.35&0.31&0.27&0.25&0.21&0.13&0.34&0.31&0.26&0.25&0.21&0.13&0.09&0.08&0.08&0.08&0.08&0.06\\ 
				&1&0.22&0.23&0.23&0.24&0.24&0.37&0.22&0.23&0.23&0.25&0.24&0.38&0.17&0.19&0.20&0.21&0.23&0.31\\ 
				&P true&0.60&0.48&0.41&0.38&0.26&0.13&0.58&0.47&0.40&0.36&0.25&0.13&0.22&0.19&0.15&0.12&0.10&0.06\\ 
				\hline 
				&& \multicolumn{18}{c}{P=10}\\  
				200&p-SBIC&0.40&0.41&0.40&0.39&0.39&0.27&0.39&0.39&0.38&0.38&0.38&0.26&0.14&0.13&0.12&0.13&0.13&0.17\\ 
				&1&0.30&0.32&0.29&0.26&0.30&0.40&0.31&0.32&0.28&0.26&0.30&0.42&0.21&0.22&0.24&0.26&0.28&0.33\\ 
				&P true&0.37&0.31&0.26&0.23&0.20&0.27&0.37&0.31&0.26&0.23&0.21&0.26&0.29&0.27&0.24&0.22&0.21&0.17\\ 
				\cline{2-20} 
				400&p-SBIC&0.28&0.29&0.30&0.28&0.22&0.19&0.28&0.29&0.30&0.29&0.22&0.19&0.12&0.12&0.12&0.11&0.11&0.11\\ 
				&1&0.18&0.20&0.23&0.25&0.28&0.36&0.17&0.19&0.21&0.23&0.26&0.38&0.23&0.25&0.27&0.28&0.28&0.33\\ 
				&P true&0.43&0.36&0.29&0.25&0.21&0.19&0.43&0.36&0.30&0.26&0.22&0.19&0.27&0.26&0.22&0.19&0.18&0.11\\ 
				\cline{2-20} 
				1000&p-SBIC&0.20&0.20&0.19&0.18&0.14&0.14&0.20&0.20&0.19&0.19&0.14&0.14&0.12&0.11&0.10&0.09&0.07&0.06\\ 
				&1&0.21&0.22&0.23&0.24&0.26&0.37&0.21&0.22&0.23&0.24&0.27&0.40&0.22&0.24&0.25&0.25&0.27&0.33\\ 
				&P true&0.44&0.37&0.30&0.26&0.20&0.14&0.43&0.36&0.29&0.25&0.20&0.14&0.31&0.27&0.21&0.18&0.15&0.06\\ 
				\hline \hline \end{tabular}}\end{center}
			\begin{tablenotes}
				\textit{Note}: Average absolute bias of bootstrap impulse–response estimates for AR($p$) designs under
				high persistence, $\sum_{i=1}^{p}\phi_i \in [0.9,0.99]$. For each horizon $h$, entries report the Monte Carlo
				mean of
				\[
				\left| \mathrm{IRF}_{\text{true},h} - \frac{1}{B}\sum_{b=1}^{B} \mathrm{IRF}^{(b)}_{\text{boot},h} \right|,
				\]
				with bootstrap series generated using the Block Wild Bootstrap (BWB), without small sample  bias correction.
				Rows vary the sample size $T$, the first-step LP lag choice (SBIC vs.\ fixed $p$), and the DGP order $P$ as indicated.
				Where multiple estimators are shown, LP results are reported for Method~1 (truncation at $H$) and, where applicable,
				Method~2 (recursion beyond $H$); “AR” denotes the autoregressive benchmark. Bias is measured in the same units as the
				response (shock normalized to one standard deviation).
				\end{tablenotes}
			\end{table}

\clearpage	

			\subsection{MA(q)-GBF(1) univariate}		
					
\begin{table}[h!]\caption{Results of MA-GBF(1) model, comparing true and estimated IRF function)}\label{tab-comp-MA_FAIR1}\begin{center}\scalebox{0.8}{\begin{tabular}{ll|cccc} \hline \hline 
				&& \multicolumn{4}{c}{Coverage}\\  
				$T$& $p/H$& 10& 20& 40& 60\\ \hline 
				200&p-SBIC&0.20&0.10&0.09&0.09\\ 
				&1&0.18&0.09&0.07&0.07\\ 
				&2&0.28&0.14&0.12&0.12\\ 
				&6&0.62&0.42&0.49&0.49\\ 
				&12&0.89&0.89&0.90&0.90\\ 
				&24&0.90&0.90&0.90&0.90\\ 
				\cline{2-6} 
				400&p-SBIC&0.18&0.10&0.09&0.09\\ 
				&1&0.14&0.07&0.06&0.06\\ 
				&2&0.23&0.11&0.09&0.09\\ 
				&6&0.54&0.33&0.40&0.40\\ 
				&12&0.90&0.85&0.84&0.84\\ 
				&24&0.91&0.90&0.90&0.90\\ 
				\cline{2-6} 
				1000&p-SBIC&0.21&0.15&0.15&0.15\\ 
				&1&0.09&0.04&0.04&0.04\\ 
				&2&0.13&0.07&0.06&0.06\\ 
				&6&0.45&0.25&0.32&0.32\\ 
				&12&0.88&0.73&0.64&0.64\\ 
				&24&0.90&0.90&0.90&0.90\\ 
				\hline \hline \end{tabular}}\end{center}
		\begin{tablenotes}
			\textit{Note}: Coverage results when the data-generating process is an MA($q$) with coefficients given by a single Gaussian basis function (GBF), “fair1” calibration (see Appendix~\ref{app:simresults}). 
			Impulse responses are estimated by fitting autoregressive models of order $p$, either selected by SBIC or fixed as indicated in the table. 
			Columns report coverage probabilities at horizons $h=\{10,20,40,60\}$; rows correspond to sample size $T$ and AR lag order $p$. 
			Results are based on 100 Monte Carlo replications.
		\end{tablenotes}
			\end{table}

\begin{table}[h!]\caption{Bias with bootstrapping local projections, MA(24)-GBF(1) model, (block wild bootstrap, BWB)}\label{tab-boot3-bias-MA-GBF1_nobiasc}\begin{center}\scalebox{0.8}{\begin{tabular}{ll|cccc|cccc|cccc} \hline \hline 
				&& \multicolumn{8}{c}{LP} & \multicolumn{4}{c}{AR}\\  
				&& \multicolumn{4}{c}{Method~1} & \multicolumn{4}{c}{Method~2}&&&&\\  
				$T$& $p/H$& 10& 20& 40& 60& 10& 20& 40& 60& 10& 20& 40& 60\\ \hline 
				200&p-SBIC&0.10&0.11&0.14&0.15&0.11&0.11&0.14&0.15&0.06&0.05&0.04&0.04\\ 
				&1&0.10&0.07&0.08&0.08&0.10&0.07&0.08&0.08&0.04&0.04&0.04&0.04\\ 
				&2&0.09&0.07&0.07&0.07&0.09&0.07&0.07&0.07&0.05&0.05&0.05&0.05\\ 
				&6&0.06&0.07&0.07&0.07&0.06&0.07&0.07&0.07&0.04&0.04&0.04&0.04\\ 
				&12&0.06&0.07&0.08&0.07&0.06&0.07&0.08&0.07&0.04&0.04&0.04&0.04\\ 
				&24&0.08&0.08&0.08&0.08&0.08&0.08&0.08&0.08&0.05&0.05&0.05&0.05\\ 
				\cline{2-14} 
				400&p-SBIC&0.08&0.11&0.14&0.15&0.08&0.12&0.14&0.15&0.05&0.05&0.04&0.04\\ 
				&1&0.07&0.05&0.06&0.06&0.07&0.05&0.06&0.06&0.04&0.04&0.04&0.04\\ 
				&2&0.06&0.05&0.05&0.05&0.06&0.05&0.05&0.05&0.04&0.04&0.04&0.04\\ 
				&6&0.05&0.05&0.05&0.05&0.05&0.05&0.05&0.05&0.04&0.04&0.04&0.04\\ 
				&12&0.04&0.04&0.05&0.05&0.04&0.04&0.05&0.05&0.03&0.03&0.03&0.03\\ 
				&24&0.04&0.04&0.05&0.05&0.04&0.04&0.05&0.05&0.03&0.03&0.04&0.04\\ 
				\cline{2-14} 
				1000&p-SBIC&0.07&0.11&0.15&0.16&0.07&0.11&0.15&0.16&0.04&0.05&0.04&0.04\\ 
				&1&0.04&0.03&0.04&0.04&0.04&0.03&0.04&0.04&0.04&0.04&0.04&0.04\\ 
				&2&0.04&0.03&0.03&0.03&0.04&0.03&0.03&0.03&0.04&0.04&0.04&0.04\\ 
				&6&0.04&0.03&0.03&0.03&0.04&0.03&0.03&0.03&0.04&0.04&0.04&0.04\\ 
				&12&0.03&0.03&0.03&0.03&0.03&0.03&0.03&0.03&0.03&0.02&0.02&0.02\\ 
				&24&0.03&0.03&0.03&0.03&0.03&0.03&0.03&0.03&0.02&0.02&0.03&0.03\\  
				\hline \hline \end{tabular}}\end{center}
			\begin{tablenotes}
				\textit{Note}: Average absolute bias at each horizon $h$ for an MA(24) data–generating process with
				Gaussian basis–function (GBF) coefficients (“fair1” calibration; see Appendix~\ref{app:simresults}).
				Bias is computed as $\left| \text{IRF}_{\text{true},h} - \frac{1}{B}\sum_{b=1}^B \text{IRF}^{(b)}_{\text{boot},h} \right|$
				for each replication and then averaged across replications. Rows vary the sample size $T$ and the first-step
				LP lag choice (SBIC or fixed $p$); columns list horizons $h$. Where reported separately, \textit{Method~1}
				constructs bootstrap series using only the first $H$ MA terms, whereas \textit{Method~2} extends beyond $H$
				via the recursion described in Appendix~\ref{app:bootalg}. Because the true IRF is exactly zero for $h>q=24$,
				bias at long horizons reflects estimation/bootstrapping noise around zero. Results use the Block Wild
				Bootstrap (BWB), no small sample  bias correction, and 100 Monte Carlo replications.
				\end{tablenotes}
			\end{table}

\clearpage
\subsection{Comparing bootstrap methods}

\begin{figure}[h!]
	\caption{Comparing bootstrap methods (T=1000, $p=\text{SBIC}$, $H=10$, coverage with percentile-$t$)}
	\label{Fig-comp-boot-ct-h10}
	\centering
	\subfloat[\tiny Panel A. $\phi=0$]{
		\includegraphics[width=0.5\textwidth]{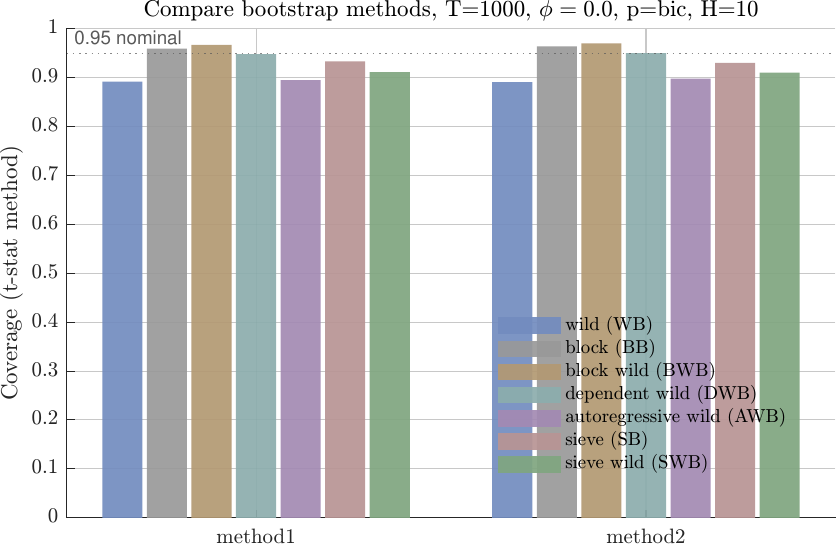}}
	\subfloat[\tiny Panel B. $\phi=0.5$]{
		\includegraphics[width=0.5\textwidth]{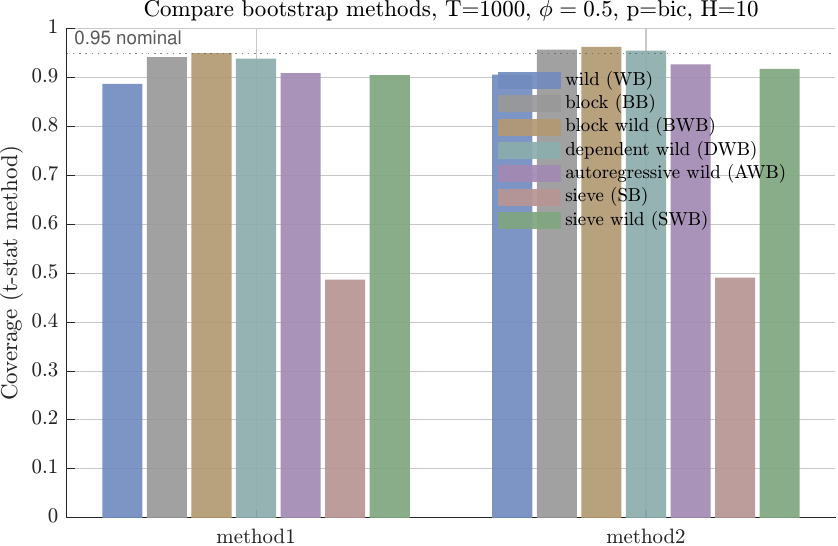}}\\
	\subfloat[\tiny Panel C. $\phi=0.95$]{
		\includegraphics[width=0.5\textwidth]{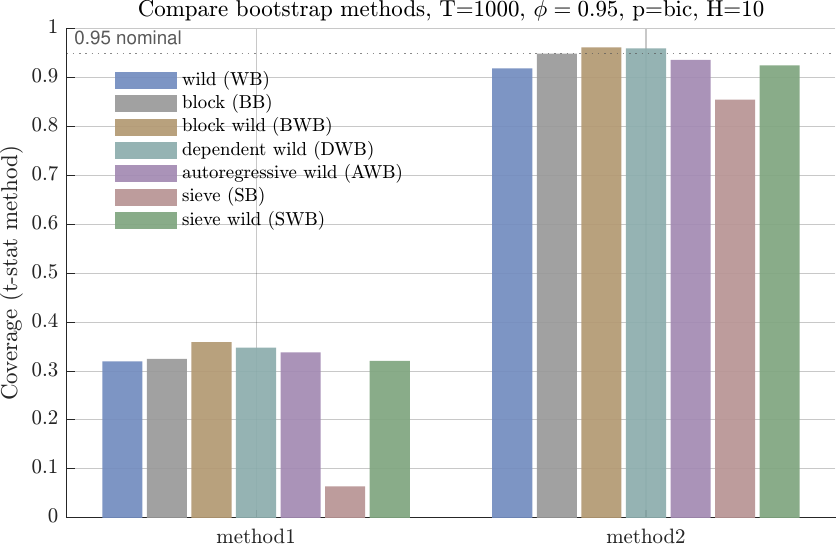}}
	\subfloat[\tiny Panel D. $\phi=1$]{
		\includegraphics[width=0.5\textwidth]{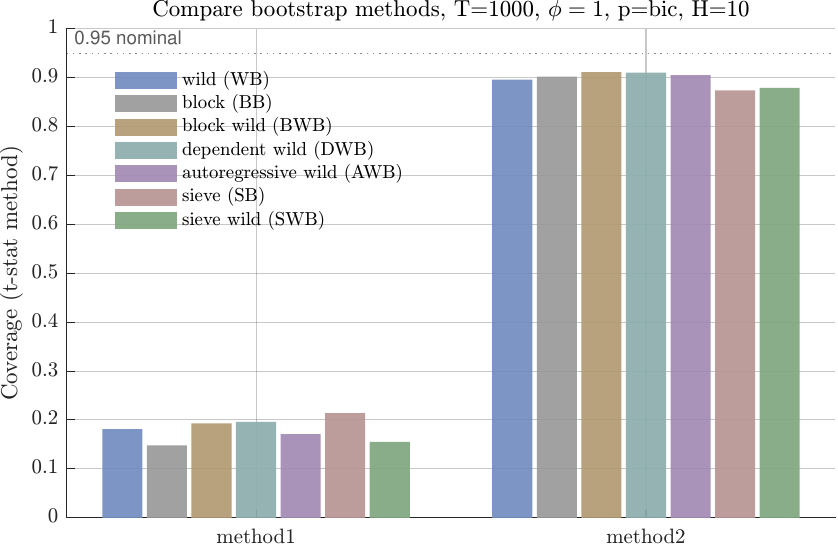}}\\
	\begin{figurenotes}
		\textit{Note}: Each subpanel reports empirical coverage of 95\% percentile-$t$ intervals
		for the bootstrap schemes shown in the legend. Bars are grouped by two estimation
		variants (``method1'' and ``method2'') with lag selection via SBIC. The gray horizontal
		line marks the 95\% nominal target; values closer to this line indicate better
		calibration. At $H=10$, differences across schemes are moderate and become more
		pronounced as persistence rises.
	\end{figurenotes}
\end{figure}

\begin{figure}[h!]
	\caption{Comparing bootstrap methods (T=1000, $p=\text{SBIC}$, $H=60$, coverage with percentile-$t$)}
	\label{Fig-comp-boot-ct-h60}
	\centering
	\subfloat[\tiny Panel A. $\phi=0$]{
		\includegraphics[width=0.5\textwidth]{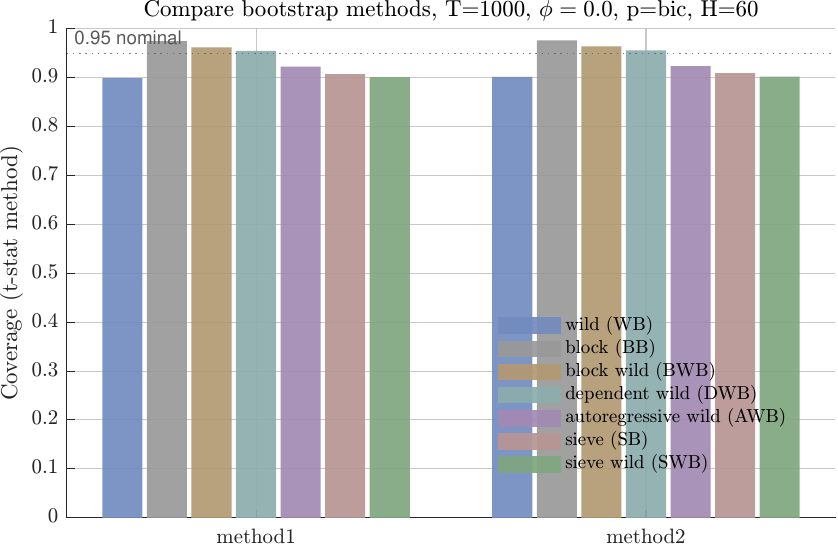}}
	\subfloat[\tiny Panel B. $\phi=0.5$]{
		\includegraphics[width=0.5\textwidth]{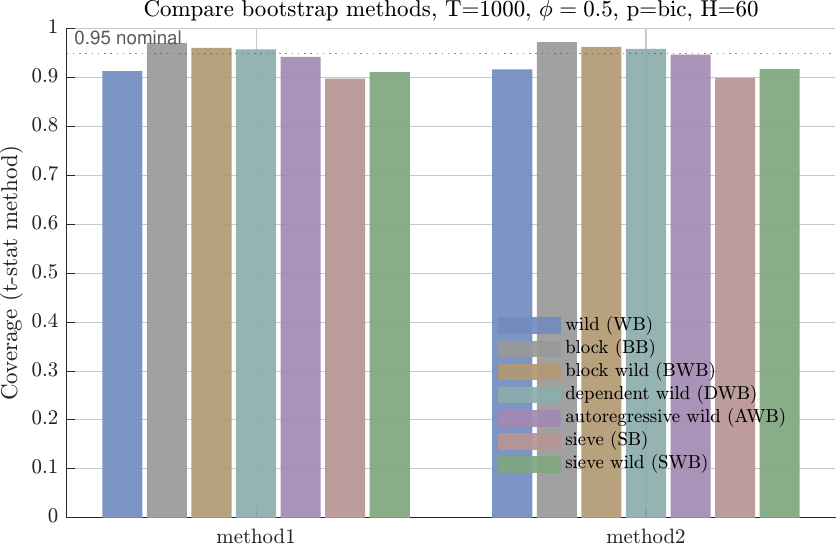}}\\
	\subfloat[\tiny Panel C. $\phi=0.95$]{
		\includegraphics[width=0.5\textwidth]{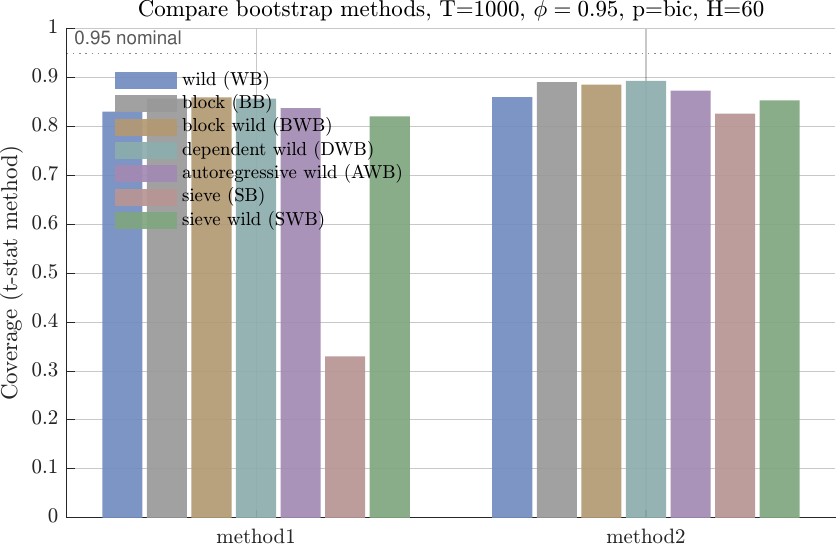}}
	\subfloat[\tiny Panel D. $\phi=1$]{
		\includegraphics[width=0.5\textwidth]{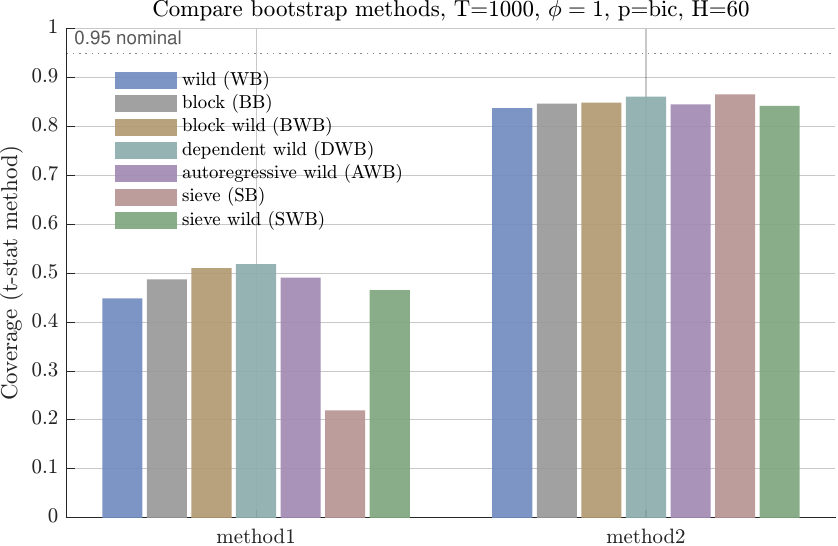}}\\
	\begin{figurenotes}
		\textit{Note}: Same setup as Figure~\ref{Fig-comp-boot-ct-h10} but for $H=60$. Differences
		across schemes widen at long horizons and high persistence ($\phi \to 1$), where some
		methods drift further from the 95\% nominal target.
	\end{figurenotes}
\end{figure}

\end{document}